\documentclass[11pt]{article}
\usepackage{graphicx,color}
\usepackage{subfigure}
\usepackage{amsmath, amsfonts, amssymb, amstext}

\def\bs{\boldsymbol}

\def\ol{\overline}

\def\bs{\boldsymbol}

\title{Semiparametric modeling of autonomous  nonlinear dynamical systems with applications}

\author{Debashis Paul\footnote{equal contributors}, Jie Peng$^*$ \& Prabir Burman
\\ \\
\textit{Department of Statistics, University of California, Davis}}

\date{}

\begin{document}



\maketitle
\begin{abstract}

In this paper, we propose a semi-parametric model for autonomous
nonlinear dynamical systems and devise an estimation procedure for
model fitting. This model incorporates subject-specific effects and
can be viewed as a nonlinear semi-parametric mixed effects model.
We also propose a computationally efficient model selection procedure.
We prove consistency of the proposed estimator under suitable
regularity conditions. We show by simulation studies that the proposed
estimation as well as model selection procedures can efficiently
handle sparse and noisy measurements. Finally, we apply the proposed
method to a plant growth data used to study growth displacement
rates within meristems of maize roots under two different experimental
conditions.

\end{abstract}


\section{Introduction}\label{sec:intro}

Continuous time dynamical systems arise, among other places, in
modeling certain biological processes. For example, in plant
science, the spatial distribution of growth is an active area of
research (Basu \textit{et al.}, 2007; Schurr, Walter and Rascher,
2006; van der Weele \textit{et al.}, 2003; Walter \textit{et al.},
2002). One particular region of interest is the root apex, which is
characterized by cell division, rapid cell expansion and cell
differentiation. A single cell can be followed over time, and thus
it is relatively easy to measure its cell division rate. However, in
a meristem\footnote{meristem is the tissue in plants consisting of
undifferentiated cells and found in zones of the plant where growth
can take place.}, there is a changing population of dividing cells.
Thus the cell division rate, which is defined as the local rate of
formation of cells, is not directly observable.
If one observes root development from an origin attached to the
apex, tissue elements appear to flow through, giving an analogy
between primary growth in plant root and fluid flow (Silk, 1994).
Thus in Sacks, Silk and Burman (1997), the authors propose to
estimate the cell division rates by a continuity equation that is
based on the principle of conservation of mass. Specifically, if we
assume a steady growth, then the cell division rate is estimated as
the gradient (with respect to distance) of cell flux -- the rate at
which cells are moving past a spatial point. Cell flux is the
product of cell number density and growth velocity field. The former
can be found by counting the number of cells per small unit file.
The latter is the rate of displacement of a particle placed along
the root and thus it is a function of distance from the root apex.
Hereafter we refer to it as the growth displacement rate. Note that,
growth displacement rate is not to be confused with ``growth rate''
which usually refers to the derivative of the growth trajectory with
respect to time. For more details, see Sacks \textit{et al.} (1997).
The growth displacement rate is also needed for understanding some
important physiological processes such as biosynthesis (Silk and
Erickson, 1979; Schurr \textit{et al.}, 2006). Moreover, a useful
growth descriptor called the ``relative elemental growth rate''
(REGR) can be calculated as the gradient of the growth displacement
rate (with respect to distance), which shows quantitatively the
magnitude of growth at each location within the organ.

\begin{figure}
\caption{Left Panel: image of root tip with meristem$^*$: 1 - meristem;
4 - root cap; 5 - elongation zone; Right Panel: an illustration of the root tip
with the displacements of three markers indicated at times $t_0, t_1, t_2, t_3$. {\tiny($^*$From wikipedia)}}
\label{figure:meristem}
\begin{center}
\begin{tabular}{ccccc}
\includegraphics[width=1.5in, angle=0]{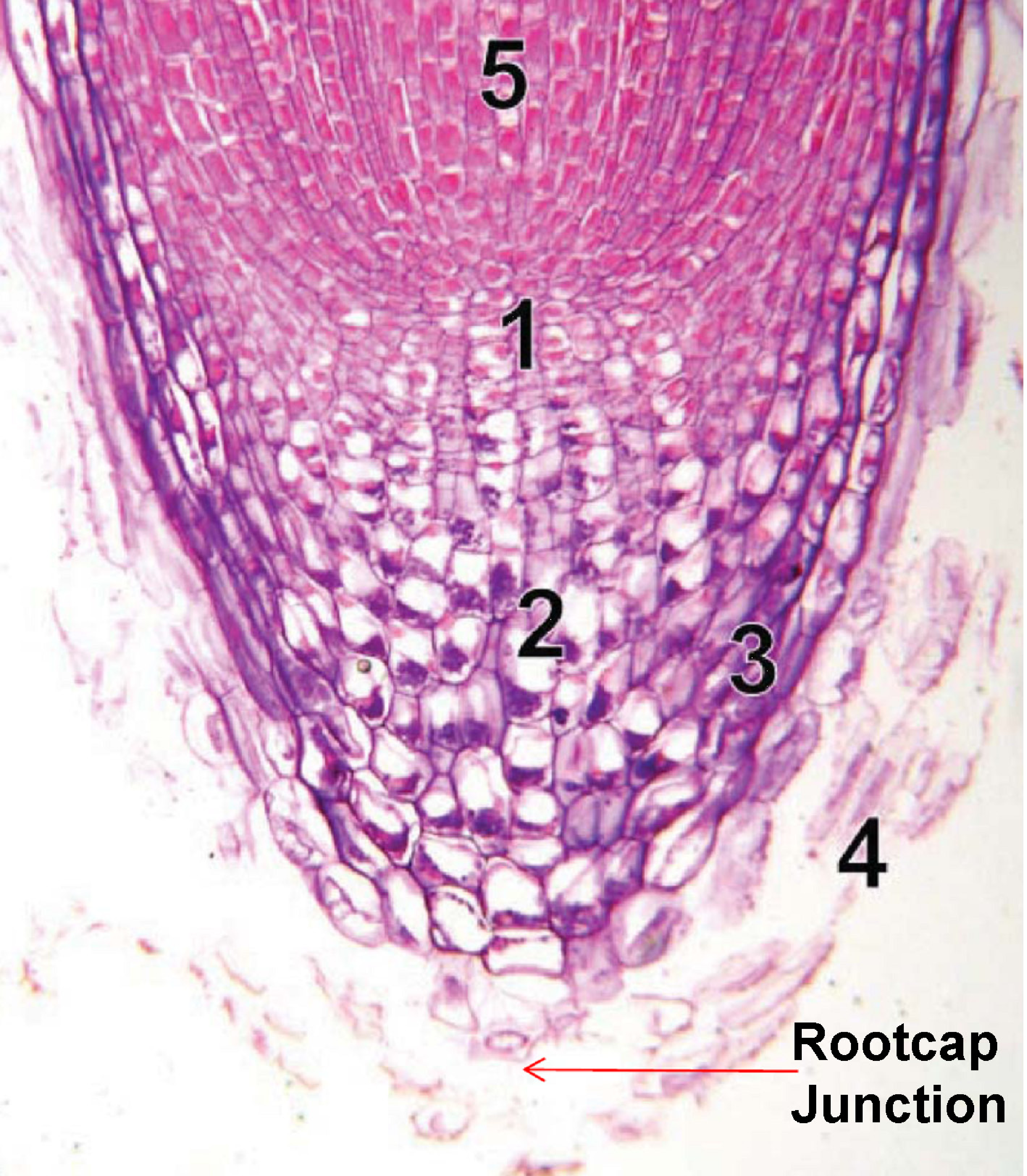} &&&&
\includegraphics[width=3in, angle=0]{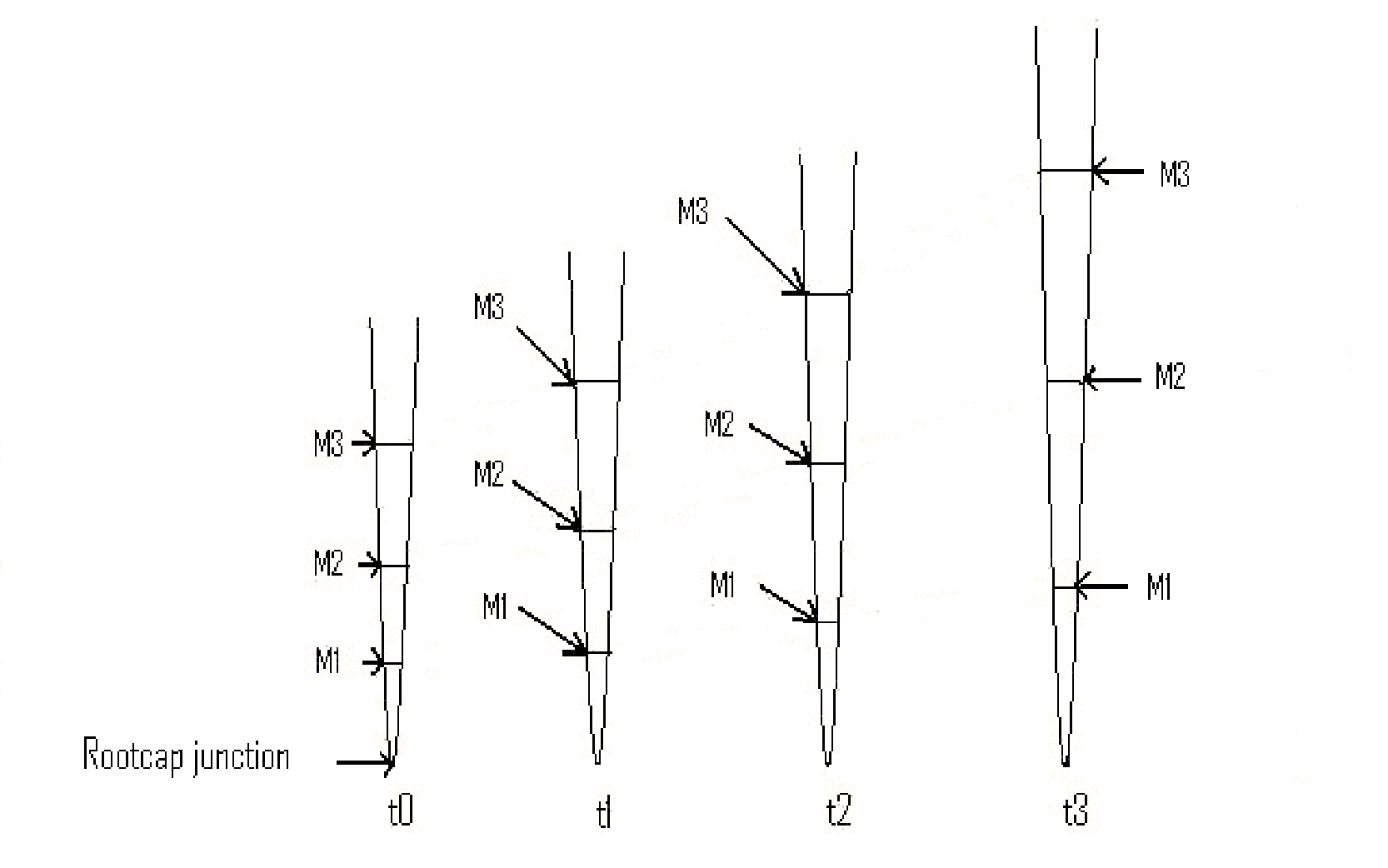}\\
\end{tabular}
\medskip
\end{center}
\end{figure}

\begin{figure}
\caption{Growth trajectories for plant data. Left panel : a plant in control
group; Right panel : a plant in treatment group} \label{figure:plant_sample}
\begin{center}
\begin{tabular}{cc}
\includegraphics[width=2.8in,height=3.1in,angle=270]{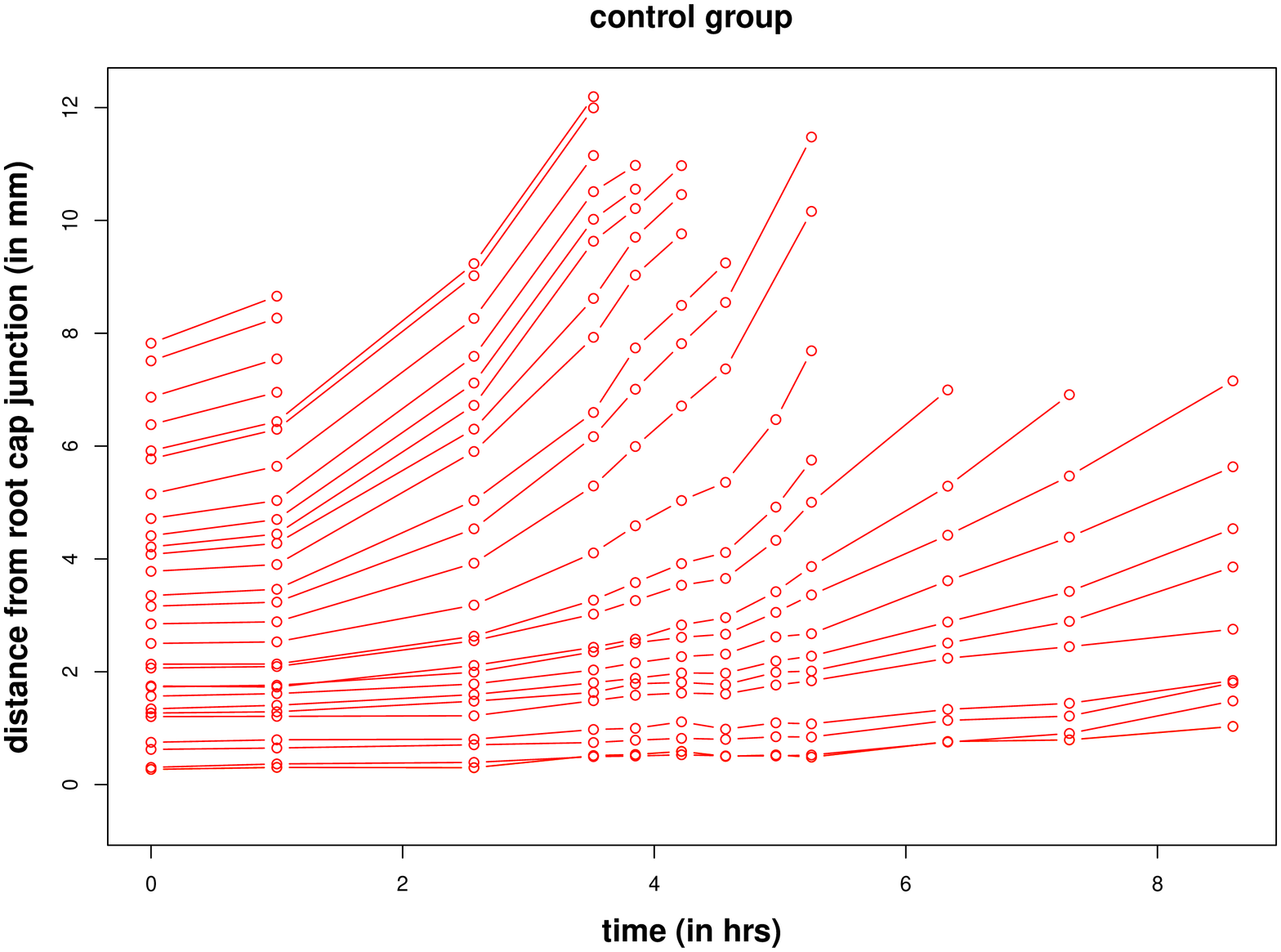} &
\includegraphics[width=2.8in,height=3.1in,angle=270]{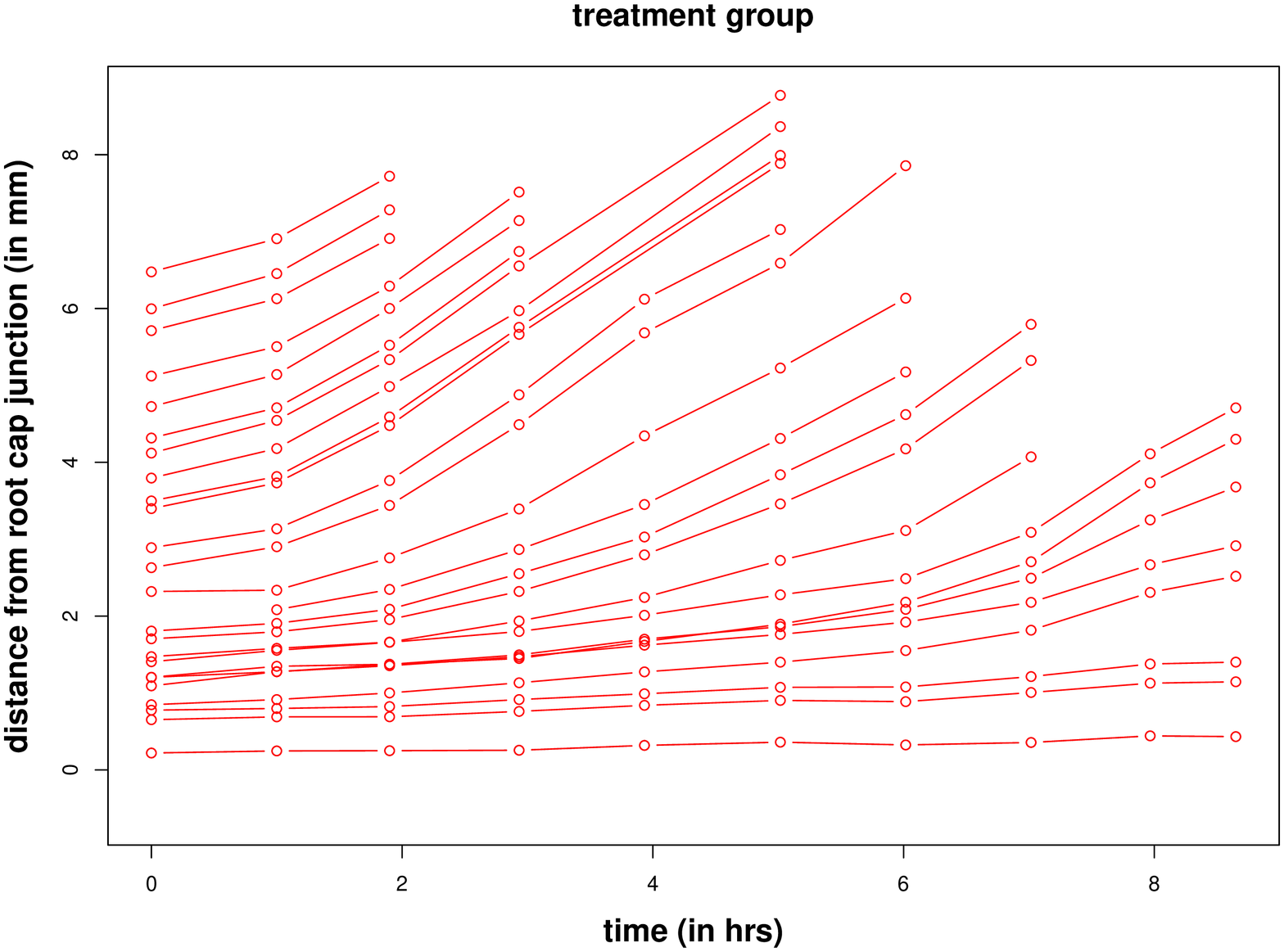}\\
\end{tabular}
\end{center}
\end{figure}

There are a lot of research aiming to understand the effect of
environmental conditions on  the growth in plant. For example, root
growth is highly sensitive to environmental factors such as
temperature, water deficit or nutrients ( Schurr \textit{et al.},
2006; Walter \textit{et al.}, 2002). For example, in Sharp, Silk and
Hsiao (1988), the authors study the effect of water potential on the
root elongation in maize primary roots. Root elongation has
considerable physiological advantages in drying soil, and therefore
knowledge of the locations and magnitudes of growth response to
water potential facilitates the quantitative understanding of the
underlying regulatory process. In Sacks \textit{et al.} (1997), an
experiment is conducted to study the effect of water stress on
cortical cell division rates through growth displacement rate within
the meristem of the primary root of maize seedlings.
In this study, for each plant, measurements are taken on the
displacement, measured as the distance in millimeters from the root
cap junction (root apex), of a number of markers on the root over a
period of $12$ hours (Fig. \ref{figure:meristem}: right panel). The
plants are divided into two groups - a control group under normal
water availability; and a treatment group under a water stress. In
Fig. \ref{figure:plant_sample}, the growth (displacement)
trajectories of one plant with $28$ markers in the control group,
and another plant with $26$ markers in the treatment group are
depicted. The meristem region of the root, where the measurements
are taken, is shown in Fig. \ref{figure:meristem} (left panel). Note
that, by definition, the growth displacement rate characterizes the
relationship between the growth trajectory and its derivative (with
respect to time). Thus it is simply the gradient function in the
corresponding dynamical system. (See Section \ref{sec:model} for
more details).


Motivated by this study, in this paper, we focus on modeling and
fitting the underlying dynamical system based on data measured over
time (referred as \textit{sample curves} or \textit{sample paths})
for a group of subjects. In particular, we are interested in the
case where there are multiple replicates corresponding to different
initial conditions for each subject. Moreover, for a given initial
condition, instead of observing the whole sample path, measurements
are taken only at a sparse set of time points together with
(possible) measurement noise. In the plant data application, each
plant is a subject. And the positions of the markers which are
located at different distances at time zero from the root cap
junction correspond to different initial conditions. There are in
total $19$ plants and $445$ sample curves in this study. The number
of replicates (i.e. markers) for each plant varies between $10$ and
$31$. Moreover, smoothness of the growth trajectories indicates low
observational noise levels and an absence of extraneous shocks in
the system. Hence, in this paper, we model the growth trajectories
through deterministic differential equations with plant-specific
effects. We refer to the (common) gradient function of these
differential equations as the baseline growth displacement rate.


We first give a brief overview of the
existing literature on fitting smooth dynamical systems in
continuous time. A large number of physical, chemical or biological
processes are modeled through systems of parametric differential
equations (cf. Ljung and Glad, 1994, Perthame, 2007, Strogatz,
2001).
Ramsay, Hooker, Campbell and Cao (2007) consider modeling a
continuously stirred tank reactor. Zhu and Wu (2007) adopt a state
space approach for estimating the dynamics of cell-virus
interactions in an AIDS clinical trial. Poyton \textit{et al.} (2006) use the principal
differential analysis approach to fit dynamical systems. Recently Chen and Wu (2008a,
2008b) propose to estimate differential equations with known functional
forms and nonparametric time-dependent coefficients.
Wu and Ding (1999) and Wu, Ding and DeGruttola (1998) propose using
nonlinear least squares procedure for fitting differential equations
that take into account subject-specific effects. In a recent work,
Cao, Fussmann and Ramsay (2008) model a nonlinear dynamical system
using splines with predetermined knots for describing the gradient
function. Most of the existing approaches assume known functional
forms of the dynamical system; and many of them
require data measured on a dense grid (e.g.,
Varah, 1982; Zhu and Wu, 2007).


For the problems that we are interested in this paper, measurements
are taken on a sparse set of points for each sample curve. Thus
numerical procedures for solving differential equations can become
unstable if we treat each sample curve separately.
Moreover, we are more interested in estimating the baseline dynamics
than the individual dynamics of each subject.
For example, in the plant study described above, we are interested in
comparing the growth displacement rates (as a function of distance from the root
cap junction) under two different experimental conditions. On
the other hand, we are not so interested in the displacement rate
corresponding to each plant.
Another important aspect in modeling data with multiple subjects is that adequate
measures need to be taken to model possible subject-specific effects, otherwise
the estimates of model parameters can have inflated variability.
Thus in this paper, we incorporate subject-specific effects into
the model while combining information across
different subjects.
In
addition, because of insufficient knowledge of the problem as is the
case for the plant growth study, in practice one often has to resort to
modeling the dynamical system nonparametrically.
For example, there is controversy among plant
scientists about whether there is a growth bump in the middle of the
meristem. There are also some natural boundary constraints of the
growth displacement rate, making it hard to specify a simple and
interpretable parametric system. (See more discussions in Section
\ref{sec:estimation}).
Therefore, in this paper, we propose to model the baseline dynamics
nonparametrrically through a basis representation approach. We use an
estimation procedure that combines nonlinear optimization techniques
with a numerical ODE solver to estimate the unknown parameters. In
addition, we derive a computationally efficient approximation of the
leave-one-curve-out cross validation score for model selection. We prove
consistency of the proposed estimators under appropriate regularity
conditions. Our asymptotic scenario involves keeping
the number of subjects fixed and allowing the number of measurements
per subject to grow to infinity. The analysis differs from
the usual nonparametric regression problems due to the structures imposed by the differential equations model.
We show by simulation studies that the proposed approach can efficiently estimate
the baseline dynamics under the setting of multiple replicates per subject with
sparse noisy measurements. Moreover, the proposed model selection procedure
is effective in maintaining a balance between fidelity to the data and to the
underlying model. Finally, we apply the proposed method to the plant data
described earlier and compare the estimated growth displacement rates under the two
experimental conditions.


The rest of paper is organized as follows. In Section \ref{sec:model},
we describe the proposed model. In Sections \ref{sec:estimation} and
\ref{sec:model_selection}, we discuss the model fitting and model
selection procedures, respectively. In Section \ref{sec:theory}, we
prove consistency of the proposed estimator. In Section
\ref{sec:simulation}, we conduct simulation studies to illustrate
finite sample performance of the proposed method. Section
\ref{sec:plant} is the application of this method to the plant data.
Technical details are in the appendices. An \texttt{R} package
\texttt{dynamics} for fitting the model described in this paper is
available upon request.

\section{Model}\label{sec:model}

In this section, we describe a class of autonomous dynamical
systems that is suitable for modeling the problems exemplified by
the plant data (Section \ref{sec:intro}). An autonomous dynamical
system has the following general form:
\begin{equation*}
X^{\prime}(t) = f(X(t)), ~~~ t \in [T_0,T_1].
\end{equation*}
Without loss of generality,
henceforth $T_0=0$ and $T_1 = 1$. Note that, the above equation means
that $X(t) = a + \int_0^t f(X(u)) du$, where $a = X(0)$ is
the initial condition. Thus in an autonomous system, the dynamics (which is characterized
by $f$) depends on time $t$ only through $X(t)$. This type of systems arises in various scientific studies
such as modelling prey-predator dynamics,
virus dynamics, or epidemiology (cf. Perthame, 2007).
Many studies in plant science such as Silk (1994), Sacks \textit{et al.} (1997),
Fraser, Silk and Rost (1990) all suggest reasonably steady growth velocity across
the meristem under both normal and water-stress conditions at an early
developmental stage. Moreover, exploratory regression analysis based
on empirical derivatives and empirical fits of the growth trajectories
indicates  that time is not a significant predictor and thus an
autonomous model is reasonable. This assumption is equivalent to the
assertion that the growth displacement rate depends only on the distance from
the root cap junction. It means that time zero does not play a role in
terms of estimating the dynamical system and there is also no additional
variation associated with individual markers.

\begin{figure}
\caption{Empirical derivatives (divided differences)
$\widehat X'(t)$ against empirical fits (averaged measurements) $\widehat X(t)$ for
treatment group.}
\label{figure:emperical_scatter}
\begin{center}
\includegraphics[width=3in,height=5in,angle=270]{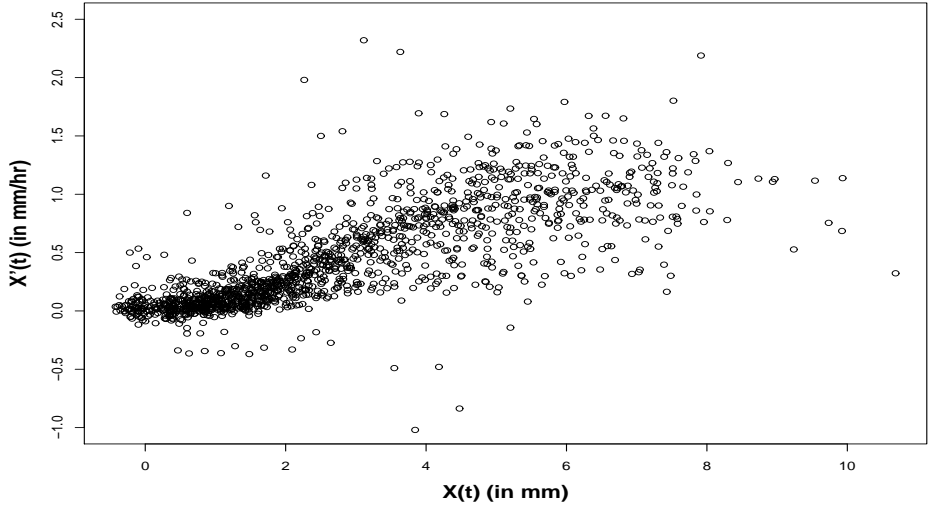}
\end{center}
\end{figure}

Figure \ref{figure:emperical_scatter} shows the scatter plot of
empirical derivatives versus empirical fits in the treatment group.
It indicates that there is an increase in the growth displacement
rate starting from a zero rate at the root cap junction, then
followed by a nearly constant rate beyond a certain location.
This means that growth stops beyond this point and the observed
displacements are due to growth in the part of the meristem closer
to the root cap junction. Where and how growth stops is of great
scientific interest. The scatter plot also indicates
excess variability towards the end which is probably caused by
plant-specific scaling effects.

Some of the features described above motivate us to consider the following
class of autonomous dynamical systems:
\begin{equation}\label{eq:basic}
X_{il}'(t) = g_i(X_{il}(t)),~~~ l=1,\cdots, N_i; i=1,\ldots,n,
\end{equation}
where $\{X_{il}(t):t \in [0,1],  l=1,\cdots, N_i; i=1,\ldots,n\}$ is
a collection of smooth curves corresponding to $n$ subjects, and
there are $N_i$ curves associated with the $i$-th subject. For
example, in the plant study, each plant is a subject
and each marker corresponds to one growth curve.
We assume that, all the curves associated with the same subject follow the same dynamics, and these are described
by the functions $\{g_i(\cdot)\}_{i=1}^n$.
We also assume that only a snapshot of each curve $X_{il}(\cdot)$ is
observed. That is, the observations are given by
\begin{equation}\label{eq:data_model}
Y_{ilj} = X_{il}(t_{ilj}) + \varepsilon_{ilj},~~j=1,\ldots,m_{il},
\end{equation}
where $0 \leq t_{il1} < \cdots < t_{ilm_{il}} \leq 1$ are the
observation times for the $l^{th}$ curve of the $i^{th}$ subject, and $\{\varepsilon_{ilj}\}$ are independently and
identically distributed noise with mean zero and variance
$\sigma^2_{\varepsilon} > 0$. In this paper, we model $\{g_i(\cdot)\}_{i=1}^n$ as:
\begin{equation}\label{eq:scale}
g_i(\cdot) = e^{\theta_i} g(\cdot),~~~i=1,\ldots,n,
\end{equation}
where
\begin{itemize}
\item[(1)]  the function $g(\cdot)$ reflects the common underlying mechanism regulating all these dynamical systems.
It is assumed to be a smooth function and is referred
as the {\it gradient function}. For the plant study, it represents
the baseline growth displacement rate for all plants within a given group (i.e., control vs. water-stress).
\item[(2)] $\theta_i's$ reflect subject-specific effects in these  systems.
The mean of $\theta_i$'s is assumed to be zero to impose identifiability. In the plant study,
$\theta_i's$ represent plant-specific scaling effects in the growth displacement rates for individual plants.
\end{itemize}

The simplicity and generality of this model
make it appealing for modeling a wide class of dynamical systems.
First, the gradient function $g(\cdot)$ can be an arbitrary smooth
function. If $g$ is nonnegative, and the initial conditions
$X_{il}(0)'s$ are also nonnegative, then the sample trajectories are
increasing functions, which encompasses growth models that are
autonomous.
Secondly, the scale parameter $e^{\theta_i}$ provides a
subject-specific tuning of the dynamics, which is flexible in
capturing variations of the dynamics in a population. In this paper,
our primary goal is to estimate the gradient function $g$
nonparametrically. For the plant data, the form of $g$ is not known
to the biologists, only its behavior at root cap junction and at
some later stage of growth are known (Silk, 1994). The fact that the
growth displacement rate increases from zero at root cap junction
before becoming a constant at a certain (unknown) distance away from
the root tip implies that a linear ODE model is apparently not
appropriate. Moreover,  popular parametric models such as the
Michaelis-Menten type either do not satisfy the boundary
constraints, and/or have parameters without clear interpretations in
the current context. On the other hand, nonparametric modeling
provides flexibility and is able to capture features of the
dynamical system which are not known to us \textit{a priori}
(Section \ref{sec:plant}). In addition, the nonparametric fit can be
used for diagnostics for lack of fit, if realistic parametric models
can be proposed.

The gradient function $g$ being smooth means that it can be well
approximated by a basis representation approach:
\begin{equation}\label{eq:basis}
g(x) = \sum_{k=1}^M \beta_k \phi_{k,M}(x)
\end{equation}
where $\phi_{1,M}(\cdot),\ldots,\phi_{M,M}(\cdot)$ are linearly
independent basis functions, chosen so that their combined support
covers the range of the observed trajectories. For example, we can
use cubic splines with a suitable set of knots. Thus, for a given
choice of the basis functions, the unknown parameters in the model
are the basis coefficients $\bs{\beta}
:=(\beta_1,\ldots,\beta_M)^T$, the scale parameters $\bs{\theta}
:=\{\theta_i\}_{i=1}^n$, and possibly the initial conditions $\bs{a}
:= \{a_{il} := X_{il}(0): l=1,\cdots,N_i\}_{i=1}^n$. Also, various
model parameters, such as the number of basis functions $M$ and the
knot sequence,  need to be selected based on the data. Therefore,
in essence, this is a nonlinear, semi-parametric, mixed effects
model.

In the plant data, $g$ is nonnegative and thus a modeling scheme
imposing this constraint  may be  more advantageous.
However, the markers are all placed  at a certain distance from the root
cap junction, where the growth displacement rate is already
positive, and the total number of measurements per plant is moderately
large. These mean that explicitly imposing nonnegativity is not
crucial for the plant data. Indeed, with the imposition of the
boundary constraints, the estimate of $g$ turns out to be
nonnegative over the entire domain of the measurements
(Section \ref{sec:plant}). In general, if $g$ is strictly positive
over the domain of interest, then we can model the logarithm of
$g$ by basis representation. Also, in this case,  the dynamical
system is stable in the sense that there is no bifurcation
phenomenon (Strogatz, 2001).

\section{Model Fitting}\label{sec:estimation}

In this section, we propose an iterative estimation procedure that imposes
regularization on the estimate of $\bs{\theta}$ and possibly
$\bs{a}$. One way to achieve this is to treat them as unknown random
parameters from some parametric distributions. Specifically, we use
the following set of working assumptions: (i) $a_{il}$'s are
independent and identically distributed as $N(\alpha,\sigma_a^2)$
and $\theta_i$'s are independent and identically distributed as
$N(0,\sigma_\theta^2)$, for some $\alpha \in \mathbb{R}$ and
$\sigma_a^2 > 0, \sigma_\theta^2>0$; (ii) the noise
$\varepsilon_{ilj}$'s are independent and identically distributed as
$N(0,\sigma_\varepsilon^2)$ for $\sigma_\varepsilon^2 > 0$; (iii)
the three random vectors $\bs{a},\ \bs{\theta}, \ \bs{\varepsilon}
:= \{\varepsilon_{ilj}\}$ are independent. Under these assumptions,
the negative joint log-likelihood of the observed data $Y
:=\{Y_{ilj}\}$, the scale parameters $\bs{\theta}$ and the initial
conditions $\bs{a}$ is, up to an additive constant and a positive
scale constant,
\begin{equation}\label{eq:log-like}
\sum_{i=1}^n \sum_{l=1}^{N_i}\sum_{j=1}^{m_{il}}
[Y_{ilj} - \widetilde X_{il}(t_{ilj};a_{il},\theta_i,\beta)]^2 + \lambda_1
\sum_{i=1}^n\sum_{l=1}^{N_i} (a_{il} - \alpha)^2 + \lambda_2 \sum_{i=1}^n \theta_i^2,
\end{equation}
where $\lambda_1 = \sigma_\varepsilon^2/\sigma_a^2$, $\lambda_2 =
\sigma_\varepsilon^2/\sigma_\theta^2$, and $\widetilde
X_{il}(\cdot)$ is the trajectory determined by $a_{il}$, $\theta_i$,
and $\bs\beta$.  This can be viewed as a hierarchical maximum
likelihood approach (Lee, Nelder and Pawitan, 2006), which is considered to be a
convenient alternative to the full (restricted) maximum likelihood
approach. Define
$$
\ell_{ilj}(a_{il},\theta_i,\bs\beta) :=[Y_{ilj} -
\widetilde X_{il}(t_{ilj};a_{il},\theta_i,\bs{\beta})]^2 + \lambda_1
(a_{il} - \alpha)^2/m_{il} + \lambda_2
\theta_i^2/\sum_{l=1}^{N_i}m_{il}~.
$$
Then the loss function in (\ref{eq:log-like}) equals $\sum_{i=1}^n
\sum_{l=1}^{N_i}\sum_{j=1}^{m_{il}}\ell_{ilj}(a_{il},\theta_i,\bs{\beta})$.
Note that the above distributional assumptions are simply working assumptions.
The expression in (\ref{eq:log-like}) can also be viewed as a regularized $\ell_2$
loss with penalties on the variability of $\bs\theta$ and
$\bs{a}$.
For the plant data, the initial conditions (markers) are chosen according to
some fixed experimental design, thus it is natural to treat them as
fixed effects. Moreover, it does not seem appropriate to shrink the
estimates toward some common value in this case. Thus in Section
\ref{sec:plant}, we set $\lambda_1=0$ when estimating $\bs{a}$. For
certain other problems, treating the initial conditions as random
effects may be more suitable. For example, Huang, Liu and Wu (2006) study a
problem of HIV dynamics where the initial conditions are
subject-specific and unobserved.

In many situations, there are boundary constraints on the
gradient function $g$. For example, according to plant science, both the
growth displacement rate and its derivative at the root cap junction
should be zero. Moreover, it should become a constant at a certain
(unknown) distance from the root cap junction. Thus for the plant
data, it is reasonable to assume that,
$g(0) = 0 = g^{\prime}(0)$ and $g^{\prime}(x)=0$ for $x \geq A$ for a given $A>0$.
The former can be implemented by an appropriate choice of the basis
functions. For the latter, we consider constraints of the form:
$\bs\beta^T \mathbf{B} \bs\beta$ for an $M \times M$ positive semi-definite matrix
$\mathbf{B}$, which can be thought of as an $\ell_2$-type constraint on some
derivative of $g$. (See Section \ref{sec:plant} for the specification of $\mathbf{B}$).
Consequently, the modified objective function becomes
\begin{equation}\label{eq:objective}
L(a,\bs\theta,\bs\beta) :=\sum_{i=1}^n
\sum_{l=1}^{N_i}
\sum_{j=1}^{m_{il}}\ell_{ilj}(a_{il},\theta_i,\bs\beta) +
\bs\beta^T \mathbf{B} \bs\beta.
\end{equation}
The proposed estimator is then the minimizer of the objective
function:
\begin{eqnarray}\label{eq:estimator}
(\widehat{\bs{a}}, \widehat{\bs{\theta}},
\widehat{\bs{\beta}}) :={\rm arg}\min_{\bs{a},\bs\theta,\bs\beta}L(\bs{a},\bs\theta,\bs\beta).
\end{eqnarray}
Note that,  here our main interest is the gradient function $g$.
Thus estimating the parameters of the dynamical system together with
the sample trajectories and their derivatives simultaneously is most efficient.
In contrast, if the trajectories and their derivatives  are first
obtained via pre-smoothing (as is done for example in
Chen and Wu (2008a, 2008b),  Varah (1982)), and then used
in a nonparametric regression framework to obtain $g$, it will be
inefficient in estimating $g$. This is because, errors introduced in the pre-smoothing step
cause loss of information which is not retrievable later on,
and also information regarding $g$ is not efficiently combined across curves.

In the following, we propose a numerical procedure
for solving (\ref{eq:estimator}) that has two main ingredients:
\begin{itemize}
\item
Given  $(\bs{a},\bs\theta,\bs\beta)$, reconstruct the
trajectories $\{\widetilde X_{il}(\cdot):
l=1,\cdots,N_i\}_{i=1}^n$ and their derivatives. This step can
be carried out using a numerical ODE solver, such as the
$4^{th}$ order Runge-Kutta method (cf. Tenenbaum and Pollard, 1985).

\item
Minimize (\ref{eq:objective}) with respect to
$(\bs{a},\bs\theta,\bs\beta)$. This amounts to a nonlinear least
squares problem (Bates and Watts, 1988). It can be carried out
using either a nonlinear least squares solver, like the
Levenberg-Marquardt method; or a general optimization procedure,
such as the Newton-Raphson algorithm.
\end{itemize}
The above procedure bears some similarity to
the local, or gradient-based, methods discussed in
Miao \textit{et al.} (2008).

We now briefly describe an optimization procedure based on the
\textit{Levenberg-Marquardt method} (cf. Nocedal and Wright, 2006).
For notational convenience, denote the current estimates by
$\boldsymbol{a}^* :=\{a_{il}^*\}$, $\boldsymbol{\theta}^*
:=\{\theta_i^*\}$ and $\bs{\beta}^*$, and define the current
residuals as: $\tilde\varepsilon_{ilj} = Y_{ilj} - \widetilde
X_{il}(t_{ilj};a_{il}^*,\theta_i^*,\bs{\beta}^*)$.
For each $i=1,\cdots, n$ and  $l=1,\cdots,N_i$, define the $m_{il}
\times 1$ column vectors
\begin{equation*}
\mathbf{J}_{il,a_{il}^*} := \left(\frac{\partial}{\partial a_{il}} \widetilde
X_{il}(t_{ilj};a_{il}^*,\theta_i^*,\bs{\beta}^*)\right)_{j=1}^{m_{il}},
~~~ \boldsymbol{\widetilde\varepsilon}_{il}=\left(\widetilde \varepsilon_{ilj}\right)_{j=1}^{m_{il}}.
\end{equation*}
For each $i=1,\cdots,n$, define the $m_{i\cdot} \times 1$ column
vectors
\begin{equation*}
\mathbf{J}_{i,\theta_i^*} = \left(\frac{\partial}{\partial \theta_i}
\widetilde X_{il}(t_{ilj};a_{il}^*,\theta_i^*,\bs{\beta}^*)
\right)_{j=1, l=1}^{m_{il},N_i}; ~~~ \boldsymbol{\widetilde\varepsilon}_{i}
= \left(\widetilde \varepsilon_{ilj}\right)_{j=1, l=1}^{m_{il},N_i},
\end{equation*}
where $m_{i\cdot}: =\sum_{l=1}^{N_i} m_{il}$ is the total number of
measurements of the $i^{th}$ cluster. Finally, for each $k=1,\cdots, M$,
define the $m_{\cdot\cdot} \times 1$ column vectors:
\begin{equation*}
\mathbf{J}_{\beta_k^*} = \left(\frac{\partial}{\partial \beta_k} \widetilde
X_{il}(t_{ilj};a_{il}^*,\theta_i^*,\bs{\beta}^*)\right)_{j=1,l=1,i=1}^{m_{il},N_i,n};
~~~ \boldsymbol{\widetilde\varepsilon}=\left(\tilde \varepsilon_{ilj}\right)_{j=1,l=1,i=1}^{m_{il},N_i,n},
\end{equation*}
where $m_{\cdot\cdot} := \sum_{i=1}^n\sum_{l=1}^{N_i} m_{il}$ is the
total number of measurements.
Note that, given $\boldsymbol{a}^*$,$\boldsymbol{\theta}^*$ and
$\bs{\beta}^*$, the trajectories  $\{\widetilde X_{il}\}'s$ and
their gradients (as well as Hessians)
can be easily evaluated on a fine grid by using
numerical ODE solvers such as the $4^{th}$ order
Runge-Kutta method as mentioned above (see Appendix A).

We break the updating step into three parts corresponding to the
three different sets of parameters.  For each set of parameters,
we first derive a first order Taylor expansion of the
curves $\{\widetilde X_{il}\}$ around the current values of these
parameters and then update them by a least squares fitting, while
keeping the other two sets of parameters fixed at the current
values. The equation for updating $\bs\beta$,
while keeping $\bs{a}^*$ and $\bs{\theta}^*$ fixed, is
$$
\left[J_{\bs{\beta}^*}^T J_{\bs{\beta}^*} +
\lambda_3~ \mbox{diag}(J_{\bs{\beta}^*}^T
J_{\bs{\beta}^*}) + \mathbf{B} \right](\bs\beta - \bs\beta^*) =
J_{\bs\beta^*}^T \widetilde{\bs\varepsilon} - \mathbf{B}\bs\beta^*,
$$
where
$J_{\bs\beta^*}:=(J_{\beta_1^*}:\cdots:J_{\beta_M^*})$
is an  $m_{\cdot\cdot}  \times M$ matrix. Here $\lambda_3$ is a sequence of
positive constants converging to zero as the number of iterations
increases. They are used to avoid possible singularities in the system of
equations.
The normal equation for updating $\theta_i$ is
\begin{equation}
\label{eq:theta_diff}
(\mathbf{J}_{i,\theta_i^*}^T \mathbf{J}_{i,\theta_i^*} +
\lambda_2)(\theta_i-\theta_i^*) = \mathbf{J}_{i,\theta_i^*}^T
\boldsymbol{\widetilde\varepsilon}_{i} - \lambda_2 \theta_i^*, ~~~i=1,\ldots,n.
\end{equation}
The equation for updating $a_{il}$ is derived similarly, while keeping $\theta_i$
and $\bs{\beta}$ fixed at $\theta_i^*$, $\bs{\beta}^*$:
\begin{equation}
\label{eq:a_diff} (\mathbf{J}_{il,a_{il}^*}^T \mathbf{J}_{il,a_{il}^*} +
\lambda_1)(a_{il}-a_{il}^*) = \mathbf{J}_{il,a_{il}^*}^T
\boldsymbol{\widetilde\varepsilon}_{il} + \lambda_1 \alpha^*_{il}, ~~~l=1,\cdots, N_i, ~~i=1,\cdots,n,
\end{equation}
where $\alpha^* =
\sum_{i=1}^n \sum_{l=1}^{N_i}a_{il}^*/N_{\cdot}, ~~
\alpha_{il}^* = \alpha^* - a_{il}^*$ with $N_{\cdot}:=\sum_{i=1}^n N_i$ being the total number of sample curves.

In summary, this procedure begins by
taking initial estimates
and then iterates by cycling through
the updating steps for $\bs{\beta}$, $\boldsymbol{\theta}$ and
$\boldsymbol{a}$ until convergence. The initial estimates can be
conveniently chosen. For example, $a_{il}^{ini}=Y_{il1}$,
$\theta_i^{ini} \equiv 0$; or $a_{il}^{ini}
\equiv \frac{1}{N_{\cdot}} \sum_{i=1}^n\sum_{l=1}^{N_i}
Y_{i1l}$.
Even though the model is
identifiable, in practice, for small $n$, there can be drift in the estimates of
$\theta_i$ and $g$ due to flatness of the objective function in some
regions. To avoid this and increase stability, we also impose the
condition that $\sum_{i=1}^n \theta_i^{\ast} = 0$. This can be
easily achieved by subtracting $\bar{\theta}^{*} :=
\frac{1}{n}\sum_{i=1}^n\theta_i^{\ast}$ from $\theta_i^{\ast}$ at
each iteration after updating $\{\theta_i\}$.


All three updating steps described above are based on the general principle of
Levenberg-Marquardt algorithm by the linearization of the curves
$\{\widetilde X_{il}\}$ (see Appendix B). However, the tuning
parameter $\lambda_3$ plays a different role than the penalty
parameters $\lambda_1$ and $\lambda_2$.
The parameter $\lambda_3$ is used to stabilize the
updates of $\bs{\beta}$ and thereby facilitate
convergence. Thus it needs to decrease to zero with increasing
iterations in order to avoid introducing bias in the estimate. There
are ways of implementing this adaptively (see e.g. Nocedal
and Wright, 2006, Ch. 10). In this paper, we use a simple
non-adaptive method: $\lambda_{3j} = \lambda_3^0 / j$ for the $j$-th
iteration, for some pre-specified $\lambda_3^0 > 0$. On the other
hand, $\lambda_1$ and $\lambda_2$ are parts of the penalized loss
function (\ref{eq:objective}). Their main role is to control the
bias-variance trade-off of the estimators, even though they also
help in regularizing the optimization procedure.
From the likelihood view point, $\lambda_1$, $\lambda_2$
are determined by the variances $\sigma^2_\varepsilon$, $\sigma^2_a$
and $\sigma^2_\theta$. After each loop over all the parameter
updates, we can estimate these variances from the current residuals and
current values of $\boldsymbol{a}$ and $\boldsymbol{\theta}$.
By assuming that $m_{il} > 2$ for each pair $(i,l)$,
\begin{eqnarray*}
\widehat\sigma_\varepsilon^2 &=&
\frac{1}{m_{\cdot\cdot}-N_{\cdot}-n-M}\sum_{i=1}^n
\sum_{l=1}^{N_i}\sum_{j=1}^{m_{il}} \widetilde
\varepsilon_{ilj}^2,\\
\widehat \sigma_a^2 &=& \frac{1}{N_{\cdot}-1} \sum_{i=1}^n
\sum_{l=1}^{N_i} (a_{il}^* - \alpha^*)^2,~~~~
\widehat\sigma_\theta^2 = \frac{1}{n-1}\sum_{i=1}^n (\theta_i^*)^2.
\end{eqnarray*}
We can then plug in the estimates $\widehat\sigma_\varepsilon^2$,
$\widehat \sigma_a^2$ and $\widehat \sigma_\theta^2$ to get new
values of $\lambda_1$ and $\lambda_2$ for the next iteration. On the
other hand, if we take the penalized loss function view point, we
can simply treat $\lambda_1$, $\lambda_2$ as fixed regularization
parameters, and then use a model selection approach to select
their values based on data. In the following sections, we refer
the method as \texttt{adaptive} if $\lambda_1$, $\lambda_2$ are
updated after each iteration; and refer the method as
\texttt{non-adaptive} if they are kept fixed throughout the
optimization.

The Levenberg-Marquardt method is quite stable and robust to the
initial estimates. However, it converges slowly in the neighborhood
of the minima of the objective function.
On the other hand, the Newton-Raphson algorithm has a very fast
convergence rate when starting from estimates that are already near
the minima. Thus, in practice the we first use the
Levenberg-Marquardt approach to obtain a
reasonable estimate, and then use the Newton-Raphson
algorithm to expedite the search of the minima. The implementation
of the Newton-Raphson algorithm of the current problem is standard
and is outlined in Appendix C.

\section{Model Selection }\label{sec:model_selection}

After specifying a scheme for the basis
functions $\{\phi_{k,M}(\cdot)\}$, we still need to determine various model
parameters such as the number of basis functions $M$, the knot sequence, etc.
In the literature AIC/BIC/AICc criteria
have been proposed for model selection while estimating dynamical
systems with nonparametric time-dependent components
(e.g. Miao \textit{et al.}, 2008).
Here, we propose an approximate leave-one-curve-out
cross-validation score for model selection. Under the current
context, the leave-one-curve-out CV score is defined as
\begin{equation}\label{eq:CV}
CV := \sum_{i=1}^n\sum_{l=1}^{N_{i}}\sum_{j=1}^{m_{il}}
\ell_{ilj}^{cv}(\widehat a_{il}^{(-il)}, \widehat
\theta_{i}^{(-il)}, \widehat{\bs{\beta}}^{(-il)})
\end{equation}
where  $\widehat \theta_{i}^{(-il)}$ and
$\widehat{\bs{\beta}}^{(-il)}$ are estimates of $\theta_{i}$ and
$\bs{\beta}$, respectively, based on the data after dropping the
$l^{th}$ curve in the $i^{th}$ cluster; and $\widehat
a_{il}^{(-il)}$ is the minimizer of $\sum_{j=1}^{m_{il}}
\ell_{ilj}(a_{il},\widehat \theta_{i}^{(-il)},
\widehat{\bs{\beta}}^{(-il)})$ with respect to $a_{il}$. The function
$\ell^{cv}_{ilj}$ is a suitable criterion function for cross validation.
Here,  we use the prediction error loss:
$$
\ell_{ilj}^{cv}(a_{il},\theta_i,\bs{\beta}) :=\left(Y_{ilj} -
\widetilde X_{il}(t_{ilj};a_{il},\theta_i,\bs{\beta})\right)^2.
$$
Calculating CV score (\ref{eq:CV}) is computationally very
demanding.
Therefore, we propose to approximate
$\widehat \theta_{i}^{(-il)}$ and $\widehat{\bs{\beta}}^{(-il)}$  by
a first order Taylor expansion around the estimates  $\widehat
\theta_{i}, \widehat{\bs{\beta}}$ based on the full data.
Consequently we derive an approximate CV score which is computationally
inexpensive. A similar approach is taken in Peng and Paul (2009) under
the context of functional principal component analysis.
Observe that, when evaluated at the estimate
$\boldsymbol{\widehat{a}}$, $\boldsymbol{\widehat{\theta}}$ and
$\widehat{\bs{\beta}}$ based on the full data,
\begin{equation}
\label{eq:CV_first}
\frac{\partial}{\partial \theta_i}\left(\sum_{l,j} \ell_{ilj}^{cv}\right)+2\lambda_2\theta_i=0, ~~~i=1,\cdots,n; ~~~
\frac{\partial}{\partial \bs{\beta}}\left(\sum_{i,l,j}
\ell_{ilj}^{cv}\right)+2\mathbf{B}\bs{\beta}=0.
\end{equation}
Whereas, when evaluated at the drop $(i,l)$-estimates: $\widehat
a_{il}^{(-il)}, \widehat \theta_{i}^{(-il)},
\widehat{\bs{\beta}}^{(-il)}$,
\begin{equation}\label{eq:CV_second}
\frac{\partial}{\partial \theta_i}\left(\sum_{l^*, j: l^* \neq l} \ell_{i l^* j}^{cv}\right)
+ 2 \lambda_2\theta_i=0; ~~~\frac{\partial}{\partial \bs{\beta}}\left(\sum_{i^*,
l^*, j: (i^*,l^*) \neq (i,l)} \ell_{i^* l^* j}^{cv} \right)+2\mathbf{B}\bs{\beta}=0.
\end{equation}
Expanding the left hand side of (\ref{eq:CV_second}) around
$\widehat{\bs{\beta}}$, we obtain
\begin{eqnarray*}
0 &\approx& \sum_{i^*, l^*, j: (i^*,l^*) \neq (i,l)} \frac{\partial}
{\partial \bs{\beta}}\ell_{i^* l^* j}^{cv}
\left|_{\widehat{\bs{\beta}}}\right. + 2\mathbf{B}\widehat{\bs{\beta}}+
\left[\sum_{i^*, l^*, j: (i^*,l^*) \neq (i,l)} \frac{\partial^2}{\partial \bs{\beta}
\partial \bs{\beta}^T}\ell_{i^* l^* j}^{cv}\left|_{\widehat{\bs{\beta}}}\right. + 2\mathbf{B} \right]
(\widehat{\bs{\beta}}^{(-il)} - \widehat{\bs{\beta}})
\\
&\approx& - \sum_{j=1}^{m_{il}} \frac{\partial \ell^{cv}_{ilj}}{\partial
\bs{\beta}}\left|_{(\widehat{a}_{il}, \widehat{\theta}_{i},\widehat{\bs{\beta}})}\right. + \left[\sum_{i^*, l^*, j:
(i^*,l^*) \neq (i,l)} \frac{\partial^2 }{\partial \bs{\beta}
\partial \bs{\beta}^T}\ell_{i^* l^* j}^{cv}
\left|_{(\widehat{a}_{i^*l^*}, \widehat{\theta}_{i^*},\widehat{\bs{\beta}})}\right. + 2\mathbf{B} \right]
(\widehat{\bs{\beta}}^{(-il)} - \widehat{\bs{\beta}}),
\end{eqnarray*}
where in the second step we invoked (\ref{eq:CV_first}) and approximated
$\{\widehat a_{il}^{(-il)}\}, \{\widehat \theta_{i}^{(-il)}\}$ by
$\{\widehat{a}_{il}\}$, $\{\widehat{\theta}_i\}$, respectively. Similar
calculations are carried out for $\widehat\theta_{i}^{(-il)}$.
Thus we obtain the following first order approximations:
\begin{eqnarray}
\label{eq:CV_est_approx}
\widehat \theta_{i}^{(-il)} &\approx& \widetilde \theta_{i}^{(-il)} := \widehat \theta_{i} +
\left[\sum_{l'=1}^{N_{i}} \sum_{j'=1}^{m_{il'}}\frac{\partial^2
\ell_{il'j'}^{cv}}{\partial \theta_{i}^2}+2\lambda_2 \right]^{-1}
\sum_{j=1}^{m_{il}} \left(\frac{\partial
\ell_{ilj}^{cv}}{\partial \theta_i}\right)\nonumber\\
 \widehat{\bs{\beta}}^{(-il)} &\approx& \widetilde{\bs{\beta}}^{(-il)}
 := \widehat{\bs{\beta}}+
\left[\sum_{i'=1}^n \sum_{l'=1}^{N_{i'}} \sum_{j'=1}^{m_{i'l'}}
\frac{\partial^2 \ell_{i'l'j'}^{cv}}{\partial \bs{\beta} \partial
\bs{\beta}^T}+2\mathbf{B}\right]^{-1} \left(\sum_{j=1}^{m_{il}}
\frac{\partial \ell_{ilj}^{cv}}{\partial
\bs{\beta}}\right).
\end{eqnarray}
These gradients and Hessians are all evaluated at
$(\widehat {\boldsymbol{a}}, \widehat {\boldsymbol{\theta}},
\widehat{\bs{\beta}})$, and thus they have already been computed (on a fine grid) in the course of
obtaining these estimates. Thus, there is almost no additional
computational cost to obtain these approximations. Now for
$i=1,\ldots,n; l=1,\cdots,N_i$, define
\begin{equation}
\widetilde a_{il}^{(-il)}= \arg\min_{a} \sum_{j=1}^{m_{il}} (Y_{ilj}
- \widetilde X_{il}(t_{ilj};a,\widetilde{\theta_i}^{(-il)},
\widetilde{\bs{\beta}}^{(-il)}))^2 + \lambda_1(a - \widehat
\alpha)^2,
\end{equation}
where $\widehat{\alpha}$ is the estimator of $\alpha$ obtained from
the full data. Finally, the approximate
leave-one-curve-out cross-validation score is
\begin{equation}
\label{eq:approx_CV}
 \widetilde{CV} := \sum_{i=1}^n \sum_{l=1}^{N_i} \sum_{j=1}^{m_{il}}\ell^{cv}_{ilj}(\widetilde
a_{il}^{(-il)},\widetilde \theta_i^{(-il)},\widetilde{\bs{\beta}}^{(-il)}).
\end{equation}


\section{Asymptotic Theory}\label{sec:theory}

In this section, we present a result on the consistency of
the proposed estimator of $g$ under suitable technical
conditions. We assume that the number of subjects $n$ is fixed; and the
number of measurements per curve $m_{il}$, and number of curves
$N_i$ per subject, increase to infinity together. When $n$ is fixed, the
asymptotic analysis is similar irrespective of whether $\theta_i$'s
are viewed as fixed effects or random effects. Hence, for simplicity,
we treat $\theta_i$'s as fixed effects and impose the identifiability
constraint $\theta_1 = 0$. Due to this restriction,
we modify the loss function (\ref{eq:log-like}) slightly by replacing the penalty
$\lambda_2 \sum_{i=1}^n \theta_i^2$ with $\lambda_2 \sum_{i=2}^n
(\theta_i - \overline{\theta})^2$
where $\overline{\theta} = \sum_{i=2}^n \theta_i/(n-1)$.
Moreover, since $n$ is finite, in practice we can relabel the subjects so that
the curves corresponding to subject 1 has the highest rate of growth,
and hence $\theta_i \leq 0$ for all $i > 1$. This relabeling is not necessary
but simplifies the arguments considerably.

Moreover, to be consistent with the
setting of the plant data, we focus on the case where the time points
for the different curves corresponding to the same subject
are the same, so that, in particular, $m_{il} \equiv m_i$. We assume that the
time points come from a common continuous distribution $F_T$. We also assume that
the gradient function $g(x)$ is positive for $x > 0$ and is defined
on a domain $D = [x_0,x_1] \subset \mathbb{R}^+$; and the initial conditions
$\{a_{il} := X_{il}(0)\}'s$ are observed (and hence $\lambda_1 =0$)
and are randomly chosen from a common continuous distribution $F_a$
with support $[x_0,x_2]$ where $x_2 < x_1$.

Before we state the regularity conditions required for proving
the consistency result, we highlight two aspects of the asymptotic analysis.
Note that, the current problem differs from
standard semiparametric nonlinear mixed effects models. First,
the estimation of $g$ is an inverse problem, since it implicitly
requires knowledge of the derivatives of the trajectories of the
ODE which are not directly observed. The degree of ill-posedness is
quantified by studying the behavior
of the expected Jacobian matrix of the sample trajectory with respect to $\bs\beta$. This
matrix would be well-conditioned under a standard nonparametric
function estimation context. However, in the current case, its
condition number goes to infinity with the dimension of the model
space $M$. Secondly, unlike in standard nonparametric
function estimation problems where the effect of the estimation
error is localized, the estimation error propagates throughout
the entire domain of $g$ through the dynamical system. Therefore,
sufficient knowledge of the behavior of $g$ at the boundaries is imperative.

We assume the following:
\begin{itemize}
\item[{\bf A1}] $g \in C^p(D)$ for some integer $p \geq 4$, where $D = [x_0,x_1] \subset \mathbb{R}^+$.


\item[{\bf A2}] $\theta_i$'s are fixed parameters with $\theta_1 =
0$.

\item[{\bf A3}] The collection of basis functions $\Phi_M := \{\phi_{1,M},\ldots,
\phi_{M,M}\}$ satisfies: (i) $\phi_{k,M} \in C^2(D)$ for all $k$;
(ii) $\sup_{x \in D} \sum_{k=1}^M |\phi_{k,M}^{(j)}(x)|^2 = O(M^{1+2j})$, for $j=0,1,2$;
(iii) for every $k$, the length of the support of $\phi_{k,M}$ is $O(M^{-1})$;
(iv) for every $M$, there is a $\bs{\beta}^*\in \mathbb{R}^M$ such that $\parallel g - \sum_{k=1}^M
\beta_k^* \phi_{k,M}\parallel_{L^\infty(D)} = O(M^{-2p})$;
$\parallel g' - \sum_{k=1}^M \beta_k^* \phi_{k,M}'\parallel_{L^\infty(D)} = O(M^{-c})$,
for some $c > 0$; $\sum_{k=1}^M \beta_k^* \phi_{k,M}''$
is Lipschitz with Lipschitz constant $O(M)$; and
$\parallel \sum_{k=1}^M \beta_k^* \phi_{k,M}''\parallel_{L^\infty(D)}
= O(1)$.

\item[{\bf A4}] $X_{il}(0)$'s are i.i.d. from
a continuous distribution $F_a$. Denote $\mbox{supp}(F_a) = [x_0,x_2]$
and let $\Theta$ be a fixed,
open interval containing the true $\theta_i$'s, denoted by $\theta_i^*$.
Then there exists a $\tau > 0$
such that for all $a \in F_a$ and
for all $\theta \in \Theta$,
the initial value problem
\begin{equation}\label{eq:x_theta_f}
x'(t) = e^{\theta} f(x(t)),~~~x(0) = a
\end{equation}
has a solution $x(t) := x(t;a,\theta, f)$ on $[0,1]$ for all
$f \in {\cal M}(g,\tau)$, where
\begin{equation*}
{\cal M}(g,\tau) := \{f \in C^1(D) : \parallel f - g
\parallel_{1,D} \leq \tau\}.
\end{equation*}
Moreover, the range of $x(\cdot; \cdot, \cdot, f)$ (as a mapping
from $[0,1] \times \mbox{supp}(F_a) \times \Theta$) is contained in $D \pm \epsilon(\tau)$
for some $\epsilon(\tau) > 0$ (with $\lim_{\tau \to 0} \epsilon(\tau) = 0$) for all $f \in {\cal M}(g,\tau)$.
Here, $\parallel \cdot
\parallel_{1,D}$ is the seminorm defined by $\parallel f
\parallel_{1,D} = \parallel f
\parallel_{L^\infty(D)} + \parallel f' \parallel_{L^\infty(D)}$.
Furthermore, the range of $x(\cdot; \cdot, 0, g)$ contains
$D$.

\item[{\bf A5}] For each $i=1,\ldots,n$, for all $l=1,\ldots,
N_i$, the time points $t_{ilj}$
($j=1,\ldots,m_i$) belong the
set $\{T_{i,j'} : 1\leq j' \leq m_i\}$. And $\{T_{i,j'}\}$ are i.i.d. from the
continuous distribution $F_T$ supported on $[0,1]$ with a density $f_T$ satisfying
$c_1 \leq f_T \leq c_2$ for some $0 < c_1 \leq c_2 < \infty$.
Moreover, $\overline{m} := \sum_{i=1}^n m_i/n \to \infty$  as $\overline{N}
:= \sum_{i=1}^n N_i/n \to \infty$. Also, both $N_i$'s and $m_i$'s
increase to infinity uniformly meaning that $\max_i N_i/\min_i N_i$
and $\max_i m_i/\min_i m_i$ remain bounded.

\item[{\bf A6}]
Define $X_{il}(\cdot;X_{il}(0),\theta_i,\bs{\beta})$
to be the solution of the initial value problem
\begin{equation}\label{eq:initial_general}
x'(t) = e^{\theta_i} \sum_{k=1}^M \beta_k
\phi_{k,M}(x(t)),~~~t\in [0,1],~~~x(0)=X_{il}(0).
\end{equation}
Let $X_{il}^{\theta_i}(\cdot;\theta_i,\bs{\beta})$ and
$X_{il}^{\bs{\beta}}(\cdot;\theta_i,\bs{\beta})$ be its partial
derivatives with respect to parameters $\theta_i$ and $\bs{\beta}$.
And let $\bs{\beta}^* \in \mathbb{R}^M$ be as in {\bf A3}.
Define $G_{*,\theta\theta}^{i} := \mathbb{E}_{\bs{\theta}^*,\bs{\beta}^*}(X_{i1}^{\theta_i}(T_{i,1};\theta_i^*,\bs{\beta}^*))^2$,
$G_{*,\beta\theta}^{i} := \mathbb{E}_{\bs{\theta}^*,\bs{\beta}^*} \left(X_{i1}^{\theta_i}(T_{i,1};\theta_i^*,\bs{\beta}^*)
X_{i1}^{\bs{\beta}}(T_{i,1};\theta_i^*,\bs{\beta}^*)\right)$
and
$G_{*,\beta\beta}^{i} := \mathbb{E}_{\bs{\theta}^*,\bs{\beta}^*} \left(X_{i1}^{\bs{\beta}}(T_{i,1};\theta_i^*,\bs{\beta}^*)
(X_{i1}^{\bs{\beta}}(T_{i,1};\theta_i^*,\bs{\beta}^*))^T\right)$,
where $\mathbb{E}_{\bs{\theta}^*,\bs{\beta}^*}$ denotes the expectation over the
joint distribution of $(X_{i1}(0),T_{i,1})$ evaluated at  $\theta_i = \theta_i^*$ and $\beta=\beta^*$.
Define $G_{*,\theta\theta} = diag(G_{*,\theta\theta}^{i})_{i=2}^n$,
$G_{*,\beta\theta}= \left[G_{*,\beta\theta}^{2}:\cdots:G_{*,\beta\theta}^{n}\right]$, and
$G_{*,\beta\beta} = \sum_{i=1}^n G_{*,\beta\beta}^{i}$.
Then, there exists a function $\kappa_M$ and a constant $c_3 \in (0,\infty)$, such that,
\begin{equation}\label{eq:G_star_condition}
\parallel (G_{*,\beta\beta})^{-1}
\parallel \leq \kappa_M ~~~\mbox{and}~~~ \parallel (G_{*,\theta\theta})^{-1} \parallel \leq c_3.
\end{equation}

\item[{\bf A7}] The noise $\varepsilon_{ilj}$'s are i.i.d. $N(0,\sigma_\varepsilon^2)$ with
$\sigma_\varepsilon^2$ bounded above.

\end{itemize}

Before stating the main result, we give a
brief explanation of these assumptions. {\bf A1} ensures enough smoothness
of the solution paths of the differential equation (\ref{eq:x_theta_f}).
It also ensures that the approximation error, when $g$ is
approximated in the basis $\Phi_M$, is of an appropriate order.
Condition {\bf A3} is satisfied when we approximate $g$
using the $(p-1)$-th order B-splines with equally spaced knots on the interval
$D$ which are normalized so that $\int_D \phi_{k,M}(x)^2 dx = 1$ for all $k$.
Note that $g_{\bs\beta^*}$ can be viewed as an optimal approximation of
$g$ in the space generated by $\Phi_M$. Condition {\bf A4} ensures that a solution
of (\ref{eq:x_theta_f}) exist for all $f$ of the form $g_{\bs{\beta}}$
with $\bs{\beta}$ sufficiently close to $\bs{\beta}^*$. This
implies that we can apply the perturbation theory of differential
equations to bound the fluctuations of the sample paths due
to a perturbation of the parameters. Condition {\bf A5} ensures that the
time-points $\{T_{i,j}\}$ cover the domain $D$ randomly and densely,
and that there is a minimum amount of information per sample curve
in the data. Condition {\bf A6} is
about the estimability of a parameter (in this case $g$) in a semiparametric
problem in the presence of nuisance parameters (in this case $\{\theta_i\}$).
Indeed, the matrix $G_{*,\beta\beta} - G_{*,\beta\theta} (G_{*,\theta\theta})^{-1} G_{*,\theta\beta}$
plays the role of the information matrix for $\bs{\beta}$ at
$(\bs{\theta}^*,\bs{\beta}^*)$. Equation (\ref{eq:G_star_condition})
essentially quantifies the degree of ill-conditionedness of the
information matrix for $\bs{\beta}$.
Note that {\bf A4} together with {\bf A6} implicitly imposes a
restriction on the magnitude of $\parallel
g'\parallel_{L_\infty(D)}$.
Condition {\bf A6} has further implications.
Unlike in parametric problems, where the information matrix is
typically well-conditioned, we have $\kappa_M \to \infty$ in our
setting (see Theorem 2 and Proposition 1 below).
Note that in situations when $g \geq 0$ and the initial conditions are
nonnegative, one can simplify {\bf A6}
considerably, since then we can obtain explicit
formulas for the derivatives of the sample paths (see Appendix A).
And then one can easily verify the
second part of equation (\ref{eq:G_star_condition}).



\vskip.1in\noindent{\bf Theorem 1:} \textit{Assume that the data follow the
model described by equations (\ref{eq:basic}), (\ref{eq:data_model})
and (\ref{eq:scale}) with $\theta_1=0$. Suppose that the true gradient
function $g$, the distributions $F_a$ and $F_T$, and the collection of
basis functions $\Phi_M$ satisfy {\bf A1}-{\bf A7}. Suppose further that
$g$ is strictly positive over $D=[x_0,x_1]$. Suppose that
$\{X_{il}(0)\}$ are known (so that $\lambda_1=0$),
$\lambda_2 = o(\alpha_N \ol{N}\ol{m} \kappa_M^{-1})$ and the sequence $M = M(\ol{N},\ol{m})$
is such that $\min\{\ol{N},\ol{m}\} \gg \kappa_M M \log(\ol{N}\ol{m})$,
$\kappa_M M^{-(p-1)} \to 0$, and $\alpha_N \max\{\kappa_M M^{1/2},
\kappa_M^{1/2} M^{3/2}\} \to 0$ as $\ol{N},\ol{m} \to \infty$, where
$\alpha_N \geq C \max\{ \sigma_\varepsilon \kappa_M^{1/2} M^{1/2}
(\ol{N}\ol{m})^{-1/2}, \kappa_M^{1/2} M^{-p}\}$ for some sufficiently
large constant $C > 0$. Then there exists a minimizer $(\widehat{\bs\theta},
\widehat{\bs\beta})$ of the objective function (\ref{eq:log-like})
such that if $\widehat g := \sum_{k=1}^M \widehat \beta_k
\phi_{k,M}$, then the following holds with probability tending to
1:}
\begin{equation}\label{eq:consistency}
\int_D |\widehat g(x) - g(x)|^2 dx \leq \alpha_N^2 + O(M^{-2p}),
~~~~\sum_{i=2}^n|\widehat \theta_i - \theta_i^*|^2 \leq \alpha_N^2.
\end{equation}


\vskip.1in As explained earlier, $\kappa_M$ is related to the
inverse of the smallest eigenvalue of the matrix
\begin{equation*}
\begin{bmatrix}
G_{*,\bs{\beta}\bs{\beta}}&
G_{*,\bs{\beta}\bs{\theta}}\\
G_{*,\bs{\theta}\bs{\beta}} &
G_{*,\bs{\theta}\bs{\theta}}
\end{bmatrix}
\end{equation*}

In order to show that our method leads to a consistent estimator
of $g$, we need to know the behavior of $\kappa_M$ as $M \to \infty$.
The following result quantifies the behavior when we choose a
B-spline basis with equally spaced knots inside the domain $D$.

\vskip.1in\noindent{\bf Theorem 2:} \textit{Suppose that
supp$(F_a) =[x_0,x_2] \subset \mathbb{R}^+$ and $g$
is strictly positive over the domain $D = [x_0,x_1]$. Suppose
also that the (normalized) B-splines of order
$\geq 2$ are used as basis functions $\{\phi_{k,M}\}$ where the knots are equally
spaced on the interval $[x_0 + \delta, x_1 - \delta]$, for some
small constant $\delta > 0$. Then
$\kappa_M = O(M^2)$.}

\vskip.1in The condition that the knots are in the interior
of the domain $D$ is justified if the function $g$ is completely known
on the set $[x_0,x_0 + \delta] \cup [x_1-\delta,x_1]$.
Then this information can be used to modulate the B-splines
near the boundaries so that all the properties listed in
{\bf A3} still hold and we have the appropriate order of the approximations. We conjecture that the
same result ($\kappa_M = O(M^2)$) still holds even if $g$ is known only up to
a parametric form near the boundaries, and a combination of the parametric
form and B-splines with equally spaced knots is used to represent it.
If instead  the distribution $F_a$ is
such that near the end points ($x_0$ and $x_2$) of the support of $F_a$, the density
behaves like $(x-x_0)^{-1+\gamma}$ and $(x_2-x)^{-1+\gamma}$,
for some $\gamma \in (0,1]$, then it can be shown that (Proposition 1)
$\kappa_M = O(M^{2+2\gamma})$.
Thus, in the worst case scenario, we can only guarantee
that $\kappa_M = O(M^4)$. In that case $g$ needs to have a higher order
of smoothness ($g \in C^{6+\epsilon}(D)$, for some $\epsilon > 1/2$), and
higher-order (at least seventh order) B-splines are needed
to ensure consistency.

\vskip.1in It can be shown that under mild conditions $\kappa_M$
should be at least $O(M^{2})$. Thus, the condition $\alpha_N
\max\{\kappa_M  M^{1/2}, \kappa_M^{1/2} M^{3/2}\}=o(1)$ can be
simplified to $\kappa_M \alpha_N M^{1/2} =o(1)$. When $\kappa_M
\asymp M^2$, Theorem 1 holds with $p=4$, so that $g \in C^4$ and
cubic B-splines can be used. Moreover, under that setting as long as
$\ol{m}/\ol{N}$ is bounded both above and below and
$\sigma_\varepsilon$ is bounded below, then $\min\{\ol{N},\ol{m}\}
\gg \kappa_M M \log(\ol{N}\ol{m})$. The following proposition
states the dependence of $\kappa_M$ on the behavior of the density
of the distribution $F_a$.

\vskip.1in\noindent{\bf Proposition 1:} \textit{Assume that the density of $F_a$
behaves like $(x - x_0)^{-1+\gamma}$ and $(x_2 - x)^{-1+\gamma}$,
near the endpoints $x_0$ and $x_2$,
for some $\gamma \in (0,1]$, and is bounded away from zero in the interior.
Then $\kappa_M = O(M^{2+2\gamma})$.}

\vskip.1in
The proof of Theorem 1 involves a second order Taylor expansion
of loss function around the \textit{optimal parameter} $(\bs{\theta}^*,
\bs{\beta}^*)$. We apply results on perturbation of differential equations
(cf. Deuflhard and Bornemann, 2002, Ch. 3) to bound the \textit{bias
terms} $|X_{il}(t_{ilj};a_{il},\theta_i,f) - X_{il}(t_{ilj};a_{il},\theta_i,g)|$
for arbitrary $\theta_i$ and functions $f, g$. The same approach also allows us
to provide bounds for various terms involving partial
derivatives of the sample paths with respect to the parameters
in the aforementioned Taylor expansion.
Proof of Theorem 2 involves an inequality
(\textit{Halerpin-Pitt inequality}) on bounding the square
integral of a function by the square integrals of its derivatives
(Mitrinovic, Pecaric and Fink, 1991, p. 8). The detailed proofs are
given in Appendix E.

\section{Simulation}\label{sec:simulation}

\begin{figure}
\caption{True and fitted gradient functions for the \texttt{sparse} case. Left panel:
$\boldsymbol{a}$ known; Right panel:  $\boldsymbol{a}$ estimated.} \label{figure:g_fit_band_sparse}
\begin{center}
\begin{tabular}{cc}
\includegraphics[width=2.8in,height=3in,angle=270]{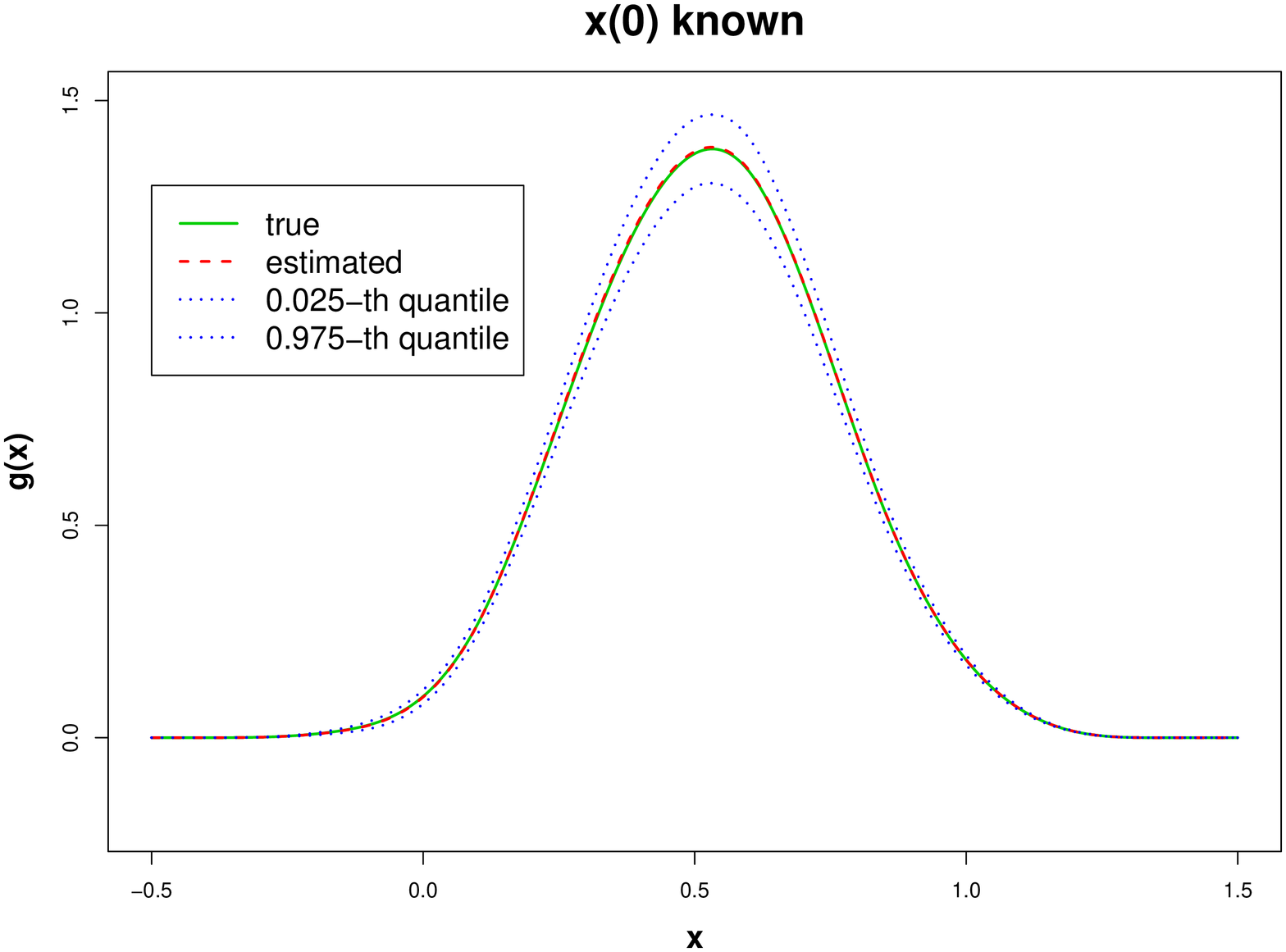} &
\includegraphics[width=2.8in,height=3in,angle=270]{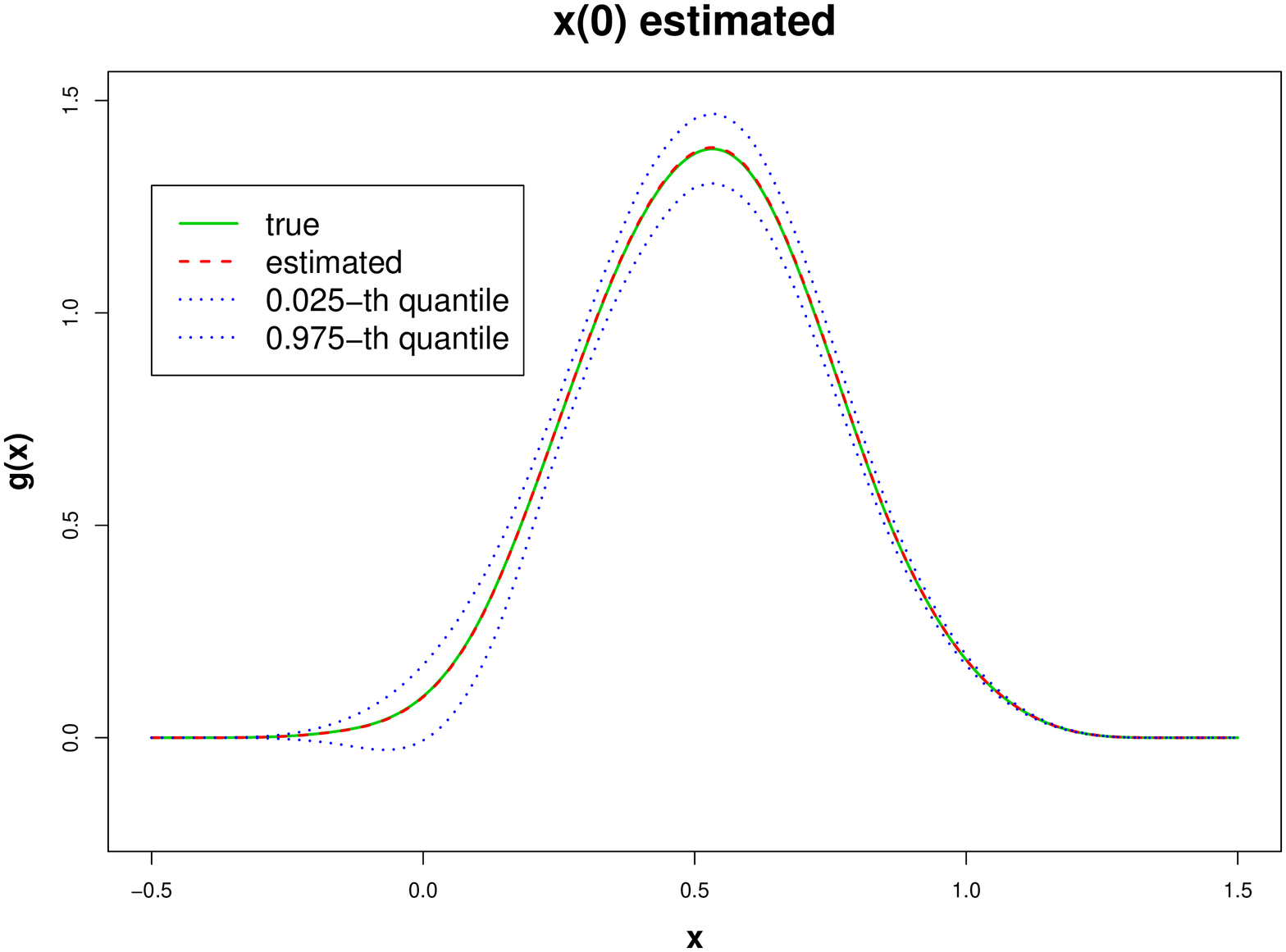}\\
\end{tabular}
\end{center}
\end{figure}

In this section, we conduct a simulation study to demonstrate the
effectiveness of the proposed estimation and model selection
procedures.
In the simulation, the true gradient function $g$ is represented
by $M_*= 4$ cubic B-spline basis functions with knots at $(0.35,0.6,0.85,1.1)$ and
basis coefficients $\bs{\beta} = (0.1, 1.2, 1.6, 0.4)^T$. It is depicted by
the solid curve in Figure \ref{figure:g_fit_band_sparse}.
We consider two different settings for the number of measurements per
curve: \texttt{moderate} case -- $m_{il}$'s are independently and identically
distributed as Uniform$[5, 20]$; \texttt{sparse} case -- $m_{il}$'s are independently
and identically distributed as Uniform$[3, 8]$. Measurement times $\{t_{ilj}\}$
are independently and identically distributed as Uniform$[0,1]$.
The scale parameters $\theta_i$'s are randomly sampled from
$N(0,\sigma_\theta^2)$ with $\sigma_\theta=0.1$; and the initial
conditions $a_{il}$'s are randomly sampled from  a $c_a
\chi_{k_a}^2$ distribution, with $c_a, k_a > 0$ chosen such that
$\alpha=0.25, \sigma_a=0.05$. Finally, the residuals
$\varepsilon_{ilj}$'s are randomly sampled from  $N(0,\sigma_\varepsilon^2)$
with $\sigma_\varepsilon=0.01$. Throughout the simulation, we set
the number of subjects $n=10$ and the number of curves per subject $N_i \equiv N=20$.
Observations $\{Y_{ilj}\}$ are generated using the model specified by
equations (\ref{eq:basic}) - (\ref{eq:basis}) in Section \ref{sec:model}.
For all the settings, $50$ independent data sets are used to evaluate the performance
of the proposed procedure.

In the estimation procedure, we consider cubic B-spline basis functions with
knots at points $0.1 + (1:M)/M$ to model $g$, where $M$ varies
from 2 to 6.
The Levenberg-Marqardt step is chosen to be \texttt{non-adaptive}, and the
Newton-Raphson step is chosen to be \texttt{adaptive}
(see Section \ref{sec:estimation} for the definition of \texttt{adaptive} and
\texttt{non-adaptive}).
We examine three different sets of initial values for
$\lambda_1$ and $\lambda_2$: (i) $\lambda_1=\sigma_{\varepsilon}^2/\sigma_a^2=0.04,
\lambda_2=\sigma_{\varepsilon}^2/\sigma_{\theta}^2=0.01$ (``true" values);
(ii)  $\lambda_1=0.01, \lambda_2=0.0025$ (``deflated" values);
(iii)  $\lambda_1=0.16, \lambda_2=0.04$ (``inflated" values).
It turns out that the estimation and model selection procedures are quite
robust to the initial choice of $(\lambda_1, \lambda_2)$,  thereby demonstrating the effectiveness
of the \texttt{adaptive} method used in the Newton-Raphson step. Thus in the following,
we only report the results when the ``true" values are used.

{
\begin{table}

\centering\caption{Convergence and model selection based on $50$ independent replicates.}
 \label{table:selection}
\begin{center}
\begin{tabular}{cc|ccccc|ccccc}\hline
    &&  \multicolumn{5}{c}
{$\mathbf{a}$ known} & \multicolumn{5}{c}{$\mathbf{a}$ estimated}\\

\cline{3-12} &Model& $2$ & $3$ & ${\bf 4}$ & $5$ & $6$ & $2$
& $3$ & ${\bf 4}$ & $5$ & $6$
\\\hline
\texttt{moderate}& Number converged & 50 & 50 & 50 & 50 & 50 & 50 & 7 & 50 & 50 & 46\\
& Number selected & 0 & 0 & 46 & 1 & 3 & 0 & 0 & 49 & 1 & 0\\\hline
\texttt{sparse} & Number converged & 50 & 50 & 50 & 50 & 50 & 50 & 5 & 49 & 44 & 38\\
& Number selected & 0 & 0 & 45 & 0 & 5 & 1 & 0 & 47 & 1 & 1\\\hline
\end{tabular}
\end{center}
\end{table}
}

\begin{table}
\centering\caption{Estimation accuracy under the true model$^*$}
\label{table:estimation}
\begin{center}
\begin{tabular}{c|c|cccc}\hline
 & & MISE($\widehat{g}$)&SD(ISE)&MSPE($\widehat{\theta}$)& SD(SPE)\\
\hline
{$\mathbf{a}$ known} & \texttt{moderate}& 0.069 & 0.072 & 0.085 & 0.095 \\
& \texttt{sparse}& 0.072 & 0.073 & 0.085 & 0.095 \\\hline
{$\mathbf{a}$ estimated} & \texttt{moderate}& 0.088 & 0.079 & 0.086 & 0.095 \\
  & \texttt{sparse}&  0.146 & 0.129 & 0.087 & 0.094\\\hline
\end{tabular}
\end{center}


\footnotesize{* All
numbers are multiplied by $100$}
\end{table}

We also compare results when (i) the initial conditions $\bs{a}$ are
known, and hence not estimated; and (ii) when $\bs{a}$ are
estimated. As can be seen from Table \ref{table:selection}, the
estimation procedure converges well and the true model
($M_{\ast}=4$) is selected most of the times for all the cases. Mean
integrated squared error (MISE) and Mean squared prediction error
(MSPE) and the corresponding standard deviations, SD(ISE) and
SD(SPE), based on 50 independent data sets, are used for measuring
the estimation accuracy of $\widehat{g}$ and $\widehat{\bs\theta}$,
respectively. Since the true model is selected most of the times, we
only report results under the true model in Table
\ref{table:estimation}. As can be seen from this table,
when the initial conditions $\bs{a}$ are known, there is not
much difference of the performance between the
\texttt{moderate case} and the \texttt{sparse} case.
On the other hand, when $\bs{a}$ are not known, the
advantages of having more measurements become much more prominent.
In Figure \ref{figure:g_fit_band_sparse}, we
have a visual comparison of the fits when the initial conditions
$\bs{a}$ are known versus when they are estimated in the
\texttt{sparse} case. In the \texttt{moderate} case, there is very
little visual difference under these two settings.
We plot the true $g$ (solid green curve), the pointwise mean of
$\widehat g$ (broken red curve), and 2.5\% and 97.5\% pointwise
quantiles (dotted blue curves) under the true model. These
plots show that both fits are almost unbiased. Also, when
$\boldsymbol{a}$ are estimated, there is greater variability in the
estimated $g$ at smaller values of $x$, partly due to scarcity of
data in that region. Overall, as can be seen from these tables and
figures, the proposed estimation and model selection procedures
perform effectively. Moreover, with sufficient
information, explicitly imposing nonnegativity in the model does not
seem to be crucial: for the \texttt{moderate} and/or ``$\bs{a}$
known'' cases the resulting estimators of $g$ are always nonnegative.

\section{Application: Plant Growth Data}\label{sec:plant}

In this section, we apply the proposed method to the plant growth
data from Sacks \textit{et al.} (1997) described in the earlier Sections.
The data consist of measurements on ten plants from a
control group and nine plants from a treatment group where the
plants are under water stress. The primary roots had grown for
approximately $18$ hours in the normal and stressed conditions
before the measurements were taken. The roots were marked at
different places using a water-soluble marker and high-resolution
photographs were used to measure the displacements of the marked
places. The measurements were in terms of distances from the root cap junction
(in millimeters) and were taken for each of these marked places,
hereafter markers, over an approximate 12-hour period while the
plants were growing.
Note that, measurements were only taken in the meristem.
Thus whenever a marker moved outside of the meristem,
its displacement would not be recorded
at later times anymore. This, together with possible
technical failures (in taking measurements), is the reason
why in Figure \ref{figure:plant_sample} some growth trajectories
were cut short.
A similar, but more sophisticated, data acquisition technique
is described in Walter et al.
(2002), who study the diurnal pattern of root growth in maize. Van
der Weele et al. (2003) describe a more advanced data
acquisition technique for measuring the expansion profile of a
growing root at a high spatial and temporal resolution. They also
propose computational methods for estimating the growth velocity
from this dense image data. Basu et al. (2007) develop a
a new image-analysis technique to study spatio-temporal patterns of
growth and curvature of roots that tracks the displacement of
particles on the root over space and time.
These methods, while providing plant scientists with valuable
information, are limited in that, they do not provide an inferential
framework and they require very dense measurements. Our method, even
though designed to handle sparse data, is potentially applicable to
these data as well.



Consider the model described in Section \ref{sec:model}.
For the control group, we have the number of curves per
subject $N_i$ varying in between $10$ and $29$; and for the water
stress group, we have $12 \leq N_i \leq 31$. The observed growth displacement
measurements $\{Y_{ilj}:j=1,\ldots,m_{il}, l=1,\ldots,N_i\}_{i=1}^n$
are assumed to follow model (\ref{eq:data_model}), where $m_{il}$ is
the number of measurements taken for the $i^{th}$ plant at its
$l^{th}$ marker, which varies  between $2$ and $17$;
and $\{t_{ilj}: j=1,\cdots,m_{il}\}$ are the times
of measurements, which are in between $[0,12]$ hours. Altogether, for the control
group there are $228$ curves with a total of $1486$ measurements and
for the treatment group there are $217$ curves with $1712$
measurements in total.  We are interested in comparing the
baseline growth displacement rate  between the treatment
and control groups.

As discussed earlier, there are natural constraints for the plant
growth dynamics. Theoretically, $g(0) = 0=g'(0)$ and $g'(x) = 0$ for $x
\geq A$ for some constant $A > 0$. For the former constraint,
we can simply omit the constant and linear terms in the
spline basis. And for the latter constraint, in the objective
function (\ref{eq:objective}) we use
$$
\bs\beta^T \mathbf{B} \bs\beta :=\lambda_R \int_A^{2A} (g'(x))^2 dx
= \lambda_R \bs\beta^T [\int_A^{2A} \phi'(x) (\phi'(x))^T
dx]\bs\beta
$$
where $\phi = (\phi_{1,M},\ldots,\phi_{M,M})^T$ and
$\lambda_R$ is a large positive number quantifying the
severity of  this constraint; and $A>0$ determines
where the growth displacement rate becomes a constant. $A$ and $\lambda_R$ are both
adaptively determined  by the model selection scheme discussed in
Section \ref{sec:model_selection}.
Moreover, as discussed earlier, since it is not appropriate to
shrink the initial conditions $\{a_{il}\}$ towards a fixed number,
we set $\lambda_1 = 0$ in the loss function (\ref{eq:objective}).

\begin{figure}
\caption{Residual versus time plots
for the treatment group. Left panel: fit by \texttt{stepwise-regression};
Right panel: fit by the proposed method.} \label{figure:res_fit_waterstress}
\begin{center}
\begin{tabular}{cc}
\includegraphics[width=3in,height=2.5in,angle=270]{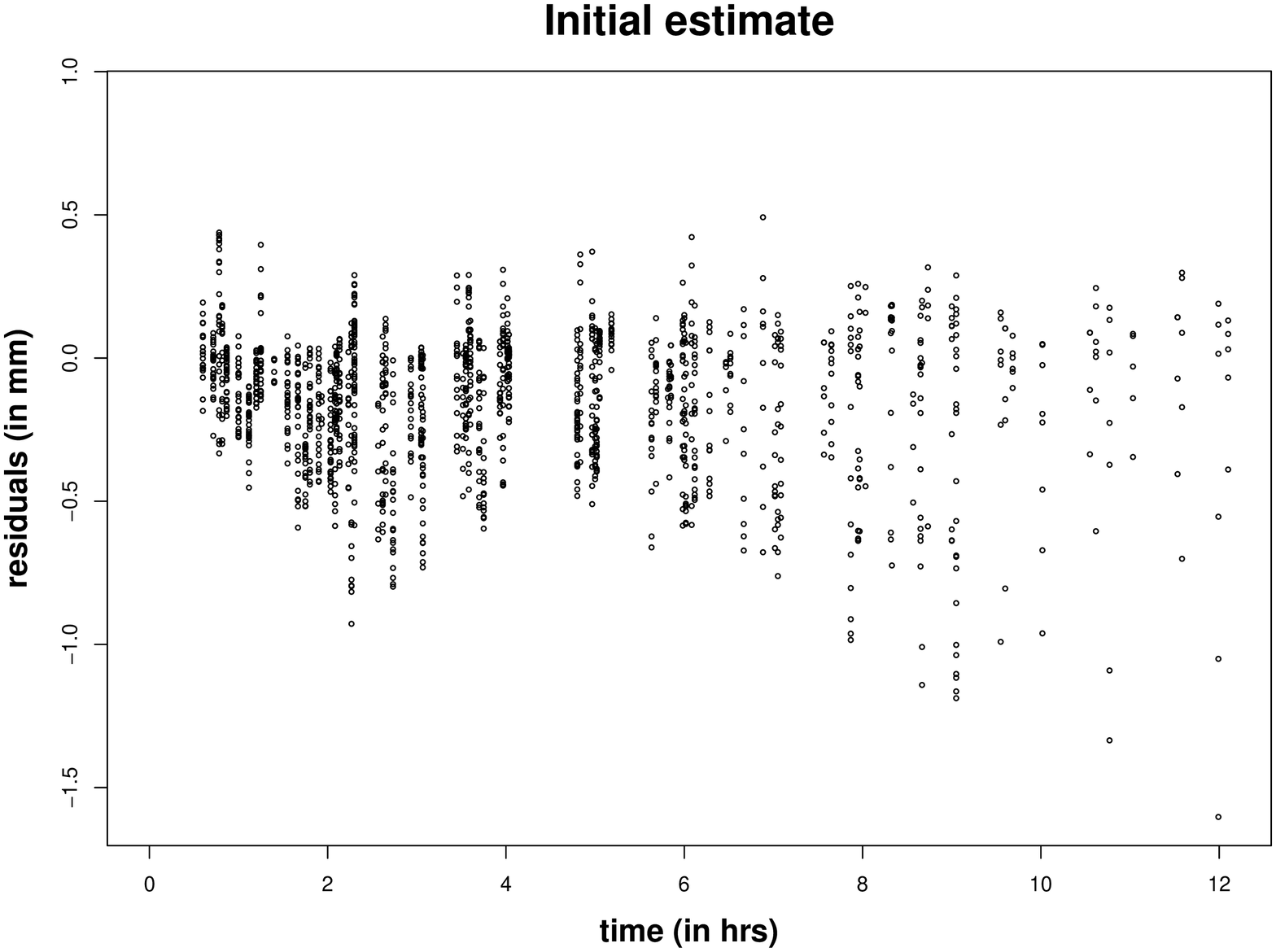} &
\includegraphics[width=3in,height=2.5in,angle=270]{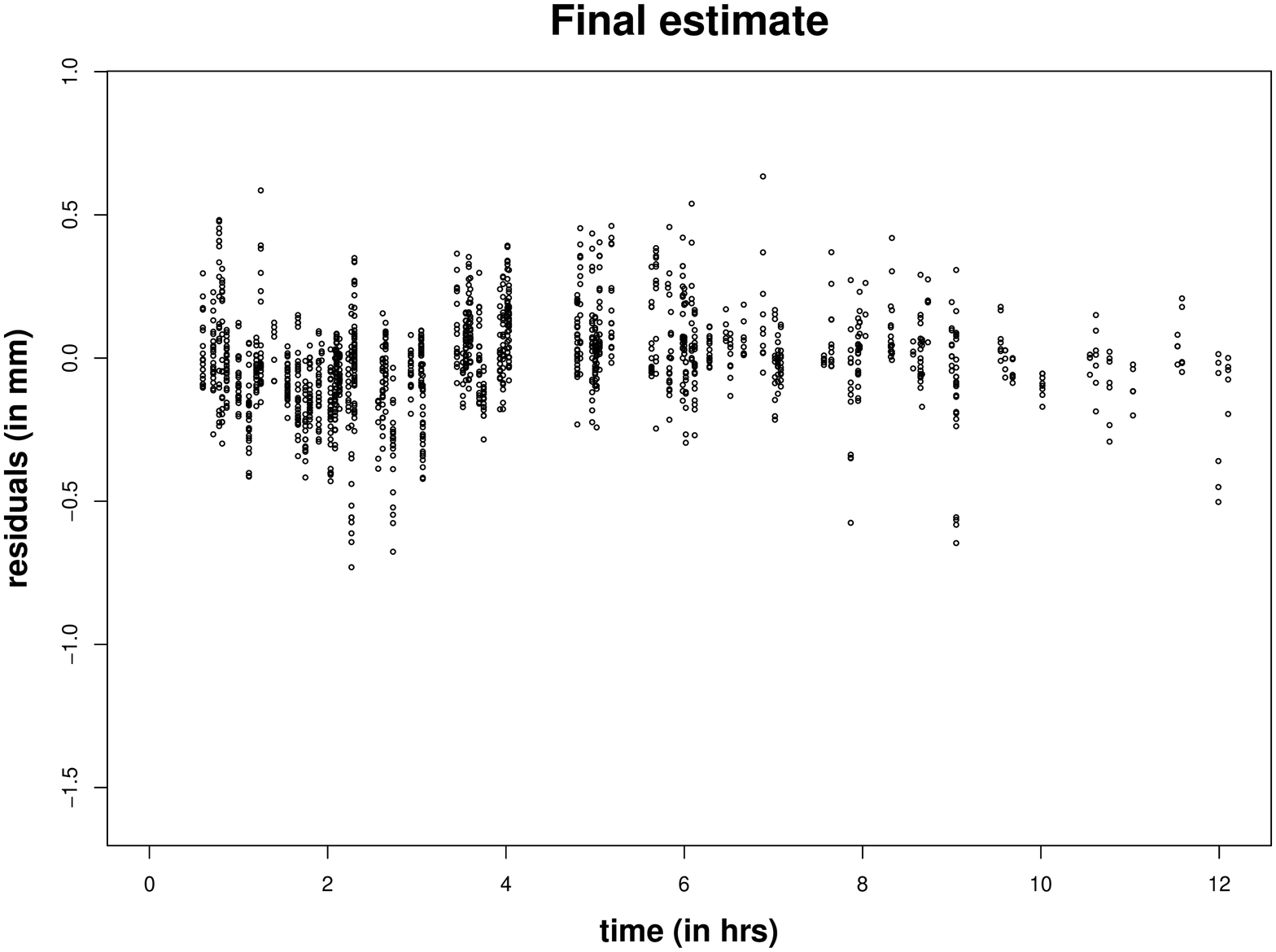}\\
\end{tabular}
\end{center}
\end{figure}

We first describe a simple regression-based method for getting a
crude initial estimate of the function $g(\cdot)$, as well as
selecting a candidate set of knots. This involves (i) computing
the re-scaled empirical derivatives $e^{-\widehat\theta_i^{(0)}}\widehat{X}^{\prime}_{ilj}$
of the sample curves from the data, where the empirical derivatives are defined
by taking divided differences: $\widehat{X}^{\prime}_{ilj} :=
(Y_{il(j+1)}-Y_{ilj})/(t_{il(j+1)}-t_{ilj})$, and $\widehat\theta_i^{(0)}$  is a
preliminary estimate of $\theta_i$; and (ii) regressing
the re-scaled empirical derivatives onto a set of basis functions
evaluated at the corresponding sample averages: $\widehat{X}_{ilj} :=
(Y_{il(j+1)}+Y_{ilj})/2$.
In this paper, we use the basis $\{x^2,x^3,(x - x_k)_+^3\}_{k=1}^K$
with a pre-specified, dense set of knots $\{x_k\}_{k=1}^K$.
Then, a model selection procedure, like the stepwise regression, with either
AIC or BIC criterion, can be used to select a set of candidate knots.
In the following, we shall refer this method as
\texttt{stepwise-regression}. The resulting estimate of $g$ and the selected
knots can then act as a starting point for the proposed
procedure. We expect this simple method to work reasonably well only when the number of
measurements per curve is at least moderately large.
Comparisons given later (Figure \ref{figure:res_fit_waterstress}) demonstrate
a clear superiority of the proposed method over this simple approach.




Next, we fit the model to the control group and the treatment group
separately. For the control group, we first fit models with $g$
represented in cubic B-splines with equally spaced knot sequence
$1+11.5(1:M)/M$ for $M=2,3,4,\cdots,12$.
At this stage, we set $\bs{\beta}^{ini} =
\mathbf{1}_M$, $\bs{\theta}^{ini} = \mathbf{0}_n$, $\bs{a}^{ini} =
(X_{il}(t_{il1}):l=1,\ldots,N_i)_{i=1}^n$.
For Levenberg-Marquardt step, we fix $\lambda_1= 0$ and
$\lambda_2 = 0.0025$; and
we update $\lambda_1,\lambda_2$ adaptively in the Newton-Raphson
step. The criterion based on the approximate CV score (\ref{eq:approx_CV}) selects the
model with $M=9$ basis functions (see Appendix D).
This is not surprising since when equally spaced knots are used,
usually a large number of basis functions are needed to fit the data adequately.
In order to get a more parsimonious model, we consider
the \texttt{stepwise-regression}
method to obtain an initial estimate of $g$ as well as finding a
candidate set of knots. We use $28$ equally spaced candidate knots on the
interval $[0.5,14]$
and use the fitted values $\{\widehat{\theta}_i\}_{i=1}^{10}$ from the previous
fit. The AIC criterion selects $11$ knots. We then consider various submodels
with knots selected from this set of $11$ knots and fit the
corresponding models again using the procedure described in
Section \ref{sec:estimation}. Specifically, we first apply the
Levenberg-Marquardt procedure with $\lambda_1$, $\lambda_2$ fixed at
$\lambda_1=0$ and $\lambda_2=(\widehat\sigma_\varepsilon^{ini})^2/(\widehat\sigma_\theta^{ini})^2=
0.042$, respectively, where $\widehat\sigma_\varepsilon^{ini}$
and $\widehat\sigma_\theta^{ini}$ are obtained from the \texttt{stepwise-regression}
fit.  Then, after convergence of $\bs{\beta}$
up to a desired precision (threshold of 0.005 for $\parallel
\bs{\beta}^{old} - \bs{\beta}^{new}\parallel$), we apply the
Newton-Raphson procedure with $\lambda_1$ fixed at zero, but
$\lambda_2$ adaptively updated from the data.
The approximate CV scores for various submodels are reported in Table
\ref{table:CV_real}.
The parameters $A$ and $\lambda_R$ are also
varied and selected by the approximate CV score. Based on
the approximate CV score, the model with knot sequence
$(3.0, 4.0, 6.0, 9.0, 9.5)$ and $(A,\lambda_R) = (9,10^5)$ is selected.
A similar procedure is applied to the treatment group.
It turns out that the model with knot sequence $(3.0, 3.5, 7.5)$
performs considerably better than other candidate models,
and hence we only report the approximate CV scores under this model in
Table \ref{table:CV_real} with various choices of $(A, \lambda_R)$.
It can be seen that, $(A, \lambda_R) =
(7,10^3)$ has the smallest approximate CV score.


{
\begin{table}
\centering \caption{Model selection for real data. Control group:
approximate CV scores for four \textit{submodels} of the model
selected by the AIC criterion in the \texttt{stepwise-regression}
step. \texttt{M1}: knots = $(3.0, 4.0, 5.0, 6.0, 9.0, 9.5)$;
\texttt{M2}: knots = $(3.0, 4.0, 5.5, 6.0, 9.0, 9.5)$; \texttt{M3}:
knots = $(3.0, 4.0, 6.0, 9.0, 9.5)$; \texttt{M4}: knots = $(3.0,
4.5, 6.0, 9.0, 9.5)$. Treatment group: approximate CV scores for the
model \texttt{M}: knots = $(3.0, 3.5, 7.5)$.} \label{table:CV_real}
\begin{center}
\begin{tabular}{cc|ccc|ccc}\hline
&  & \multicolumn{3}{c}{$\lambda_R = 10^3$} & \multicolumn{3}{c}{$\lambda_R = 10^5$}\\
\hline
Control &  Model & $A = 8.5$ & $A = 9$ & $A = 9.5$ & $A = 8.5$ & $A = 9$ & $A = 9.5$ \\
\hline
   & \texttt{M1}  & 53.0924 & 53.0877 & 53.1299 & 54.6422 & 53.0803 & 53.1307 \\
&  \texttt{M2}  & 53.0942 & 53.0898 & 53.1374  & 54.5190 & 53.0835 & 53.1375 \\
 & \texttt{M3} & 53.0300 & 53.0355  & 53.0729 & 53.8769 & {\bf 53.0063} & 53.0729 \\
 & \texttt{M4} & 53.0420 & 53.0409 & 53.0723 & 54.0538 & 53.0198 & 53.0722 \\\hline\hline
Treatment  &  Model &  $A = 7$ & $A = 7.5$ & $A = 8$ & $A = 7$ & $A = 7.5$ & $A = 8$ \\
\hline
& \texttt{M} & {\bf 64.9707}  & 64.9835 & 64.9843 & $65.5798^*$  & 64.9817 & 64.9817  \\\hline
\end{tabular}
\end{center}
\footnotesize{* no convergence}
\end{table}
}

\begin{figure}
\caption{Fitted gradient functions under the selected models for control and treatment (water-stress) groups, respectively.
}
\label{figure:compare_link_natural}
\begin{center}
\includegraphics[width=3.5in,height=5in,angle=270]{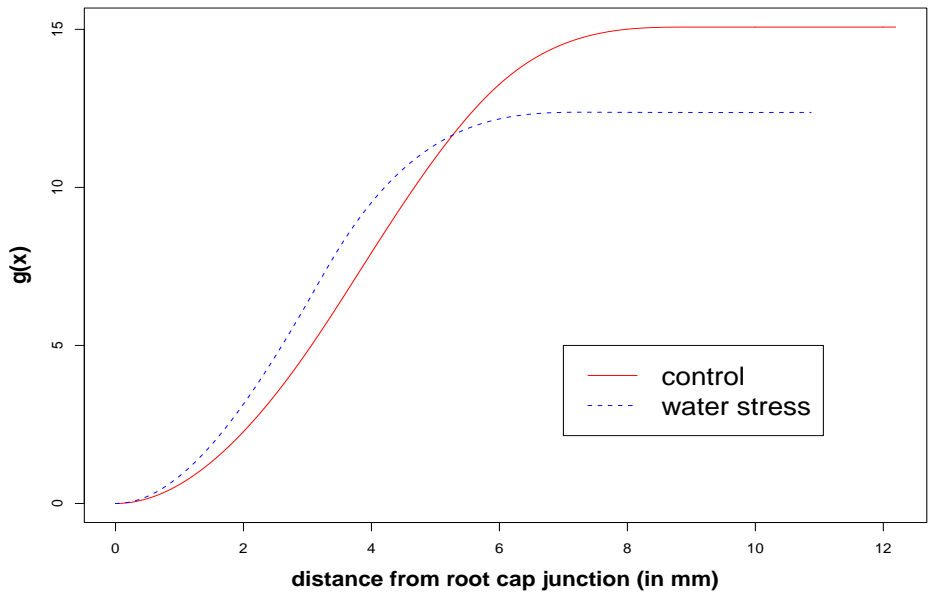}
\end{center}
\end{figure}

Figure \ref{figure:compare_link_natural} shows the estimated
gradient functions $\widehat{g}$ under the selected models for the
control and treatment groups, respectively. First of all,
there is no growth bump observed for either group. This plot also
indicates that different dynamics are at play for the two groups.
In the part of the meristem closer to the root cap junction
(distance within $\sim$ 5.5mm), the growth displacement rate for the
treatment group is higher than that for the control group. This is
probably due to the greater cell elongation rate under water stress
condition in this part of the meristem so that the root can reach
deeper in the soil to get enough water. This is a known phenomenon
in plant science. The growth displacement rate for the treatment
group flattens out beyond a distance of about 6 mm from the root cap
junction. The same phenomenon happens for the control group, however
at a further distance of about 8 mm from the root cap junction.
Also, the final constant growth displacement rate of the control
group is higher than that of the treatment group. This is due to the
stunting effect of water stress on these plants, which results in an
earlier stop of growth and a slower cell division rate. Figure
\ref{figure:compare_regr_natural} shows the estimated relative
elemental growth rates (i.e., $\widehat{g}'$) for these two groups.
Relative elemental growth rate (REGR) relates the magnitude of
growth directly to the location along the meristem. For both groups,
the growth is fastest in the middle part of the meristem ($\sim$ 3.8
mm for control group and $\sim$ 3.1 for treatment group), and then
growth dies down pretty sharply and eventually stops. Again, we
observe a faster growth in the part of the meristem closer to the
root cap junction for the water stress group and the growth dies
down more quickly compared to the control group. The shape of the
estimated $g$ may suggest that it might be modeled by a logistic
function with suitably chosen location and scale parameters, even
though the scientific meaning of these parameters is unclear and the
boundary constraints are not satisfied exactly. As discussed
earlier, there is insufficient knowledge from plant science to
suggest a functional form beforehand. This points to one major
purpose of nonparametric modeling, which is to provide insight and
to suggest candidate parametric models for further study.

Figure \ref{figure:res_fit_waterstress} shows the residual versus
time plot for the treatment group. The plot for the control group is
similar and thus is omitted.  This plot shows that the procedure
based on minimizing the objective function (\ref{eq:objective}) has
much smaller and more evenly spread residuals (SSE $=64.50$) than
the fit by \texttt{stepwise-regression} (SSE $= 147.57$), indicating
a clear benefit of the more sophisticated approach. Overall, by
considering the residual plots and CV scores, the estimation and
model selection procedures give reasonable fits under both
experimental conditions. Note that, for the first six hours, the
residuals (right panel of Figure \ref{figure:res_fit_waterstress})
show some time-dependent pattern, which is not present for later
times. Since throughout the whole $12$ hour period, the residuals
remain small compared to the scale of the measurements, the
autonomous system approximation seems to be adequate for practical
purposes. Modeling growth dynamics through nonautonomous systems may
enable scientists to determine the stages of growth that are not
steady across a region of the root. This is a topic of future
research.

{
\begin{figure}
\centering
\begin{center}
\includegraphics[width=3.5in,height=5in,angle=270]{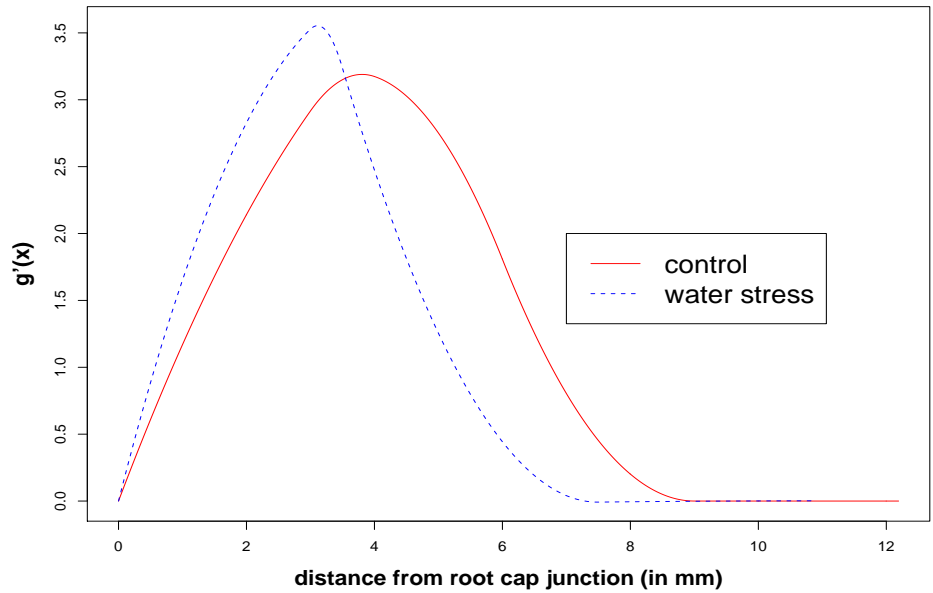}
\end{center}
\caption{Fitted relative elemental growth rate (REGR) under the selected models for control and treatment groups, respectively.}
\label{figure:compare_regr_natural}
\end{figure}
}

\section*{Acknowledgement}

Peng and Paul are partially supported by NSF-DMS grant 0806128. The
authors would like to thank Professor Wendy Silk of the Department
of Land, Air and Water Resources, University of California, Davis,
for providing the data used in the paper and for helpful discussions
on the scientific aspects of the problem.

\section*{References}

\begin{enumerate}

\item
Bates, D. M. and Watts, D. G. (1988). \textit{Nonlinear
Regression and Its Applications}. Wiley, New York.

\item
Basu, P., Pal, A., Lynch, J. P. and Brown, K. M. (2007). A
novel image-analysis technique for kinematic study of growth and
curvature. \textit{Plant Physiology} {\bf 145}, 305-316.

\item
Cao, J., Fussmann, G. F. and  Ramsay, J. O. (2008). Estimating a
predator-prey dynamical model with the parameter cascades
method. \textit{Biometrics} {\bf 64}, 959-967.

\item
Chen, J. and Wu, H. (2008a). Estimation of time-varying parameters
in deterministic dynamic models with application to HIV infections.
To appear in \textit{Statistica Sinica} {\bf 18}, 987-1006.

\item
Chen, J. and Wu, H. (2008b). Efficient local estimation for
time-varying coefficients in deterministic dynamic models with
applications to HIV-1 dynamics. \textit{Journal of American
Statistical Association} {\bf 103}, 369-384.

\item
Chicone, C. (2006). \textit{Ordinary Differential Equations with
Applications}. Springer.

\item
de Boor, C. (1978). \textit{A Practical Guide to Splines}. Springer–Verlag.

\item
Deuflhard, P. and Bornemann, F. (2002). \textit{Scientific
Computing with Ordinary Differential Equations}. Springer.

\item
Fraser, T. K., Silk, W. K. and Rost, T. L. (1990). Effects of
low water potential on cortical cell length in growing regions
of maize roots. \textit{Plant Physiology} {\bf 93}, 648-651.


\item
Huang, Y., Liu, D. and Wu, H. (2006). Hierarchical Bayesian
methods for estimation of parameters in a longitudinal HIV
dynamic system. \textit{Biometrics} {\bf 62}, 413-423.


\item
Lee, Y., Nelder, J. A. and Pawitan, Y. (2006). \textit{Generalized
Linear Models with Random Effects : Unified Analysis via
H-likelihood}. Chapman \& Hall/CRC.

\item
Li, L., Brown, M. B., Lee, K.-H., and Gupta, S. (2002).
Estimation and inference for a spline-enhanced population
pharmacokinetic model. \textit{Biometrics} {\bf 58}, 601-611.

\item
Ljung, L. and Glad, T. (1994). \textit{Modeling of Dynamical
Systems}. Prentice Hall.

\item
Miao, H., Dykes, C., Demeter, L. M. and Wu, H. (2008). Differential
equation modeling of HIV viral fitness experiments : model
identification, model selection, and multimodel inference.
\textit{Biometrics} (to appear).

\item
Mitrinovic, D. S., Pecaric, J. E. and Fink, A. M. (1991).
\textit{Inequalities Involving Functions and Their Integrals and
Derivatives}. Kluwer Academic Publishers.


\item
Nocedal, J. and Wright, S. J. (2006). \textit{Numerical
Optimization, 2nd Ed.} Springer.

\item
Peng, J. and Paul, D. (2009). A geometric approach to maximum
likelihood estimation of the functional principal components
from sparse longitudinal data. To appear in \textit{Journal of
Computational and Graphical Statistics}.
(\texttt{arXiv:0710.5343v1}). Also available at

\texttt{http://anson.ucdavis.edu/$\sim$jie/pd-cov-likelihood-technical.pdf}

\item
Perthame, B. (2007). \textit{Transport Equations in Biology}.
Birkh\"{a}user.

\item
Poyton, A. A., Varziri, M. S.,  McAuley, K. B.,  McLellan, P. J. and
Ramsay, J. O.  (2006).  Parameter estimation in continuous dynamic
models using principal differential analysis. \textit{Computers \&
Chemical Engineering} {\bf 30}, 698-708.

\item
Ramsay, J. O., Hooker, G., Campbell, D. and Cao, J. (2007).
Parameter estimation for differential equations: a generalized
smoothing approach. \textit{Journal of the Royal Statistical
Society, Series B} {\bf 69}, 741-796.

\item
Sacks, M. M., Silk, W. K. and Burman, P. (1997). Effect of
water stress on cortical cell division rates within the apical
meristem of primary roots of maize. \textit{Plant Physiology}
{\bf 114}, 519-527.

\item
Schurr, U., Walter, A. and Rascher, U. (2006). Functional
dynamics of plant growth and photosynthesis -- from steady-state
to dynamics -- from homogeneity to heterogeneity. \textit{Plant,
Cell and Environment} {\bf 29}, 340-352.

\item
Sharp, R. E., Silk, W. K. and Hsiao, T. C. (1988). Growth of the
maize primary root at low water potentials. \textit{Plant
Physiology} {\bf 87}, 50-57.

\item
Silk, W. K., and Erickson, R. O. (1979). Kinametics of plant
growth. \textit{\em Journal of Theoretical Biology} {\bf 76},
481-501.

\item
Silk, W. K. (1994). Kinametics and dynamics of primary growth.
\textit{Biomimectics} {\bf 2}(3), 199-213.

\item
Strogatz, S. H. (2001). \textit{Nonlinear Dynamics and Chaos:
With Applications to Physics, Biology, Chemistry and
Engineering}. Perseus Books Group.

\item
Tenenbaum, M., and Pollard, H. (1985). \textit{Ordinary
Differential Equations}. Dover.


\item
Van der Weele, C. M., Jiang, H. S., Krishnan K. P., Ivanov, V.
B., Palaniappan, K., and Baskin, T. I. (2003). A new algorithm
for computational image analysis of deformable motion at high
spatial and temporal resolution applied to root growth. roughly
uniform elongation in the meristem and also, after an abrupt
acceleration, in the elongation zone. \textit{Plant Physiology}
{\bf 132}, 1138-1148.

\item
Varah, J. M. (1982). A spline least squares method for numerical
parameter estimation in differential equations. \textit{SIAM Journal
on Scientific Computing} {\bf 3}, 28-46.

\item
Walter, A., Spies, H., Terjung, S., K\"{u}sters, R.,
Kirchgebner, N. and Schurr, U. (2002). Spatio-temporal dynamics
of expansion growth in roots: automatic quantification of
diurnal course and temperature response by digital image
sequence processing. \textit{Journal of Experimental Botany}
{\bf 53}, 689-698.

\item
Wu, H., Ding, A. and DeGruttola, V. (1998). Estimation of HIV
dynamic parameters. \textit{Statistics in Medicine} {\bf 17},
2463-2485.

\item
Wu, H. and Ding, A. (1999). Population HIV-1 dynamics in vivo :
applicable models and inferential tools for virological data
from AIDS clinical trials. \textit{Biometrics} {\bf 55},
410-418.


\item
Zhu, H. and Wu, H. (2007). Estimating the smooth time-varying
parameters in state space models. \textit{Journal of Computational
and Graphical Statistics} {\bf 20}, 813-832.

\end{enumerate}

\clearpage

\section*{Appendix A : Reconstruction of $X_{il}(\cdot)$
and its derivatives}\label{sec:recon}

In this section, we describe how to evaluate the $(i,l)$-th sample
trajectory $X_{il}(\cdot)$ and its derivatives given $\bs{\beta}, \theta_i$ and
$a_{il}$ on a fine grid. For notational simplicity, we omit the dependence of the
trajectories $X_{il}(\cdot)$ on the parameters
$(\bs{a},\bs{\theta},\bs{\beta})$, and drop the subscript
$M$ from $\phi_{k,M}$.

Note that, $X_{il}(\cdot)$ satisfies the first order ODE
\begin{equation}\label{eq:X_il_ODE}
\frac{d}{dt} X_{il}(t) =  e^{\theta_i} \sum_{k=1}^M \beta_k \phi_k(X_{il}(t)) , ~~~X_{il}(0)=a_{il}, ~~~t\in [0,1].
\end{equation}
Or equivalently
\begin{equation}\label{eq:tilde_X_i}
 X_{il}(t) = a_{il} + \int_{0}^t e^{\theta_i} \sum_{k=1}^M
\beta_k \phi_k( X_{il}(s)) ds, ~~~t\in [0,1].
\end{equation}
We first describe a numerical procedure ($4^{th}$ order Runge-Kutta method)
for constructing the sample trajectories $\widetilde X_{il}(t)$ and their derivatives
(with respect to the parameters) on a pre-specified fine grid.

\subsection*{Runge-Kutta method: the general procedure}

Suppose that a family of first order ODE is described in terms of
the parameters generically denoted by $\bs{\eta} =
(\eta_1,\eta_2)$, where $\eta_1$ denotes the initial condition and
$\eta_2$ can be vector-valued:
\begin{equation}\label{eq:ode_general}
\frac{d}{dt}f(t) = G(t,f(t),\eta_2), ~~~f(0) =
\eta_1, ~~~t \in [0,1].
\end{equation}
where  $G(t,x,\eta_2)$ is a smooth function.
Denote the solution for this family of ODE as $f(t,\bs{\eta})$.
Given the function $G$ and the parameter $\bs{\eta}$,
$f(t,\bs{\eta})$ can be solved numerically by an ODE solver. One
of the  commonly used approaches to solve such an initial value
problem is the $4^{th}$ order Runge-Kutta method. For a
pre-specified small value $h>0$, the $4^{th}$ order Runge-Kutta
method proceeds as follows:
\begin{enumerate}
\item Initial step: define $y_0=\eta_1$ and $t_0=0$;

\item Iterative step: in the $m+1$ step (for $ 0 \leq  m <[1/h]$),
define $y_{m+1} = y_m + \frac{h}{6}(k_1 + 2k_2 + 2k_3 + k_4),$ and
$t_{m+1} = t_m + h$, where
\begin{eqnarray*}
k_1 &=& G(t_m,y_m,\eta_2)\\
k_2 &=& G\left(t_m + \frac{h}{2},y_m + \frac{h}{2}k_1,\eta_2\right) \\
k_3 &=& G\left(t_m + \frac{h}{2},y_m + \frac{h}{2}k_2,\eta_2\right) \\
k_4 &=& G(t_m + h,y_m + h k_3,\eta_2).
\end{eqnarray*}
\item Final step: set $f(t_{m}, \bs{\eta})=y_m$ for $m=0,\cdots,
[1/h]$.
\end{enumerate}
Thus, at the end we obtain an evaluation (approximation) of
$f(\cdot,\bs{\eta})$ on the grid points $\{0,h,2h,\cdots, \}$.

Note that $f(t,\bs{\eta})$ satisfies,
\begin{equation}\label{eq:f_eta}
f(t,\bs{\eta}) = \eta_1 + \int_0^t G(s,f(s,\bs{\eta}),\eta_2)
ds,~~t\geq 0.
\end{equation}
Partially differentiating $f(t,\bs{\eta})$ with respect to
$\bs{\eta}$ and taking derivatives inside the integral, we obtain
\begin{eqnarray}
\frac{\partial}{\partial \eta_1} f(t,\bs{\eta}) &=& 1 + \int_0^t
\frac{\partial}{\partial \eta_1}f(s,\bs{\eta})
G_f(s,f(s,\bs{\eta}),\eta_2) ds,
\label{eq:f_eta_1_grad}\\
\frac{\partial}{\partial \eta_2} f(t,\bs{\eta}) &=& \int_0^t
\left[\frac{\partial}{\partial \eta_2}f(s,\bs{\eta})
G_f(s,f(s,\bs{\eta}),\eta_2) +
G_\eta(s,f(s,\bs{\eta}),\eta_2)\right] ds, \label{eq:f_eta_2_grad}
\end{eqnarray}
where $G_f$ and $G_\eta$ denote the partial derivatives of $G$
with respect to its second and third arguments, respectively. In
equations (\ref{eq:f_eta_1_grad}) and (\ref{eq:f_eta_2_grad}), if
we view the $f(\cdot,\bs{\eta})$ inside $G_f, G_\eta$ as known ,
$\frac{\partial}{\partial \eta_1} f(t,\bs{\eta})$ is the solution
of the first order ODE
\begin{eqnarray*}
\frac{d}{dt}p(t) = H(t,p(t),\eta_2), ~~~p(0) = 1, ~~~t \in [0,1],
\end{eqnarray*}
where $H(t,x,\eta_2)=xG_f(t,f(t,\bs{\eta}),\eta_2)$. Similarly,
$\frac{\partial}{\partial \eta_2} f(t,\bs{\eta})$ is the  solution
of the first order ODE with $p(0)=0$ and
$H(t,x,\eta_2)=xG_f(t,f(t,\bs{\eta}),\eta_2) +
G_\eta(t,f(t,\bs{\eta}),\eta_2)$. Thus, given the function $G$ and
the parameter $\bs{\eta}$, a general strategy for numerically
computing $f(\cdot,\bs{\eta})$ and its gradient
$\frac{\partial}{\partial \bs{\eta}} f(\cdot,\bs{\eta})$ on a fine
grid is to first use the Runge-Kutta method to approximate the
solution to (\ref{eq:f_eta}), and then using that approximate
solution in place of $f(\cdot,\bs{\eta})$ in equations
(\ref{eq:f_eta_1_grad}) and (\ref{eq:f_eta_2_grad}) to compute the
gradients by another application of the Runge-Kutta method. Note
that, if we evaluate $f(\cdot,\bs{\eta})$ on the grid points
$\{0,h,2h,\cdots\}$, by the above procedure, we will obtain the
gradients $\frac{\partial}{\partial \bs{\eta}} f(\cdot,\bs{\eta})$
on a rougher grid: $\{0,2h,4h,\cdots\}$.

\subsection*{Derivatives of the sample paths $\{X_{il}(\cdot)\}$ with respect to $(\boldsymbol{a},\boldsymbol{\theta},\boldsymbol{\beta})$}

Differentiating (\ref{eq:X_il_ODE}) with respect to the parameters, we have
\begin{eqnarray}
X_{il}^{a_{il}}(t) := \frac{\partial  X_{il}(t)}{\partial a_{il}} &=& 1 + \int_{0}^t \frac{\partial  X_{il}(s)}{\partial a_{il}} e^{\theta_i}
\sum_{k=1}^M \beta_k \phi_k'( X_{il}(s)) ds \\
X_{il}^{\theta_i}(t)
:= \frac{\partial   X_{il}(t)}{\partial \theta_i} &=& \int_{0}^t \left[\frac{\partial  X_{il}(s)}{\partial \theta_i}  e^{\theta_i} \sum_{k=1}^M
\beta_k \phi_k'( X_{il}(s)) + e^{\theta_i} \sum_{k=1}^M \beta_k \phi_k(
X_{il}(s))\right] ds \\
X_{il}^{\beta_r}(t) := \frac{\partial   X_{il}(t)}{\partial \beta_r} &=& \int_{0}^t \left[\frac{\partial  X_{il}(s)}{\partial \beta_r} e^{\theta_i} \sum_{k=1}^M
\beta_k \phi_k'( X_{il}(s)) + e^{\theta_i} \phi_r( X_{il}(s))\right] ds,
\end{eqnarray}
for $i=1,\cdots,n; l=1,\cdots, N_i; r=1,\cdots,M$. In another word, these functions  satisfy the
differential equations:
\begin{eqnarray}
\frac{d}{dt}  X_{il}^{a_{il}}(t) &=&   X_{il}^{a_{il}}(t) e^{\theta_i}\sum_{k=1}^M \beta_k \phi_k'( X_i(t)), ~~~
 X_{il}^{a_{il}}(0)=1,
\label{eq:tilde_X_a_i}\\
\frac{d}{dt}  X_{il}^{\theta_i}(t) &=& X_{il}^{\theta_i}(t)
 e^{\theta_i} \sum_{k=1}^M \beta_k \phi_k'(
X_{il}(t)) + e^{\theta_i} \sum_{k=1}^M \beta_k \phi_k( X_{il}(t)), ~~~  X_{il}^{\theta_i}(0) = 0,
\label{eq:tilde_X_theta_i}\\
\frac{d}{dt}  X_{il}^{\beta_r}(t) &=& X_{il}^{\beta_r}(t) e^{\theta_i} \sum_{k=1}^M \beta_k \phi_k'( X_{il}(t)) + e^{\theta_i} \phi_r(
X_{il}(t)), ~~~ X_{il}^{\beta_r}(0)= 0. \label{eq:tilde_X_beta}
\end{eqnarray}
Using similar arguments,
it follows that the Hessian of $X_{il}(\cdot)$ with respect to
$\bs{\beta}$, given by the matrix $(X_{il}^{\beta_r,\beta_{r'}})_{r,r'=1}^M$, where $ X_{il}^{\beta_r,\beta_{r'}}(t) :=
\frac{\partial^2}{\partial \beta_r
\partial \beta_{r'}}  X_{il}(t)$, satisfies the system of
ODEs, for $r,r'=1,\cdots,M$: {\small
\begin{eqnarray}\label{eq:tilde_X_beta_Hessian}
\frac{d}{dt}  X_{il}^{\beta_r,\beta_{r'}}(t) &=& e^{\theta_i}\left[ X_{il}^{\beta_r,\beta_{r'}}(t) \sum_{k=1}^M \beta_k \phi_k'( X_{il}(t))  +
X_{il}^{\beta_r}(t) \phi_{r'}'( X_{il}(t)) +
X_{il}^{\beta_{r'}}(t) \phi_{r}'( X_{il}(t))\right.\nonumber\\
&+&\left.  X_{il}^{\beta_r}(t) X_{il}^{\beta_{r'}}(t) \sum_{k=1}^M \beta_k \phi_k^{\prime\prime}( X_{il}(t)) \right],
~~{X}_{il}^{\beta_r,\beta_{r'}}(0)=0.
 \end{eqnarray}
}
The Hessian of $ X_{il}(\cdot)$ with respect to $\theta_i$, given by $ X_{il}^{\theta_i,\theta_i}$, satisfies the ODE
\begin{eqnarray}\label{eq:tilde_X_theta_Hessian}
\frac{d}{dt}   X_{il}^{\theta_i,\theta_i}(t) &=& e^{\theta_i} \left[\sum_{k=1}^M \beta_k \phi_k( X_{il}(t)) + ( X_{il}^{\theta_i,\theta_i}(t)  +
2 X_{il}^{\theta_i}(t)) \sum_{k=1}^M \beta_k \phi_k'( X_{il}(t)) \right.
\nonumber\\
&+& \left. ( X_{il}^{\theta_i}(t))^2 \sum_{k=1}^M \beta_k \phi_k^{\prime \prime}( X_{il}(t)) \right], ~~ X_{il}^{\theta_i,\theta_i}(0) = 0.
\end{eqnarray}
The Hessian of $ X_{il}(\cdot)$ with respect to $a_{il}$, given by $ X_{il}^{a_{il},a_{il}}$, satisfies the ODE
\begin{eqnarray}\label{eq:tilde_X_a_Hessian}
\frac{d}{dt}   X_{il}^{a_{il},a_{il}}(t) &=& e^{\theta_i} \left[ X_{il}^{a_{il},a_{il}}(t) \sum_{k=1}^M \beta_k
\phi_k'( X_{il}(t)) \right. \nonumber\\
&+& \left. ( X_{il}^{a_{il}}(t))^2 \sum_{k=1}^M \beta_k \phi_k^{\prime \prime}( X_i(t)) \right], ~~ X_{il}^{a_{il},a_{il}}(0) = 0.
\end{eqnarray}
Also, for future reference (even though it is not used in the proposed algorithm),
we calculate the mixed partial derivative of $ X_{il}(\cdot)$
with respect to $\theta_i$ and $\beta_r$ as $ X_{il}^{\theta_i,\beta_r}(t)
:= \frac{\partial^2  X_{il}(t)} {\partial \theta_i \partial \beta_r}$
which satisfies the ODE
\begin{eqnarray}\label{eq:tilde_X_theta_beta_Hessian}
\frac{d}{dt}  X_{il}^{\theta_i,\beta_r}(t) &=&
 X_{il}^{\theta_i,\beta_r}(t)  e^{\theta_i} \sum_{k=1}^M \beta_k
\phi_k'( X_{il}(t)) + e^{\theta_i} \left[  X_{il}^{\beta_r}(t)\sum_{k=1}^M \beta_k \phi_k'( X_{il}(t))
+ \phi_r( X_{il}(t))   \right. \nonumber\\
&+& \left.  X_{il}^{\theta_i}(t) \phi_r'( X_{il}(t))
 +  X_{il}^{\theta_i}(t)  X_{il}^{\beta_r}(t) \sum_{k=1}^M
\beta_k \phi_k^{\prime\prime}( X_{il}(t)) \right], ~ X_{il}^{\theta_i,\beta_r}(0) = 0.
\end{eqnarray}
Thus, the approach described above shows that as long as we have
evaluated (approximated) the function $ X_{il}(\cdot)$ at the grid points $\{0 +
mh/2: m=0,1,\ldots, 2/h\}$, we shall be able to approximate the
gradients $ X_{il}^{a_{il}}(\cdot)$, $ X_{il}^{\theta_i}(\cdot)$ and $\{
X_{il}^{\beta_r}\}_{r=1}^M$ at the grid points $\{0 + mh: m=0,1,\ldots, 1/h\}$,
and the Hessians $X_{il}^{a_{il},a_{il}}$, $X_{il}^{\theta_i,\theta_i}$ and $(X_{il}^{\beta_r,\beta_{r'}})_{r,r'=1}^M$at the grid points
$\{0 + 2mh: m=0,1,\ldots, 1/{(2h)}\}$, by successively applying the $4^{th}$ order Runge-Kutta method.

\subsection*{Expression when $g$ is positive}

Note that (\ref{eq:tilde_X_a_i}), (\ref{eq:tilde_X_theta_i}) and (\ref{eq:tilde_X_beta})
are linear differential equations. For the
\textit{growth model} we have $g$ positive and the initial conditions $a_{il}$ also can be taken to be  positive.
If the function $g_{\bs{\beta}} := \sum_{k=1}^M \beta_k \phi_k$
is also positive on the domain of $\{a_{il}\}$'s,
then the trajectories $X_{il}(t)$ are nondecreasing in $t$ (in fact strictly increasing if $g_{\bs{\beta}}$
is strictly positive). In this case, and more generally,
whenever the solutions exist on a time interval $[0,1]$ and $g_{\bs{\beta}}$
is twice continuously differentiable (so that the solution paths
for $ X_{il}^{a_{il}}$, $ X_{il}^{\theta_{il}}$, $ X_{il}^{\beta_r}$,
$ X_{il}^{a_{il},a_{il}}$, $ X_{il}^{\theta_i,\theta_i}$ and $
X_{il}^{\beta_r,\beta_{r'}}$ are $C^{1}$ functions on $[0,1]$) the
gradients of the trajectories can be solved explicitly:
\begin{eqnarray}
 X_{il}^{a_{il}}(t) &=& \frac{g_{\bs{\beta}}
 (X_{il}(t))}{g_{\bs{\beta}}( X_{il}(0))};\label{eq:tilde_X_a_i_closed}\\
 X_{il}^{\theta_i}(t) &=&  e^{\theta_i} t
g_{\bs{\beta}}( X_{il}(t)); \label{eq:tilde_X_theta_i_closed}\\
 X_{il}^{\beta_r}(t) &=& g_{\bs{\beta}}( X_{il}(t))
\int_{ X_{il}(0)}^{ X_{il}(t)} \frac{\phi_r(x)}{(g_{\bs{\beta}}(x))^2} dx. \label{eq:tilde_X_beta_closed}
\end{eqnarray}
In the following, we verify equation(\ref{eq:tilde_X_beta_closed}).
The proofs for others are similar and thus omitted. We can express
\begin{eqnarray*}
 X_{il}^{\beta_r}(t) &=& e^{\theta_i} \int_{0}^{t}
\phi_r( X_{il}(s)) \exp\left(e^{\theta_i} \int_s^t
g_{\bs{\beta}}'( X_{il}(u))du\right)ds \\
&=& e^{\theta_i} \int_{0}^{t} \phi_r( X_{il}(s)) \exp\left(\int_s^t \frac{g_{\bs{\beta}}'( X_{il}(u))}{g_{\bs{\beta}}( X_{il}(u))}  X_{il}'(u)
du\right)ds~~~(\mbox{using}~  X_{il}'(u) = e^{\theta_i}
g_{\bs{\beta}}( X_{il}(u))~)\\
&=& e^{\theta_i} \int_{0}^{t} \phi_r( X_{il}(s)) \exp(\log g_{\bs{\beta}}( X_{il}(t)) - \log g_{\bs{\beta}}(
X_{il}(s)))ds\\
&=& g_{\bs{\beta}}( X_{il}(t)) \int_{0}^t \frac{\phi_r( X_{il}(s))}{(g_{\bs{\beta}}(
X_{il}(s)))^2}  X_{il}'(s)ds\\
&=& g_{\bs{\beta}}( X_{il}(t))  \int_{ X_{il}(0)}^{ X_{il}(t)} \frac{\phi_r(x)}{(g_{\bs{\beta}}(x))^2} dx.
\end{eqnarray*}

Using analogous calculations, we can obtain the Hessians in closed form as well.
Thus, solutions to (\ref{eq:tilde_X_a_Hessian}),
(\ref{eq:tilde_X_theta_Hessian}) and (\ref{eq:tilde_X_beta_Hessian}) become
\begin{eqnarray}
 X_{il}^{a_{il},a_{il}}(t) &=& \frac{g_{\bs{\beta}}(
X_{il}(t))}{(g_{\bs{\beta}}( X_{il}(0)))^2} [g_{\bs{\beta}}'( X_{il}(t))- g_{\bs{\beta}}'(
X_{il}(0))]; \label{eq:tilde_X_a_Hessian_closed}\\
 X_{il}^{\theta_i,\theta_i} (t) &=&
e^{\theta_i}g_{\bs{\beta}}( X_{il}(t))[t + e^{\theta_i} t^2 g_{\bs{\beta}}'( X_{il}(t))]; \label{eq:tilde_X_theta_Hessian_closed}
\end{eqnarray}
\begin{eqnarray}\label{eq:tilde_X_beta_Hessian_closed}
&&  X_{il}^{\beta_r,\beta_{r'}}(t) \nonumber\\
&=& g_{\bs{\beta}}( X_{il}(t)) \int_{ X_{il}(0)}^{ X_{il}(t)} \frac{1}{g_{\bs{\beta}}(x)} \left[\phi_r'(x)(F_{r'}(x) - F_{r'}( X_{il}(0)))  +
\phi_{r'}'(x)(F_{r}(x) - F_{r}(
X_{il}(0)))\right]dx \nonumber\\
&& + (F_{r}( X_{il}(t)) - F_{r}( X_{il}(0))) (F_{r'}( X_{il}(t)) - F_{r'}( X_{il}(0))) g_{\bs{\beta}}( X_{il}(t)) g_{\bs{\beta}}'(
X_{il}(t))\nonumber\\
&& - e^{-\theta_i} g_{\bs{\beta}}( X_{il}(t)) \int_{ X_{il}(0)}^{ X_{il}(t)}
\frac{g_{\bs{\beta}}'(x)}{(g_{\bs{\beta}}(x))^3}\left[\phi_r(x)(F_{r'}(x) - F_{r'}( X_{il}(0)))  + \phi_{r'}(x)(F_{r}(x) -
F_{r}( X_{il}(0)))\right]dx,\nonumber\\
&&
\end{eqnarray}
where, for $x_1 < x_2$,
\begin{equation*}
F_r(x_2)- F_r(x_1) = \int_{x_1}^{x_2} \frac{\phi_r(y)}{(g_{\bs{\beta}}(y))^2} dy, ~~~1\leq r \leq M.
\end{equation*}
We can express $X_{il}^{\beta_r,\beta_{r'}}(t)$ alternatively as
\begin{eqnarray}\label{eq:tilde_X_beta_Hessian_alt}
X_{il}^{\beta_r,\beta_{r'}}(t) &=& e^{\theta_i} g_{\bs{\beta}}(X_{il}(t)) \int_0^t \frac{1}{g_{\bs{\beta}}(X_{il}(s))} \left[
X_{il}^{\beta_r}(s) \phi_{r'}'(X_{il}(s)) +  \phi_{r}'(X_{il}(s)) X_{il}^{\beta_{r'}}(s)\right]dt
\nonumber\\
&& + e^{\theta_i} g_{\bs{\beta}}(X_{il}(t)) \int_0^t  \frac{1}{g_{\bs{\beta}}(X_{il}(s))} X_{il}^{\beta_r}(s)X_{il}^{\beta_{r'}}(s)
g_{\bs{\beta}}''(X_{il}(s)) ds.
\end{eqnarray}
Similarly, we have the representation
\begin{eqnarray}\label{eq:tilde_X_beta_theta_Hessian_alt}
X_{il}^{\theta_i,\beta_r}(t) &=& e^{\theta_i} g_{\bs{\beta}}(X_{il}(t)) \int_0^t \frac{1}{g_{\bs{\beta}}(X_{il}(s))} X_{il}^{\theta_i}(s)
\phi_r'(X_{il}(s)) ds
\nonumber\\
&& + e^{\theta_i} g_{\bs{\beta}}(X_{il}(t)) \int_0^t \frac{1}{g_{\bs{\beta}}(X_{il}(s))} \left[X_{il}^{\beta_r}(s) g_{\bs{\beta}}'(X_{il}(s)) +
\phi_r(X_{il}(s)) + X_{il}^{\theta_i}(s)
X_{il}^{\beta_r}(s) g_{\bs{\beta}}''(X_{il}(s))\right] ds. \nonumber\\
&&
\end{eqnarray}

\section*{Appendix B : Levenberg-Marquardt method \label{sec:LM}}

The Levenberg-Marquardt method  is a method for solving the nonlinear least squares problem:
\begin{equation*}
\label{eq:nls}
\min_{\bs{\gamma}} S(\bs{\gamma}) ~~~\mbox{where}~~ S(\bs{\gamma}) =
\sum_{i=1}^n[y_i - f_i(\bs{\gamma})]^2,
\end{equation*}
where $f_i(\bs{\gamma})$'s are nonlinear functions of the parameter
$\bs{\gamma} \in \mathbb{R}^p$. The key idea is to linearly
approximate $f_i(\bs{\gamma}+\bs{\delta}) \approx f_i(\bs{\gamma}) +
J_i^T \bs{\delta}$, for a small $\bs{\delta} \in \mathbb{R}^p$,
where $J_i$ is the Jacobian of $f_i$ at $\bs{\gamma}$. Denote
$$
\mathbf{y} = (y_1,\ldots,y_n)^T, ~~~\mathbf{f}(\bs{\gamma}) =
(f_1(\bs{\gamma}),\ldots,f_n(\bs{\gamma}))^T,
$$
and $\mathbf{J}$ to be the $n \times p$ matrix with rows
$J_1^T,\ldots,J_n^T$. The resulting linearized least squares problem
involves, for given $\bs{\gamma}$ solving for $\bs{\delta}$ the
equation
\begin{equation}\label{eq:LM_general}
(\mathbf{J}^T \mathbf{J} + \lambda~ \mbox{diag}(\mathbf{J}^T
\mathbf{J})) \bs{\delta} = \mathbf{J}^T(\mathbf{y} -
\mathbf{f}(\bs{\gamma})),
\end{equation}
for a regularization parameter $\lambda > 0$. Note that, this
solution bears similarity with the ridge regression estimate.
However, the formulation in (\ref{eq:LM_general}) is according to
the observation by Marquardt that if each component of the gradient
is scaled according to the curvature then there is a larger movement
in the directions where the gradient is smaller. In practice, the
regularization parameter $\lambda$ is chosen adaptively to
facilitate convergence.

\section*{Appendix C : Newton-Raphson procedure  \label{sec:NR}}

We briefly describe the key steps of the Newton-Raphson procedure
for optimizing the objective function (\ref{eq:objective}).
As in the implementation of the Levenberg-Marquardt algorithm,
we break the iterative procedure in three steps.
The update of $\bs{a}$ is still performed by the
Levenberg-Marquardt algorithm (\ref{eq:a_diff}), while keeping
$\bs{\theta}$ and $\bs{\beta}$ fixed at the current values.
However, we employ Newton-Raphson to update $\bs{\theta}$ and
$\bs{\beta}$. Fixing $\bs{a}$, $\bs{\beta}$ at the current
estimates $\bs{a}^*$ and $\bs{\beta}^*$, respectively, we
update $\theta_{i}$'s from the current estimates $\theta_{i}^*$ by
\begin{eqnarray}\label{eq:NR_theta_plant}
\theta_{i}^{new} &=& \theta_{i}^* - \left[\sum_{l=1}^{N_{i}}
\sum_{j=1}^{m_{il}}\frac{\partial^2 \ell_{ilj}}{\partial
\theta_{i}^2} \right]^{-1}\sum_{l=1}^{N_{i}} \sum_{j=1}^{m_{il}}
\frac{\partial \ell_{ilj}}{\partial \theta_{i}},
\end{eqnarray}
where the quantities on the right hand side are all evaluated at
$(\bs{a}^*,\bs{\theta}^*, \bs{\beta}^*)$, and

$$
\frac{\partial \ell_{ilj}}{\partial \theta_{i}}=-2\widetilde \varepsilon_{ilj} \frac{\partial}{\partial \theta_{i}}
\widetilde X_{il}(t_{ilj}) +2\frac{\lambda_2
\theta_{i}}{\sum_{l=1}^{N_{i}} m_{il}}
$$

$$
\frac{\partial^2 \ell_{ilj}}{\partial
\theta_{i}^2} =-2\widetilde \varepsilon_{ilj}\frac{\partial^2} {\partial^2
\theta_{i}} \widetilde X_{il}(t_{ijl})+2(\frac{\partial}{\partial
\theta_{i}} \widetilde X_{il}(t_{ijl}))^2+2
\frac{\lambda_2}{\sum_{l=1}^{N_i}m_{il}}.
$$

 Similarly, the
Newton-Raphson update for $\bs{\beta}$ is given by
\begin{eqnarray}\label{eq:NR_beta}
\bs{\beta}^{new} &=& \bs{\beta}^* - \left[\sum_{i=1}^n
\sum_{l=1}^{N_{i}} \sum_{j=1}^{m_{il}} \frac{\partial^2
\ell_{ilj}}{\partial \bs{\beta} \partial \bs{\beta}^T}+2\mathbf{B}\right]^{-1}
\left(\sum_{i=1}^n \sum_{l=1}^{N_{i}} \sum_{j=1}^{m_{il}}
\frac{\partial \ell_{ilj}}{\partial \bs{\beta}}+2\mathbf{B}\bs{\beta}^*\right),
\end{eqnarray}
where the quantities on the right hand side are again evaluated at
$(\bs{a}^*,\bs{\theta}^*, \bs{\beta}^*)$, and
\begin{eqnarray*}
\frac{\partial \ell_{ilj}}{\partial \bs{\beta}}
&=& - 2\widetilde \varepsilon_{ilj} \frac{\partial}{\partial \bs{\beta}}
\widetilde X_{il}(t_{ilj})\\
\frac{\partial^2 \ell_{ilj}}{\partial \bs{\beta} \partial
\bs{\beta}^T} &=& 2 \frac{\partial}{\partial \bs{\beta}}
\widetilde X_{il}(t_{ilj})\left(\frac{\partial}{\partial \bs{\beta}}
\widetilde X_{il}(t_{ilj})\right)^T- 2\widetilde \varepsilon_{ilj}\frac{\partial^2}{\partial
\bs{\beta} \partial \bs{\beta}^T} \widetilde X_{il}(t_{ilj}).
\end{eqnarray*}

\section*{Appendix D : Cubic B-spline fits to the plant data}

We first consider the control group. In Table \ref{table:CV_control_Bspline}, we report the results using B-spline basis
with knots at $1+11.5(1:M)/M$ for $M=2,3,\cdots,12$; and using $\sigma_\varepsilon^{ini} = 0.05$, and $\sigma_\theta ^{ini}=
1$ as initial estimates.
In the B-spline fitting, we set the penalty matrix $\boldsymbol{B}$ to be the zero matrix, that is $\lambda_R=0$.
In the Newton-Raphson step, both
$\lambda_1$ and $\lambda_2$ are estimated adaptively from the data.
However, the Levenberg-Marquardt step is non-adaptive, that is it uses the initial values of $\lambda_1$
and $\lambda_2$ throughout. From Table
\ref{table:CV_control_Bspline}, for $M$= 2 to 8 there is no convergence.
For  $M$ = 9 to 12, the approximate CV scores are quite similar and the
minimum is achieved at $M=9$.


\begin{table}

\caption{Approximate leave-one-curve-out CV scores for \textit{control group}.
Cubic B-spline basis with knot sequence
$1+11.5(1:M)/M$; and $\sigma_\varepsilon^{ini}=0.05$,
$\sigma_\theta^{ini} = 1$. (* = no convergence)}\label{table:CV_control_Bspline}

\begin{center}

\begin{tabular}{c|ccc}
\hline
   $M$  &   \# L-M & \# N-R & CV score  \\
\hline
2* & 216 & 1000 & 489.01162 \\
3* & 445 & 1000 & 416.69848 \\
4* & 458 & 1000 & 91.21308 \\
5* & 546 & 1000 & 74.12581 \\
6* & 337 & 1000 & 58.25487 \\
7* & 279 & 1000 & 53.69243 \\
8* & 190 & 1000 & 53.37721 \\
9  & 233 & 195 & {\bf 53.16987} \\
10 & 147 & 120 & 53.26008 \\
11 &  94 &  79 & 53.26125 \\
12 &  78 &  54 & 53.41077 \\
\hline
\end{tabular}
\end{center}

\end{table}

\begin{table}

\caption{Approximate leave-one-curve-out CV scores for \textit{treatment group}.
Cubic B-spline basis with knot sequence
$1+9.5(1:M)/M$; and $\sigma_\varepsilon^{ini}=0.05$,
$\sigma_\theta^{ini} = 1$. (* = no convergence)}\label{table:CV_waterstress_Bspline}

\begin{center}

\begin{tabular}{c|ccc}
\hline
   $M$  &  \# L-M & \# N-R & CV score \\
\hline
2* & 228 & 1000 & 348.65867 \\
3* & 426 & 1000 & 422.03137 \\
4* & 233 & 1000 & 96.66250  \\
5* & 257 & 1000 & 71.77904  \\
6* & 539 & 1000 & 65.85252 \\
7  & 336 & 277 & 64.25370 \\
8  & 197 & 143 & 63.91828 \\
9  & 125 &  83 & {\bf 63.83346} \\
10 &  94 &  38 & 63.90003 \\
11*  & --  & --  & -- \\
12*  & --  & --  & -- \\
\hline
\end{tabular}
\end{center}


\end{table}

We then consider the fits for the treatment group.
The results using B-splines with knots at $1+9.5(1:M)/M$ for $M=2,3,\cdots,12$;
and using $\sigma_\varepsilon^{ini} =
0.05$, and $\sigma_\theta^{ini} = 1$ are reported in
Table \ref{table:CV_waterstress_Bspline}. We again set $\lambda_R=0$ (that is no penalty).
As for $M$ = 2 to 6, there is no
convergence. For $M$=7 to 10, the CV scores are similar and the minimum
is achieved again at $M$=9. For $M=11$ and $12$, the method breaks down due
to numerical instability.


\section*{Appendix E : Proof details}

In this section we provide the proofs of the key asymptotic results.

\subsection*{Proof of Theorem 1}

For convenience, we introduce the following notations:
\begin{equation*}
\ell_{i.}(\theta_i,\bs{\beta}) := \sum_{l=1}^{N_i}\sum_{j=1}^{m_{i}}\ell_{ilj}(a_{il},\theta_i,\bs{\beta})
\qquad \mbox{and} \qquad
\ell_{..}(\bs{\theta},\bs{\beta}) := \sum_{i=1}^n \ell_{i.}(\theta_i,\bs{\beta}).
\end{equation*}
Here, $\bs\theta :=(\theta_2,\ldots,\theta_n)^T$ since
$\theta_1 \equiv 0$.
For $\alpha > 0$, define
\begin{equation}\label{eq:omega_alpha}
\Omega(\alpha) :=\{(\bs{\theta},\bs{\beta}) : \bs\theta = \bs\theta^* + \alpha \bs\eta, \bs\beta
= \bs\beta^* + \alpha \bs\delta,~\bs\eta \in \mathbb{R}^{n-1},\bs\delta \in \mathbb{R}^M,~\mbox{s.t.}~
\parallel \bs\eta\parallel^2 + \parallel \bs\delta\parallel^2 = 1\}.
\end{equation}
We use $X_{ilj}^g$ to denote $X_{il}(T_{i,j};a_{il},g)$
where $X_{il}(\cdot)$ is the solution of the equation
$x'(t) = e^{\theta_i^*} g(x(t))$ with $x(0) = a_{il}$. We use $X_{il}(\cdot;\bs{\theta},\bs{\beta})$
to denote the solution of (\ref{eq:initial_general}) when $X_{il}(0) = a_{il}$, and $X_{ilj}(\bs{\theta},\bs{\beta}) :=
X_{il}(T_{i,j};\bs{\theta},\bs{\beta})$. We define
$X_{il}^{\theta_i}(\cdot;\bs{\theta},\bs{\beta})$, $X_{il}^{\beta_r}(\cdot;\bs{\theta},\bs{\beta})$,
$X_{il}^{\theta_i,\theta_i}(\cdot;\bs{\theta},\bs{\beta})$, $X_{il}^{\theta_i,\beta_r}(\cdot;\bs{\theta},\bs{\beta})$
and $X_{il}^{\beta_r,\beta_{r'}}(\cdot;\bs{\theta},\bs{\beta})$
as the partial
derivatives and mixed partial derivatives of $X_{il}(\cdot;\bs{\theta},\bs{\beta})$
with respect to $\theta_i$, $\beta_r$, $(\theta_i,\theta_i)$, $(\theta_i,\beta_r)$ and
$(\beta_r,\beta_{r'})$, respectively. Notations such as $X_{ilj}^{\theta_i}(\bs{\theta},\bs{\beta})$
are used to mean $X_{il}^{\theta_i}(T_{i,j};\bs{\theta},\bs{\beta})$. We use $g_{\bs{\beta}}$ to denote
the function $\sum_{k=1}^M \beta_k \phi_k$ (for convenience henceforth dropping the subscript $M$ from
$\phi_{k,M}$) and denote its first and second derivatives by $g_{\bs{\beta}}'$ and $g_{\bs{\beta}}''$,
respectively. Finally, we use $\parallel \cdot \parallel_\infty$ to mean $\parallel \cdot
\parallel_{L^{\infty}(D)}$, and denote the operator norm of a matrix and $l_2$ norm
of a vector by $\parallel \cdot \parallel$. We use $\bs{T}$ to denote $\{T_{i,j}:j=1,\ldots,m_i;i=1,\ldots,n\}$
and $\bs\varepsilon$ to denote $\{\varepsilon_{ilj}:j=1,\ldots,m_i;l=1,\ldots,N_i;i=1,\ldots,n\}$.

Let $\bs{\eta} \in \mathbb{R}^{n-1}$ and $\bs{\delta} \in \mathbb{R}^M$ be arbitrary vectors
satisfying $\parallel \bs{\eta}\parallel^2 + \parallel \bs{\delta}\parallel^2 = 1$.
Define $J_{n-1} := I_{n-1} - \frac{1}{n-1} \mathbf{1}_{n-1}\mathbf{1}_{n-1}^T$ and
observe that $\sum_{i=2}^n(\theta_i - \ol\theta)^2 = \bs\theta^T J_{n-1} \bs\theta$.
Define
\begin{eqnarray*}
W_{\bs\beta} &:=& \sum_{i=1}^n\sum_{l=1}^{N_i}\sum_{j=1}^{m_{i}} (Y_{ilj} - X_{il}(T_{i,j};a_{il},\theta_i^*,\bs{\beta}^*))
\frac{\partial X_{il}}{\partial \bs{\beta}}(T_{i,j};a_{il},\theta_i^*,\bs{\beta}^*)
\end{eqnarray*}
and $W_{\bs\theta}$ to be an $(n-1)\times 1$ vector with $(i-1)$-th coordinate
$$
\sum_{l=1}^{N_i}\sum_{j=1}^{m_{i}} (Y_{ilj} - X_{il}(T_{i,j};a_{il},\theta_i^*,\bs{\beta}^*))
\frac{\partial X_{il}}{\partial \theta_i}(T_{i,j};a_{il},\theta_i^*,\bs{\beta}^*)
$$
for $i=2,\ldots,n$. Also let $W := (W_{\bs\beta}^T,W_{\bs\theta}^T)^T$.
Then by a second order Taylor expansion, we have,
\begin{eqnarray}\label{eq:basic_expansion}
&& \ell_{..}(\bs{\theta}^* +\alpha_N \bs{\eta},
\bs{\beta}^* +\alpha_N \bs{\delta}) - \ell_{..}(\bs{\theta}^*,\bs{\beta}^*)
\nonumber\\
&=& \lambda_2 \alpha_N (2 (\bs{\theta}^*)^T J_{n-1} \bs\eta  + \alpha_N \bs\eta^T J_{n-1} \bs\eta)
+ 2\alpha_N [\bs\delta^T , \bs\eta^T]
\begin{bmatrix}  W_{\bs\beta} \\
W_{\bs\theta} \\
 \end{bmatrix}  \nonumber\\
&& + \alpha_N^2 [\bs\delta^T , \bs\eta^T ]
\begin{bmatrix}
{\cal G}_{\beta\beta}(\ol{\bs{\theta}},\ol{\bs{\beta}}) & {\cal G}_{\beta\theta}(\ol{\bs{\theta}},\ol{\bs{\beta}})\\
{\cal G}_{\theta\beta}(\ol{\bs{\theta}},\ol{\bs{\beta}}) & {\cal G}_{\theta\theta}(\ol{\bs{\theta}},\ol{\bs{\beta}})\\
\end{bmatrix}
\begin{bmatrix} \bs\delta \\  \bs\eta \\
\end{bmatrix}
\nonumber\\
&& - \alpha_N^2 [\bs\delta^T, \bs\eta^T  ]
\begin{bmatrix}
{\cal H}_{\beta\beta}(\ol{\bs{\theta}},\ol{\bs{\beta}}) & {\cal H}_{\beta\theta}(\ol{\bs{\theta}},\ol{\bs{\beta}})\\
{\cal H}_{\theta\beta}(\ol{\bs{\theta}},\ol{\bs{\beta}}) & {\cal H}_{\theta\theta}(\ol{\bs{\theta}},\ol{\bs{\beta}})\\
\end{bmatrix}
\begin{bmatrix} \bs\delta \\  \bs\eta \\ \end{bmatrix},
\end{eqnarray}
where $(\ol{\bs{\theta}},\ol{\bs{\beta}})$
satisfies $\parallel \ol{\bs{\beta}}-\bs{\beta}^*\parallel \leq \alpha_N$ and
$\parallel \ol{\bs{\theta}}-\bs{\theta}^*\parallel \leq \alpha_N$.
Note that $(\ol{\bs{\theta}},\ol{\bs{\beta}})$ depends on $(\boldsymbol{a},\boldsymbol{T})$ and $(\bs\eta,\bs\delta)$,
but not on $\bs\varepsilon$.
In the above, ${\cal G}_{\beta\beta}(\bs{\theta},\bs{\beta})$ is the $M \times M$ matrix
\begin{equation*}
\sum_{i=1}^n\sum_{l=1}^{N_i} \sum_{j=1}^{m_{i}} \left(\frac{\partial X_{il}}{\partial \bs{\beta}}(T_{i,j};a_{il},\theta_i,\bs{\beta})\right)
\left(\frac{\partial X_{il}}{\partial \bs{\beta}}(T_{i,j};a_{il},\theta_i,\bs{\beta})\right)^T;
\end{equation*}
${\cal G}_{\theta\beta}(\bs{\theta},\bs{\beta})$ is the $(n-1) \times M$ matrix
with $(i-1)$-th row
\begin{equation*}
\sum_{l=1}^{N_i} \sum_{j=1}^{m_{i}}
\frac{\partial X_{il}}{\partial \theta_i}(T_{i,j};a_{il},\theta_i,\bs{\beta})
\left(\frac{\partial X_{il}}{\partial \bs{\beta}}(T_{i,j};a_{il},\theta_i,\bs{\beta})\right)^T,~~~~i=2,\ldots,n;
\end{equation*}
${\cal G}_{\theta\theta}(\bs{\theta},\bs{\beta})$ is the $(n-1) \times (n-1)$ diagonal matrix
with the $(i-1)$-th diagonal entry
\begin{equation*}
\sum_{l=1}^{N_i} \sum_{j=1}^{m_{i}} \left(\frac{\partial X_{il}}{\partial \theta_i}(T_{i,j};a_{il},\theta_i,\bs{\beta})\right)^2,
~~~~i=2,\ldots,n;
\end{equation*}
${\cal G}_{\beta\theta}(\bs{\theta},\bs{\beta}) = {\cal G}_{\theta\beta}(\bs{\theta},\bs{\beta})^T$;
${\cal H}_{\beta\beta}(\bs{\theta},\bs{\beta})$ is the $M \times M$ matrix
\begin{equation*}
\sum_{i=1}^n\sum_{l=1}^{N_i} \sum_{j=1}^{m_{i}} (Y_{ilj} - X_{il}(T_{i,j};a_{il},\theta_i,\bs{\beta}))
\frac{\partial^2 X_{il}}{\partial \bs{\beta}\partial \bs{\beta}^T}(T_{i,j};a_{il},\theta_i,\bs{\beta});
\end{equation*}
${\cal H}_{\theta\beta}(\bs{\theta},\bs{\beta})$ is the $(n-1) \times M$ matrix with $(i-1)$-th row
\begin{equation*}
\sum_{l=1}^{N_i} \sum_{j=1}^{m_{i}} (Y_{ilj} - X_{il}(T_{i,j};a_{il},\theta_i,\bs{\beta}))
\frac{\partial^2 X_{il}}{\partial \theta_i \partial \bs{\beta}^T}(T_{i,j};a_{il},\theta_i,\bs{\beta}),
~~~~i=2,\ldots,n;
\end{equation*}
${\cal H}_{\theta\theta}(\bs{\theta},\bs{\beta})$ is the $(n-1) \times (n-1)$ matrix with
$(i-1)$-th diagonal entry
\begin{equation*}
\sum_{l=1}^{N_i} \sum_{j=1}^{m_{i}} (Y_{ilj} - X_{il}(T_{i,j};a_{il},\theta_i,\bs{\beta}))
\frac{\partial^2 X_{il}}{\partial \theta_i^2}(T_{i,j};a_{il},\theta_i,\bs{\beta}), ~~~~i=2,\ldots,n;
\end{equation*}
and ${\cal H}_{\beta\theta}(\bs{\theta},\bs{\beta}) = {\cal H}_{\theta\beta}(\bs{\theta},\bs{\beta})^T$.
Let ${\cal G}_{*,\theta\theta}$, ${\cal G}_{*,\beta\theta}$, ${\cal G}_{*,\theta\beta}$ and
${\cal G}_{*,\beta\beta}$ denote the expectations of
${\cal G}_{\theta\theta}(\bs{\theta}^*,\bs{\beta}^*)$, ${\cal G}_{\beta\theta}(\bs{\theta}^*,\bs{\beta}^*)$,
${\cal G}_{\theta\beta}(\bs{\theta}^*,\bs{\beta}^*)$ and ${\cal G}_{\beta\beta}(\bs{\theta}^*,\bs{\beta}^*)$
with respect to $(\bs{a},\bs{T})$. For future reference, we define the $(M+n-1)\times (M+n-1)$
symmetric matrix ${\cal G}(\bs\theta,\bs\beta)$ as
$$
{\cal G}(\bs\theta,\bs\beta) = \begin{bmatrix} {\cal G}_{\beta\beta}(\bs{\theta},\bs{\beta}) & {\cal G}_{\beta\theta}(\bs{\theta},\bs{\beta})\\
{\cal G}_{\theta\beta}(\bs{\theta},\bs{\beta}) & {\cal G}_{\theta\theta}(\bs{\theta},\bs{\beta})\\
\end{bmatrix}.
$$
We define ${\cal H}(\bs\theta,\bs\beta)$ and ${\cal G}_*$ analogously.

The following decomposition of the residuals is used throughout:
\begin{equation}\label{eq:error_decomp}
Y_{ilj} - X_{ilj}(\bs{\theta},\bs{\beta})
= \varepsilon_{ilj} + (X_{ilj}^g - X_{ilj}(\bs{\theta}^*,\bs{\beta}^*))
+ (X_{ilj}(\bs{\theta}^*,\bs{\beta}^*) - X_{ilj}(\bs{\theta},\bs{\beta})).
\end{equation}
Without loss of generality in the following
we assume that $\alpha_N M^{3/2} \to 0$, so that in
particular the bounds (\ref{eq:X_diff_bound}) - (\ref{eq:X_diff_beta_beta_bound}) are valid.
The proof of Theorem 1 then follows from the following sequence of lemmas.

\vskip.1in\noindent{\bf Lemma A.1 :} \textit{Let $\bs{\gamma} =
(\bs{\delta}^T,\bs{\eta}^T)^T$, and $W$ be as defined earlier. Then,
with probability tending to 1, uniformly in $\bs{\gamma}$ such that $\parallel \bs{\gamma}
\parallel = 1$, we have
\begin{eqnarray}\label{eq:grad_bound}
|\bs\gamma^T W|
&=& \left[O(\sigma_\varepsilon M^{1/2} \sqrt{\log(\ol{N}\ol{m})})
+ O(M^{-p} (\ol{N}\ol{m})^{1/2})\right] \sqrt{\bs{\gamma}^T{\cal G}(\bs{\theta}^*,\bs{\beta}^*)\bs{\gamma}}.
\end{eqnarray}
}

\vskip.1in\noindent{\bf Lemma A.2 :} \textit{With $\bs{\gamma}$ as in Lemma A.1, uniformly
over $\bs{\gamma}$, we have
\begin{eqnarray}\label{eq:G_diff_norm_bound}
&& \bs{\gamma}^T
\begin{bmatrix}
{\cal G}_{\beta\beta}(\ol{\bs{\theta}},\ol{\bs{\beta}}) - {\cal G}_{\beta\beta}(\bs{\theta}^*,\bs{\beta}^*)&
{\cal G}_{\beta\theta}(\ol{\bs{\theta}},\ol{\bs{\beta}}) - {\cal G}_{\beta\theta}(\bs{\theta}^*,\bs{\beta}^*) \\
{\cal G}_{\theta\beta}(\ol{\bs{\theta}},\ol{\bs{\beta}}) - {\cal G}_{\theta\beta}(\bs{\theta}^*,\bs{\beta}^*)&
{\cal G}_{\theta\theta}(\ol{\bs{\theta}},\ol{\bs{\beta}}) - {\cal G}_{\theta\theta}(\bs{\theta}^*,\bs{\beta}^*)\\
\end{bmatrix} \bs{\gamma} \nonumber\\
&=& O(\alpha_N M^{3/2}(\ol{N}\ol{m})^{1/2} ) \sqrt{\bs{\gamma}^T {\cal G}(\bs{\theta}^*,\bs{\beta}^*) \bs{\gamma}}
+ O(\alpha_N^2 M^3 \ol{N}\ol{m}).
\end{eqnarray}
}

\vskip.1in\noindent{\bf Lemma A.3 :} \textit{There exists a constant $c_7 > 0$ such that,
with $\bs{\gamma}$ as in Lemma A.1, uniformly over $\bs{\gamma}$, we have
\begin{equation}\label{eq:G_gamma_quad_bound}
\bs{\gamma}^T {\cal G}(\bs{\theta}^*,\bs{\beta}^*) \bs{\gamma}
\geq \bs{\gamma}^T {\cal G}_* \bs{\gamma} (1-o_P(1)) \geq c_7 \kappa_M^{-1} \ol{N}\ol{m} (1-o_P(1)).
\end{equation}
}

\vskip.1in\noindent{\bf Lemma A.4 :} \textit{With $\bs{\gamma}$ as in Lemma A.1,
with probability tending to 1, uniformly
over $\bs{\gamma}$,
\begin{eqnarray}\label{eq:H_diff_norm_bound}
&& \bs{\gamma}^T
\begin{bmatrix}
{\cal H}_{\beta\beta}(\ol{\bs{\theta}},\ol{\bs{\beta}}) - {\cal H}_{\beta\beta}(\bs{\theta}^*,\bs{\beta}^*)&
{\cal H}_{\beta\theta}(\ol{\bs{\theta}},\ol{\bs{\beta}}) - {\cal H}_{\beta\theta}(\bs{\theta}^*,\bs{\beta}^*) \\
{\cal H}_{\theta\beta}(\ol{\bs{\theta}},\ol{\bs{\beta}}) - {\cal H}_{\theta\beta}(\bs{\theta}^*,\bs{\beta}^*)&
{\cal H}_{\theta\theta}(\ol{\bs{\theta}},\ol{\bs{\beta}}) - {\cal H}_{\theta\theta}(\bs{\theta}^*,\bs{\beta}^*)\\
\end{bmatrix} \bs{\gamma} \nonumber\\
&=& O(\alpha_N M^{3/2}(\ol{N}\ol{m})^{1/2}) \sqrt{\bs{\gamma}^T {\cal G}_* \bs{\gamma}} + O(\alpha_N M^{1/2} \ol{N}\ol{m})
+ O(\alpha_N^2 M^3 \ol{N}\ol{m}) \nonumber\\
&& + O(\alpha_N M^{5/2-p} \ol{N}\ol{m}) + O(\sigma_\varepsilon \alpha_N M^{3} (\ol{N}\ol{m})^{1/2}\sqrt{\log(\ol{N}\ol{m})}).
\end{eqnarray}
}

\noindent
Finally, using (\ref{eq:error_decomp}), (\ref{eq:X_path_bias}),
and (\ref{eq:X_diff_theta_theta_bound})-(\ref{eq:X_diff_beta_beta_bound}), we have
\begin{eqnarray}\label{eq:H_star_bound}
&& \max\{\parallel {\cal H}_{\beta\beta}(\bs{\theta}^*,\bs{\beta}^*) \parallel,
\parallel {\cal H}_{\beta\theta}(\bs{\theta}^*,\bs{\beta}^*) \parallel,
\parallel {\cal H}_{\theta\theta}(\bs{\theta}^*,\bs{\beta}^*) \parallel\} \nonumber\\
&=& O_P(\sigma_\varepsilon M (\ol{N}\ol{m})^{1/2}) + O(M^{-(p-1)} \ol{N}\ol{m}).
\end{eqnarray}
Combining (\ref{eq:G_diff_norm_bound}) - (\ref{eq:H_star_bound}),
from (\ref{eq:basic_expansion}), with probability
tending to 1, uniformly in $\bs\gamma$,
\begin{eqnarray}
&& \ell_{..}(\bs{\theta}^* +\alpha_N\bs\eta, \bs{\beta}^* +\alpha_N \bs\delta) - \ell_{..}(\bs{\theta}^*,\bs{\beta}^*)
\nonumber\\
&\geq& \alpha_N^2 \bs{\gamma}^T {\cal G}(\bs{\theta}^*,\bs{\beta}^*) \bs{\gamma}  \nonumber\\
&& - \alpha_N \left(O(\sigma_\varepsilon M^{1/2} \sqrt{\log(\ol{N}\ol{m})})
+ O(M^{-p} (\ol{N}\ol{m})^{1/2}) + O(\alpha_N^2 M^{3/2}(\ol{N}\ol{m})^{1/2} ) \right)
\sqrt{\bs{\gamma}^T{\cal G}(\bs{\theta}^*,\bs{\beta}^*)\bs{\gamma}}
\nonumber\\
&& - \alpha_N \lambda_2 \parallel \bs{\theta}^*\parallel  O(1)
- \alpha_N^2 O(\alpha_N M^{3/2}(\ol{N}\ol{m})^{1/2}) \sqrt{\bs{\gamma}^T {\cal G}_* \bs{\gamma}}
\nonumber\\
&& - \alpha_N^2 O((\alpha_N M^{1/2} + \alpha_N^2 M^3 + \alpha_N M^{5/2-p} +  M^{-(p-1)}) \ol{N}\ol{m}) \nonumber\\
&& -  \alpha_N^2 O((\sigma_\varepsilon \alpha_N M^3
+ \sigma_\varepsilon M)(\ol{N}\ol{m})^{1/2}\sqrt{\log(\ol{N}\ol{m})})
\nonumber\\
&\geq& c_4 \kappa_M^{-1} \alpha_N^2 \ol{N}\ol{m} (1-o_P(1))
\end{eqnarray}
where $c_4 >0$ is some constant. The last step uses Lemma A.3 and
the following fact:
\begin{itemize}
\item[[{\bf Q}]]
For any positive definite matrix $A$, with $\parallel A^{-1}
\parallel \leq \kappa$, if $2c\sqrt{\kappa} < 1$, then
for all $\mathbf{x}$ such that $\parallel \mathbf{x} \parallel
=1$
\begin{equation*}
\mathbf{x}^T A \mathbf{x} - c \sqrt{\mathbf{x}^T A
\mathbf{x}} \geq \frac{1}{2} \mathbf{x}^T A \mathbf{x}
\end{equation*}
\end{itemize}
Thus, with probability tending to 1, there is a local minimum
$(\widehat{\bs\theta},\widehat{\bs\beta})$ of the objective function
(\ref{eq:log-like}) with $\parallel \widehat{\bs\theta} -
\bs{\theta}^*\parallel^2 + \parallel \widehat{\bs\beta} -
\bs{\beta}^*\parallel^2 \leq \alpha_N^2$. This completes the proof
of Theorem 1.

\subsection*{Proof of Theorem 2}

We make use of the following inequality due to Halperin and Pitt
(Mitrinovic, Pecaric and Fink, 1991, page 8):
\textit{If $f$ is locally absolutely continuous and $f^{\prime\prime}$ is in
$L_{2}([0,A])$, then for any $\epsilon>0$ the following inequality holds
\begin{equation*}
\int_{0}^{A}f^{\prime2}\leq K(\epsilon)\int_{0}^{A}f^{2}+\epsilon
\int_{0}^{A}f^{\prime\prime2}%
\end{equation*}
where $K(\epsilon)=1/\epsilon+12/A^{2}$.}

Define  $X_i(t,x)$ as
the sample path $X_{il}(t;a_{il},\theta_i^*,\bs{\beta}^*)$ when $a_{il}=x$.
Since $\theta_1 =0$ and $X_{il}^{\bs\beta}(\cdot;\bs\theta,\bs\beta)$
is given by (\ref{eq:tilde_X_beta_closed}) (Appendix A), in order
to prove Theorem 2, it is enough to find a lower bound on
\begin{equation*}
\min_{\parallel \mathbf{b}\parallel =1} \int \int_0^1 \left[\int
_{0}^{t}g_{\mathbf{b}}(X_1(u,x))/g_{\bs{\beta}^*}(X_1(u,x))du \right]^{2}f_T(t) dt dF_a(x)
\end{equation*}
where $g_{\mathbf{b}}(u) = \mathbf{b}^T \bs{\phi}(u)$. By {\bf A5}, without loss
of generality we can take the density $f_T(\cdot)$ to be uniform on $[0,1]$.
Let
\begin{equation*}
R(t,x) := \int_{0}^{t}g_{\mathbf{b}}(X_1(u,x))/g_{\bs{\beta}^*}(X_{1}(u,x))du.
\end{equation*}
Then,
\begin{eqnarray*}
r(t,x) &:=& \frac{\partial}{\partial t}R(t,x)
= \frac{g_{\mathbf{b}}(X_1(t,x))}{g_{\bs{\beta}^*}(X_{1}(t,x))} \\
r'(t,x) &:=& \frac{\partial}{\partial t} r(t,x)
= \left[\frac{g_{\mathbf{b}}'(X_1(t,x))}{g_{\bs{\beta}^*}(X_{1}(t,x))} -
\frac{g_{\mathbf{b}}(X_1(t,x))g_{\bs{\beta}^*}'(X_{1}(t,x))}{g_{\bs{\beta}^*}^2(X_{1}(t,x))}\right]
X_1'(t,x)\\
&=& \left[\frac{g_{\mathbf{b}}'(X_1(t,x))}{g_{\bs{\beta}^*}(X_{1}(t,x))} -
\frac{g_{\mathbf{b}}(X_1(t,x))g_{\bs{\beta}^*}'(X_{1}(t,x))}{g_{\bs{\beta}^*}^2(X_{1}(t,x))}\right]
g_{\bs{\beta}^*}(X_{1}(t,x))
\end{eqnarray*}
From this, and the fact that the coordinates of $\bs{\phi}'(u)$ are
of the order $O(M^{3/2})$,
coordinates of $\bs{\phi}(u)$ are of the order $O(M^{1/2})$, and
all these functions are supported on intervals of length $O(M^{-1})$,
we obtain that, uniformly in $x$,
\begin{equation}\label{eq:r_prime_int}
\int_0^1 (r'(t,x))^2  dt = O(M^2).
\end{equation}
Application of Halperin-Pitt inequality with $f(x) = \int_0^1 R(t,x)^2 dt$ yields
\begin{equation}\label{eq:HP_main}
\int\int_0^1 (r(t,x))^2 dt dF_a(x)
\leq(1/\epsilon+12)\int \int_0^1 (R(t,x))^2 dt dF_a(x) +
\epsilon \int \int_0^1 (r'(t,x))^2 dt dF_a(x).
\end{equation}
Take $\epsilon = k_0 M^{-2}$ for some $k_0 > 0$, then by (\ref{eq:r_prime_int}),
\begin{eqnarray*}
\int \int_{0}^{1} (R(t,x))^2 dt dF_a(x) &\geq&
k_1 M^{-2}\int \int_0^1 (r(t,x))^2 dt dF_a(x)
- k_2 M^{-2},
\end{eqnarray*}
for constants $k_1,k_2 > 0$ dependent on $k_0$. Rewrite
$\int \int_0^1 (r(t,x))^2 dt dF_a(x)$ as
\begin{equation}\label{eq:smallest_eigen_bound}
\int \int_x^{X_1(1,x)} \frac{g_{\mathbf{b}}^2(v)}{g_{\bs{\beta}^*}^3(v)} dv dF_a(x)
= \int g_{\mathbf{b}}^2(v) h(v) dv
\end{equation}
where $h(v) = g_{\bs{\beta}^*}^{-3}(v) \int \mathbf{1}_{\{x \leq v
\leq X_1(1,x)\}} dF_a(x)$. If the knots are equally spaced on $[x_0
+ \delta, x_1 - \delta]$ for some constant $\delta > 0$ is bounded
below, then $\inf_{v \in D_0} h(v)$ is bounded below (even as $M \to
\infty$) where $D_0$ is the union of the supports of
$\{\phi_{k,M}\}_{k=1}^M$, which contained in $[x_0 + \delta/2,x_1 -
\delta/2]$ for $M$ sufficiently large). In this case, $\int
\int_x^{X_1(1,x)} \frac{g_{\mathbf{b}}^2(v)}{g_{\bs{\beta}^*}^3(v)}
dv \geq k_3$ for some constant $k_3 > 0$. Thus, by appropriate
choice of $\epsilon$, we have $\int \int_{0}^{1} (R(t,x))^2 dt
dF_a(x) \geq k_4 M^{-2}$ for some $k_4 > 0$, which yields $\kappa_M
= O(M^2)$.


\subsection*{Proof of Proposition 1}

The proof is based on the following lemmas.

\vskip.1in\noindent{\bf Lemma A.5:} \textit{Let $\mathcal{P}_{d}$ be the class of
all polynomials $p(x) = \sum_{j=0}^d \beta_j x^j$ of degree $d$ on $[0,1]$
such that $|p|_{\infty}=1.$ Then there exists a
constant $c>0$ such that
\begin{equation*}
|p|_{\infty}\geq c\max_{0\leq j\leq d}|\beta_{j}|.
\end{equation*}}

\vskip.1in\noindent{\bf Lemma A.6:} \textit{Let $\mu$ be a measure on the
interval $[0,1]$ with the property that for any $L>0$, there exists a constant
$C(L)>0$ such that for any interval $A \subset [0,1]$, $\mu(B)/\mu(A)\geq C(L)$ for all intervals
$B\subset A$ with $length(B)/length(A)\geq L$. Then for any polynomial $p$ of
degree $d$ on $[0,1],$ there exists a constant $c>0$ such that
\begin{equation*}
\int_{A}p^{2}d\mu\geq c\sup_{u\in A}|p(u)|^{2}\mu(A) .
\end{equation*}}

For the next lemma, assume that the knots are $t_{1}=\cdot\cdot
\cdot=t_{d+1}=0,t_{M+1}=\cdot\cdot\cdot=t_{M+d+1}=1$ and $0<t_{d+2}<\cdot
\cdot\cdot<t_{M}<1$. Note that we have placed extra knots at $0$ and $1$ in
order to obtain a B-spline basis. Let $\bs\psi :=\{\psi_{j}:j=1,...,M\}$ be the (unnormalized)
$B$-spline basis with the knots $\{t_{j}:j=d+2,...,M\}$. Let $\bs{\beta} \in \mathbb{R}^{M}$,
and consider the spline $s(x):=\sum_{j=1}^M \beta_{j}\psi_{j}(x)$. Then on the interval
$A_{i} :=[t_{i},t_{i+1}]$,  $s(x) =\sum_{i-d\leq j\leq i}\beta_{j}\psi_{j}(x)$ with
$\sum_{i-d\leq j\leq i}\psi_{j}(x)=1$.

\vskip.1in\noindent{\bf Lemma A.7:} \textit{Assume that $\mu$ is a measure on $[0,1]$
satisfying the properties of
Lemma A.6 above. Consider the vector $\bs{\psi}$ of $B$-splines on $[0,1]$ of degree $d$
with well-conditioned knots at $t_{d+2},...,t_{M}$,
i.e., the sequence
$\{M(t_{i+1}-t_{i}):i=d+1,...,M\}$ remains bounded between two positive
constants for any $M$. Then there exist constants $c_{12},c_{13}>0$
(which do not depend on $t_{d+2},...,t_{M})$ such that all the
eigenvalues of the matrix $\int\bs{\psi}\bs{\psi}^T d\mu$ are between $c_{12}
\min_{d+1\leq i\leq M}\mu(A_{i})$ and $c_{13}\max_{d+1\leq i\leq M}\mu(A_{i})$.}

\vskip.1in\noindent{\bf Lemma A.8:} \textit{Let $h$ be a bounded nonnegative
function on $[0,1]$ which is bounded away from zero except perhaps near $0$ and $1$. Assume that
$\lim_{x\rightarrow 0}x^{-\gamma}h(x)$ and $\lim_{x\rightarrow 1}(1-x)^{-\gamma}h(x)$
are positive constants for some $0<\gamma \leq 1$. Let $\bs{\psi}$ be a
(unnormalized) B-spline basis (as in Lemma A.7). Then all the eigenvalues of $\int\bs{\psi}\bs{\psi}^T h dx$
are bounded between $c_{10}M^{-1-\gamma}$ and $c_{11}M^{-1}$ for some positive
constants $c_{10},c_{11}>0$.}

\vskip.1in\noindent
Observe that under the stated condition on the density of $F_a$
in the proposition, the function
$h(v)$ appearing in (\ref{eq:smallest_eigen_bound}) has the same behavior
as stated in Lemma A.8 (after a change of location and scale). Proposition 1
now follows from using Halperin-Pitt inequality as in (\ref{eq:HP_main}),
but now taking $\epsilon \sim M^{-2-\gamma}$.

\subsection*{Rate bounds}\label{subsec:rates}

In this subsection, we summarize approximations of various quantities that
are useful in proving Lemmas A.1-A.4.
First, by {\bf A3} we have the following:
\begin{equation}\label{eq:g_beta_estimates}
\parallel g_{\bs{\beta}}^{(j)} - g_{\bs{\beta}^*}^{(j)} \parallel_\infty
= O(\alpha_N M^{j+1/2}) ~~\mbox{if}~ \parallel \bs{\beta} - \bs{\beta}^*
\parallel \leq \alpha_N, ~~j=0,1,2.
\end{equation}
Next, from {\bf A3} and {\bf A4}, for $M$ large enough, solutions
$\{X_{il}(t;\bs{\theta},\bs{\beta}):t \in [0,1]\}$ exist for all
$(\bs{\theta},\bs{\beta})$ such that $\max\{\parallel
\bs{\theta} - \bs{\theta}^*\parallel,\parallel \bs{\beta} - \bs{\beta}^*\parallel\}
\leq \alpha_N$. This also implies that the solutions $X_{il}^{\theta_i}(\cdot;\bs{\theta},\bs{\beta})$,
$X_{il}^{\beta_r}(\cdot;\bs{\theta},\bs{\beta})$, $X_{il}^{\theta_i,\theta_i}(\cdot;\bs{\theta},\bs{\beta})$,
$X_{il}^{\theta_i,\beta_r}(\cdot;\bs{\theta},\bs{\beta})$
and $X_{il}^{\beta_r,\beta_{r'}}(\cdot;\bs{\theta},\bs{\beta})$ exist on
$[0,1]$ for all $(\bs{\theta},\bs{\beta})$ such that
$\max\{\parallel \bs{\theta} - \bs{\theta}^*\parallel,\parallel \bs{\beta} - \bs{\beta}^*\parallel\}
\leq \alpha_N$, since the latter are linear differential equations
where the coefficient functions depend on $X_{il}(t;\bs{\theta},\bs{\beta})$ (see Appendix A).
Moreover, by \textit{Gronwall's lemma} (Lemma F.1), (\ref{eq:g_beta_estimates})
and the fact that $\parallel g_{\bs{\beta}^*}^{(j)}\parallel_\infty
= O(1)$ for $j=0,1,2$ (again by {\bf A3}), all these solutions
are bounded for all $\theta_i$ and $a_{il}$,  by compactness of supp$(F_a)$.

Hence, if $\alpha_N M^{3/2} = o(1)$, then using Corollary F.2
(in Appendix F),
the fact that $\parallel g_{\bs{\beta}^*}^{(j)}\parallel_\infty
= O(1)$ for $j=0,1,2$, and the expressions
for the ODEs for the partial and mixed partial derivatives (see
Appendix A), after some algebra we obtain the following (almost surely):
\begin{equation}\label{eq:X_path_bias}
\parallel X_{il}(\cdot;\bs{\theta}^*,\bs{\beta}^*) - X_{il}^g(\cdot)\parallel_\infty
= O(M^{-p}).
\end{equation}
The same technique can be used to prove the following (almost surely):
\begin{eqnarray}
\parallel X_{il}(\cdot;\bs{\theta},\bs{\beta}) - X_{il}(\cdot;\bs{\theta}^*,\bs{\beta}^*)\parallel_\infty
&=& O(\alpha_N M^{1/2})   \label{eq:X_diff_bound}\\
\parallel X_{il}^{\theta_i}(\cdot;\bs{\theta},\bs{\beta}) - X_{il}^{\theta_i}(\cdot;\bs{\theta}^*,\bs{\beta}^*)
\parallel_\infty
&=& O(\alpha_N M^{3/2}) \label{eq:X_diff_theta_bound}\\
\max_{1\leq r\leq M} \parallel X_{il}^{\beta_r}(\cdot;\bs{\theta}^*,\bs{\beta}^*) \parallel_\infty
&=& O(M^{-1/2}) \label{eq:X_beta_bound}\\
\max_{1\leq r\leq M} \parallel X_{il}^{\beta_r}(\cdot;\bs{\theta},\bs{\beta}) - X_{il}^{\beta_r}(\cdot;\bs{\theta}^*,\bs{\beta}^*)
\parallel_\infty
&=& O(\alpha_N M) \label{eq:X_diff_beta_bound}\\
\parallel X_{il}^{\theta_i,\theta_i}(\cdot;\bs{\theta},\bs{\beta}) - X_{il}^{\theta_i,\theta_i}(\cdot;\bs{\theta}^*,\bs{\beta}^*)
\parallel_\infty
&=& O(\alpha_N M^{5/2}) \label{eq:X_diff_theta_theta_bound}\\
\max_{1\leq r\leq M} \parallel X_{il}^{\theta_i,\beta_r}(\cdot;\bs{\theta}^*,\bs{\beta}^*) \parallel_\infty
&=& O(M^{1/2}) \label{eq:X_theta_beta_bound}\\
\max_{1\leq r\leq M} \parallel X_{il}^{\theta_i,\beta_r}(\cdot;\bs{\theta},\bs{\beta})
- X_{il}^{\theta_i,\beta_r}(\cdot;\bs{\theta}^*,\bs{\beta}^*) \parallel_\infty
&=& O(\alpha_N M^2) \label{eq:X_diff_theta_beta_bound}\\
\max_{1\leq r,r'\leq M} \parallel X_{il}^{\beta_r,\beta_{r'}}(\cdot;\bs{\theta}^*,\bs{\beta}^*)
\parallel_\infty &=& O(1) \label{eq:X_beta_beta_bound}\\
\max_{1\leq r,r'\leq M} \parallel X_{il}^{\beta_r,\beta_{r'}}(\cdot;\bs{\theta},\bs{\beta}) -
X_{il}^{\beta_r,\beta_{r'}}(\cdot;\bs{\theta}^*,\bs{\beta}^*) \parallel_\infty
&=& O(\alpha_N M^{3/2}) \label{eq:X_diff_beta_beta_bound}
\end{eqnarray}
whenever $\max\{\parallel \bs{\theta} - \bs{\theta}^*\parallel,
\parallel \bs{\beta} - \bs{\beta}^*\parallel\} \leq \alpha_N$.

To illustrate the key arguments, we prove (\ref{eq:X_beta_bound})
and (\ref{eq:X_diff_beta_bound}). By (\ref{eq:tilde_X_beta_closed}),
and the fact that $\parallel \phi_{r} \parallel_\infty = O(M^{1/2})$
and is supported on an interval of length $O(M^{-1})$,
(\ref{eq:X_beta_bound}) follows; in fact it holds for all $(\bs{\theta},\bs{\beta})
\in \Omega(\alpha_N)$ with $\Omega(\alpha)$ as defined in (\ref{eq:omega_alpha}).
Next, note that the function
$\phi_{r}$  is Lipschitz with Lipschitz constant
$O(M^{3/2})$ and is supported
on an interval of length $O(M^{-1})$. Since (\ref{eq:tilde_X_beta}) (in Appendix A)
is a linear differential equation, using Corollary F.2 with
\begin{eqnarray*}
&& \delta f(t,x) \\
&=& x\left[e^{\theta_i}g_{\bs{\beta}}'(X_{il}(t;\bs{\theta},
\bs{\beta})) - e^{\theta_i^*} g_{\bs{\beta}^*}'(X_{il}(t;\bs{\theta}^*,\bs{\beta}^*))
\right]
+~ e^{\theta_i}\phi_{r}(X_{il}(t;\bs{\theta},\bs{\beta})) -
e^{\theta_i^*}\phi_{r}(X_{il}(t;\bs{\theta}^*,\bs{\beta}^*))\\
&=& (e^{\theta_i} - e^{\theta_i^*}) \left[ x g_{\bs{\beta}}'(X_{il}(t;\bs{\theta},
\bs{\beta})) + \phi_{r}(X_{il}(t;\bs{\theta},\bs{\beta}))\right]
+ x e^{\theta_i^*} (g_{\bs{\beta}}'(X_{il}(t;\bs{\theta},
\bs{\beta})) - g_{\bs{\beta}}'(X_{il}(t;\bs{\theta}^*,\bs{\beta}^*)))\nonumber\\
&& + x e^{\theta_i^*} (g_{\bs{\beta}}'(X_{il}(t;\bs{\theta}^*,\bs{\beta}^*)))
- g_{\bs{\beta}^*}'(X_{il}(t;\bs{\theta}^*,\bs{\beta}^*)))
+ e^{\theta_i^*} (\phi_{r}(X_{il}(t;\bs{\theta},\bs{\beta}))
- \phi_{r}(X_{il}(t;\bs{\theta}^*,\bs{\beta}^*))),
\end{eqnarray*}
we obtain (\ref{eq:X_diff_beta_bound}) by using (\ref{eq:X_diff_bound}) and the following facts:
on $[0,1]$, $|X_{il}^{\beta_r}(t)| = O(M^{-1/2})$
for all $(\bs{\theta},\bs{\beta}) \in \Omega(\alpha_N)$;
$\parallel g_{\bs{\beta}}''\parallel_\infty = O(\alpha_N M^{5/2})$;
$\parallel g_{\bs{\beta}}' - g_{\bs{\beta}^*}'
\parallel_\infty = O(\alpha_N M^{3/2})$; and $\alpha_N M^{3/2} = o(1)$.

\subsection*{Proof of lemmas}

\noindent{\bf Proof of Lemma A.1 :}  Using
(\ref{eq:error_decomp}), write
\begin{eqnarray*}
D_1(\bs{\gamma}) &:=& \bs{\gamma}^T W
~=~\sum_{i=1}^n \sum_{l=1}^{N_i} \sum_{j=1}^{m_i} (\varepsilon_{ilj} + \Delta_{ilj}) \bs{\gamma}^T \mathbf{v}_{ilj},
\end{eqnarray*}
where $\Delta_{ilj} := X_{ilj}^g -
X_{ilj}(\bs{\theta}^*,\bs{\beta}^*)$, and $\mathbf{v}_{ilj}$ is the
$(M+n-1)\times 1$ vector with the first $M$ coordinates given by
$\mathbf{v}_{ilj}^{\beta} :=
X_{il}^{\bs{\beta}}(T_{i,j};a_{il},\theta_i^*,\bs{\beta}^*)$, and
the last $(n-1)$ coordinates given by $\mathbf{v}_{ilj}^{\theta} :=
X_{il}^{\theta_i}(T_{i,j};a_{il},\theta_i^*,\bs{\beta}^*)
\mathbf{e}_{i-1}$, where $\mathbf{e}_i$ is the $i$-th canonical
basis vector in $\mathbb{R}^{n-1}$, and $\mathbf{e}_0 :=
\bs{0}_{n-1}$. Notice that ${\cal G}(\bs\theta^*,\bs\beta^*) =
\sum_{i=1}^n \sum_{l=1}^{N_i} \sum_{j=1}^{m_i} \mathbf{v}_{ilj}
\mathbf{v}_{ilj}^T$. Thus, by Cauchy-Schwarz inequality, and the
fact that $\max_{i,l,j} |\Delta_{ilj}| = O(M^{-p})$ (by
(\ref{eq:X_path_bias})) we have, uniformly in $\bs\gamma$,
\begin{eqnarray*}
|\sum_{i=1}^n \sum_{l=1}^{N_i} \sum_{j=1}^{m_i} \Delta_{ilj} \bs{\gamma}^T \mathbf{v}_{ilj}| &=&
O(M^{-p} (\ol{N}\ol{m})^{1/2}) \sqrt{\bs{\gamma}^T{\cal G}(\bs{\theta}^*,\bs{\beta}^*)\bs{\gamma}}.
\end{eqnarray*}
Since $\varepsilon_{ilj}$ are i.i.d. $N(0,\sigma_\varepsilon^2)$, we
also have
$$
\sum_{i=1}^n \sum_{l=1}^{N_i} \sum_{j=1}^{m_i} \varepsilon_{ilj} \bs{\gamma}^T \mathbf{v}_{ilj} \sim
N(0,\sigma_\varepsilon^2 \bs{\gamma}^T{\cal G}(\bs{\theta}^*,\bs{\beta}^*)\bs{\gamma})
$$
conditional on $(\bs{a},\mathbf{T})$. Since the (conditional)
Gaussian process
$$
f(\bs\gamma) := \frac{\sum_{i=1}^n \sum_{l=1}^{N_i}
\sum_{j=1}^{m_i} \varepsilon_{ilj} \bs{\gamma}^T
\mathbf{v}_{ilj}}{\sigma_\varepsilon\sqrt{\bs{\gamma}^T{\cal
G}(\bs{\theta}^*,\bs{\beta}^*)\bs{\gamma}}}
$$
is a smooth function over $\mathbb{S}^{M+n-1}$ (the unit sphere
centered at 0 in $\mathbb{R}^{M+n-1}$), and since by assumption $M =
O((\ol{N}\ol{m})^d)$ for some $d > 0$, using a covering of the
sphere $\mathbb{S}^{M+n-1}$ by balls of radius $\epsilon_M \sim
(\ol{N}\ol{m})^{-D}$ for an appropriately chosen $D > 0$, and using
the fact that $P(N(0,1) > t) \leq t^{-1} {(2\pi)}^{-1/2}
\exp(-t^2/2)$, for $t > 0$, we conclude that uniformly in
$\bs\gamma$,
$$
\sum_{i=1}^n \sum_{l=1}^{N_i} \sum_{j=1}^{m_i} \varepsilon_{ilj} \bs{\gamma}^T \mathbf{v}_{ilj}
= O(\sigma_\varepsilon M^{1/2}\sqrt{\log(\ol{N}\ol{m})}) \sqrt{\bs{\gamma}^T{\cal G}(\bs{\theta}^*,\bs{\beta}^*)\bs{\gamma}}
$$
except on a set with probability converging to zero. This completes
the proof of the lemma.

\vskip.1in\noindent{\bf Proof of Lemma A.2 :} Define $\mathbf{u}_{ilj}$ the same way
as $\mathbf{v}_{ilj}$
is defined in the proof of Lemma A.1, with $(\bs\theta^*,\bs\beta^*)$ replaced by $(\ol{\bs\theta},\ol{\bs\beta})$.
Express
$D_2(\bs{\gamma}) := \bs{\gamma}^T({\cal G}(\ol{\bs{\theta}},\ol{\bs{\beta}}) - {\cal G}(\bs{\theta}^*,\bs{\beta}^*))
\bs{\gamma}$ as
\begin{eqnarray*}
&& \bs{\gamma}^T \left[\sum_{i=1}^n\sum_{l=1}^{N_i} \sum_{j=1}^{m_i} \mathbf{u}_{ilj}\mathbf{u}_{ilj}^T -
\sum_{i=1}^n\sum_{l=1}^{N_i} \sum_{j=1}^{m_i} \mathbf{v}_{ilj}\mathbf{v}_{ilj}^T \right]\bs{\gamma}\nonumber\\
&=& \bs{\gamma}^T \left[\sum_{i=1}^n\sum_{l=1}^{N_i} \sum_{j=1}^{m_i} ((\mathbf{u}_{ilj} - \mathbf{v}_{ilj}) \mathbf{v}_{ilj}^T
+ (\mathbf{u}_{ilj} - \mathbf{v}_{ilj}) (\mathbf{u}_{ilj} - \mathbf{v}_{ilj})^T + \mathbf{v}_{ilj}
(\mathbf{u}_{ilj} - \mathbf{v}_{ilj})^T)\right]\bs{\gamma}.
\end{eqnarray*}
Then, by Cauchy-Schwarz inequality and (\ref{eq:X_diff_theta_bound}) and (\ref{eq:X_diff_beta_bound}),
and the arguments used in the proof of Lemma A.1,
\begin{eqnarray*}
|D_2(\bs{\gamma})| &=& O(\alpha_N M^{3/2}(\ol{N}\ol{m})^{1/2}) \sqrt{\bs{\gamma}^T {\cal G}(\bs{\theta}^*,\bs{\beta}^*) \bs{\gamma}}
+ O(\alpha_N^2 M^3 \ol{N}\ol{m} ).
\end{eqnarray*}

\vskip.1in\noindent{\bf Proof of Lemma A.3 :} Define
$D_3(\bs{\gamma}) := \bs{\gamma}^T ({\cal
G}(\bs{\theta}^*,\bs{\beta}^*)  - {\cal G}_*) \bs{\gamma}$. Then
$$
D_3(\bs\gamma) = \sum_{i=1}^n\sum_{l=1}^{N_i} \sum_{j=1}^{m_i}
u_{ilj}(\bs\gamma) + \sum_{i=1}^n\sum_{l=1}^{N_i} \sum_{j=1}^{m_i} w_{ilj}(\bs\gamma),
$$
where $u_{ilj}(\bs\gamma) = \bs\gamma^T (\nabla_{i}
X_{ilj}\nabla_{i} X_{ilj}^T  - \mathbb{E}[(\nabla_{i}
X_{ilj}\nabla_{i} X_{ilj}^T)|a_{il}])\bs\gamma$ and
$w_{ilj}(\bs\gamma) = \bs\gamma^T (\mathbb{E}[(\nabla_{i}
X_{ilj}\nabla_{i} X_{ilj}^T)|a_{il}]  - \mathbb{E}[\nabla_{i}
X_{ilj}\nabla_{i} X_{ilj}^T)])\bs\gamma$, where, for notational
simplicity,
$$
\nabla_i X_{ilj} = \begin{bmatrix} X_{il}^{\bs{\beta}}(T_{i,j}) \\ X_{il}^{\theta_i}(T_{i,j}) \mathbf{e}_{i-1} \\
\end{bmatrix}, \qquad i=1,\ldots,n.
$$
Note that, the random variables $u_{ilj}(\bs\gamma)$ have zero
conditional mean (given $a_{il}$), are uniformly bounded and the
variables $Z_{ij}(\bs\gamma) := \sum_{l=1}^{N_i} u_{ilj}(\bs\gamma)$
are independent. Similarly, the random variables $\{w_{ilj}\}_{i,j}$
have zero mean are uniformly bounded and the variables
$\sum_{j=1}^{m_i} w_{ilj}(\bs\gamma)$ are independent. Indeed, for
each fixed $(i,l)$, the variables $\{w_{ilj}\}_{j=1}^{m_i}$ are
identical since $T_{i,j}$ are i.i.d. Moreover, the collections
$\{u_{ilj}(\bs\gamma)\}$ and $\{w_{ilj}(\bs\gamma)\}$ are
differentiable functions of $\gamma$. Define ${\cal G}_*(\bs{a}) :=
\mathbb{E}({\cal G}_*(\bs\theta^*,\bs\beta^*)|\bs{a})$. Then, since
$Z_{ij}(\bs\gamma)$ are uniformly bounded by $K_1 \ol{N}$ for some
constant $K_1 > 0$, and are independent given $\bs{a}$, we have
\begin{eqnarray*}
\mbox{Var}(\sum_{i=1}^n\sum_{j=1}^{m_i} Z_{ij}(\bs\gamma) |\bs{a}) &=&
\sum_{i=1}^n\sum_{j=1}^{m_i} \mathbb{E}[(Z_{ij}(\bs\gamma))^2|\bs{a}] \nonumber\\
&\leq& \sum_{i=1}^n\sum_{j=1}^{m_i} N_i \sum_{l=1}^{N_i} \mathbb{E}[u_{ilj}^2(\bs\gamma)|a_{il}] \nonumber\\
&\leq& K_2\ol{N} \sum_{i=1}^n\sum_{l=1}^{N_i}\sum_{j=1}^{m_i}\mathbb{E}[|u_{ilj}(\bs\gamma)||a_{il}] ~\leq~
2K_2\ol{N} \bs\gamma^T{\cal
G}_*(\bs{a}) \bs\gamma .
\end{eqnarray*}
In the above, second inequality uses  $(\sum_{i=1}^N x_i)^2 \leq
N\sum_{i=1}^N x_i^2$, and the last follows from fact that
$u_{ilj}(\bs\gamma)$ is a difference of two nonnegative quantities,
the second one being the conditional expectation of the first one
given $\bs{a}$. Thus, applying Bernstein's inequality, for every $v
> 0$, for every $\bs\gamma \in \mathbb{S}^{M+n-1}$,
\begin{equation*}
\mathbb{P}(|\sum_{i=1}^n\sum_{l=1}^{N_i} \sum_{j=1}^{m_i}
u_{ilj}(\bs\gamma)| > v|\bs{a}) \leq 2\exp\left(-\frac{v^2/2}{2K_2 \ol{N}\bs\gamma^T{\cal
G}_*(\bs{a}) \bs\gamma + K_1 \ol{N} v/3}\right).
\end{equation*}
Thus, using  an entropy argument as in the proof of Lemma A.1, we
conclude that given $\delta > 0$ there exist positive constants
$C_1(\delta)$ and $C_2(\delta)$ such that on the set $\{\bs{a} |
\bs\gamma^T{\cal G}_*(\bs{a}) \bs\gamma \geq C_2(\delta) \ol{N} M
\log(\ol{N}\ol{m})\}$,
\begin{equation}\label{eq:sum_u_bound}
\mathbb{P}\left(\sup_{\bs\gamma \in \mathbb{S}^{M+n-1}} \frac{|\sum_{i=1}^n\sum_{l=1}^{N_i} \sum_{j=1}^{m_i}
u_{ilj}(\bs\gamma)|}{\sqrt{\bs\gamma^T{\cal
G}_*(\bs{a}) \bs\gamma}} > C_1(\delta) (\ol{N}M
\log(\ol{N}\ol{m}))^{1/2} ~|~\bs{a}\right) \leq (\ol{N}\ol{m})^{-\delta}.
\end{equation}

On the other hand, using an inversion formula for block matrices,
\begin{eqnarray}\label{eq:G_star_inv_bound}
\parallel
{\cal G}_*^{-1}
\parallel
&=&
\parallel
\begin{bmatrix}
 {\cal G}_{*,\beta\beta} & {\cal G}_{*,\beta\theta} \\
{\cal G}_{*,\theta\beta}  & {\cal G}_{*,\theta\theta} \\
\end{bmatrix}^{-1}
\parallel \nonumber\\
&=&
\parallel
\begin{bmatrix}
C_*^{-1} & - C_*^{-1} {\cal G}_{*,\beta\theta} ({\cal G}_{*,\theta\theta} )^{-1} \\
-({\cal G}_{*,\theta\theta} )^{-1}  {\cal G}_{*,\theta\beta}  C_*^{-1} & ({\cal G}_{*,\theta\theta} )^{-1}
+ ({\cal G}_{*,\theta\theta} )^{-1}  {\cal G}_{*,\theta\beta}  C_*^{-1} {\cal G}_{*,\beta\theta}
({\cal G}_{*,\theta\theta} )^{-1}
\end{bmatrix}
\parallel  \nonumber\\
&=& O(\kappa_M (\ol{N}\ol{m})^{-1}),
\end{eqnarray}
where $C_* := {\cal G}_{*,\beta\beta} - {\cal G}_{*,\beta\theta}
({\cal G}_{*,\theta\theta})^{-1} {\cal G}_{*,\theta\beta}$. The last
equality in (\ref{eq:G_star_inv_bound}) is because {\bf A6} together
with (\ref{eq:X_beta_bound}) implies in particular that $\parallel
{\cal G}_{*,\beta\theta} ({\cal G}_{*,\theta\theta})^{-1}\parallel =
O(1)$. Now, from the facts that
$$
\bs\gamma^T {\cal G}_* \bs\gamma \geq K_3 \frac{\ol{N}\ol{m}}{\kappa_M}
~~(\mbox{by}~(\ref{eq:G_star_inv_bound})) \qquad\mbox{and}\qquad  \min\{\ol{N},\ol{m}\} \gg \kappa_M M
\log(\ol{N}\ol{m}),
$$
for some constant $K_3 >0$, so that $\bs\gamma^T{\cal G}_* \bs\gamma
\gg \ol{m} M \log(\ol{N}\ol{m})$, and using arguments similar to
those leading to (\ref{eq:sum_u_bound}) we have, for some
$C_3(\delta) > 0$,
\begin{equation}\label{eq:sum_w_bound}
\mathbb{P}\left(\sup_{\bs\gamma \in \mathbb{S}^{M+n-1}} \frac{|\sum_{i=1}^n\sum_{l=1}^{N_i} \sum_{j=1}^{m_i}
w_{ilj}(\bs\gamma)|}{\sqrt{\bs\gamma^T{\cal
G}_* \bs\gamma}} > C_3(\delta) (\ol{m}M
\log(\ol{N}\ol{m}))^{1/2} \right) \leq (\ol{N}\ol{m})^{-\delta}.
\end{equation}
Now, observing that $\sum_{i=1}^n\sum_{l=1}^{N_i} \sum_{j=1}^{m_i}
w_{ilj}(\bs\gamma) = \bs\gamma^T{\cal G}_*(\bs{a}) \bs\gamma -
\bs\gamma^T{\cal G}_* \bs\gamma$, using fact [{\bf Q}], and
combining with (\ref{eq:sum_u_bound}) and (\ref{eq:sum_w_bound}), we
obtain that, there is a $C_4(\delta) > 0$ such that
\begin{equation}\label{eq:D_3_gamma_prob_limit}
\mathbb{P}\left(\sup_{\bs\gamma \in \mathbb{S}^{M+n-1}}\frac{|D_3(\bs\gamma)|}
{\sqrt{\bs\gamma^T{\cal G}_* \bs\gamma}} \geq C_4(\delta) (\ol{m}^{1/2}+\ol{N}^{1/2})
(M \log(\ol{N}\ol{m}))^{1/2} \right) = O((\ol{N}\ol{m})^{-\delta}).
\end{equation}
(Note that, if the time points $\{t_{ilj}\}$ were independently and
identically distributed for different curves $(i,l)$, then quantity
$(\ol{m}^{1/2}+\ol{N}^{1/2}) (M \log(\ol{N}\ol{m}))^{1/2}$ in
(\ref{eq:D_3_gamma_prob_limit}) can be replaced by $(M
\log(\ol{N}\ol{m}))^{1/2}$). From (\ref{eq:D_3_gamma_prob_limit}) it
follows that
\begin{eqnarray*}
\bs{\gamma}^T {\cal G}(\bs{\theta}^*,\bs{\beta}^*) \bs{\gamma}
&\geq&  \bs\gamma^T {\cal G}_* \bs\gamma (1-o_P(1))
~\geq~ c_6 \kappa_M^{-1} \ol{N}\ol{m}(1-o_P(1))
\end{eqnarray*}
for some constant $c_6 >  0$ and for sufficiently large $\ol{N}$.

\vskip.1in\noindent{\bf Proof of Lemma A.4 :} Using (\ref{eq:tilde_X_beta_Hessian_alt}) and
(\ref{eq:tilde_X_beta_theta_Hessian_alt}) in Appendix A, for any $t$,
we can express the $(M+n-1) \times (M+n-1)$
matrix with blocks $X_{il}^{\bs{\beta},\bs{\beta}^T}(t) := (( X_{il}^{\beta_r,\beta_{r'}}(t)))_{r,r'=1}^M$,
$\mathbf{e}_{i-1} X_{il}^{\theta_i,\bs{\beta}^T}(t)$, $X_{il}^{\bs{\beta},\theta_i}(t)\mathbf{e}_{i-1}^T$
and $X_{il}^{\theta_i,\theta_i}(t)$, as $U_{il}(t) + U_{il}(t)^T + V_{il}(t)$ where
\begin{eqnarray}\label{eq:X_Hess_repr}
U_{il}(t) &=& e^{\theta_i} g_{\bs{\beta}}(X_{il}(t))\int_0^t \frac{1}{g_{\bs{\beta}}(X_{il}(s))}
\begin{bmatrix} X_{il}^{\bs{\beta}}(s) \\ X_{il}^{\theta_i}(s) \mathbf{e}_{i-1} \\
\end{bmatrix}
\begin{bmatrix} \bs{\phi}'(X_{il}(s))\\ \bs{0}_{n-1} \\ \end{bmatrix}^T ds
\end{eqnarray}
and $\parallel V_{il}(t) \parallel = O(1)$ uniformly in $t$, $i$ and $l$.
Note that, in the above description, all the sample paths and their
derivatives are evaluated at $(\bs\theta^*,\bs\beta^*)$.

Observe that, since $Y_{ilj} - X_{il}(T_{i,j};\bs{\theta}^*,\bs{\beta}^*) = \varepsilon_{ilj} + \Delta_{ilj}$,
where $\Delta_{ilj}$ is as in the proof of Lemma A.1, we have
\begin{eqnarray}\label{eq:H_diff_expansion}
&& {\cal H}(\ol{\bs{\theta}},\ol{\bs{\beta}}) - {\cal H}(\bs{\theta}^*,\bs{\beta}^*) \nonumber\\
&=& \sum_{i=1}^n\sum_{l=1}^{N_i}\sum_{j=1}^{m_i} (X_{ilj}(\bs{\theta}^*,\bs{\beta}^*)
- X_{ilj}(\ol{\bs{\theta}},\ol{\bs{\beta}}))
\begin{bmatrix}
X_{ilj}^{\bs{\beta},\bs{\beta}^T}(\ol{\bs{\theta}},\ol{\bs{\beta}}) &
X_{ilj}^{\bs{\beta},\theta_i}(\ol{\bs{\theta}},\ol{\bs{\beta}})\mathbf{e}_{i-1}^T  \\
\mathbf{e}_{i-1} X_{ilj}^{\theta_i,\bs{\beta}^T}(\ol{\bs{\theta}},\ol{\bs{\beta}})
& X_{ilj}^{\theta_i,\theta_i}(\ol{\bs{\theta}},\ol{\bs{\beta}})\mathbf{e}_{i-1}\mathbf{e}_{i-1}^T
\end{bmatrix}\nonumber\\
&& + \sum_{i=1}^n\sum_{l=1}^{N_i}\sum_{j=1}^{m_i} (\varepsilon_{ilj} + \Delta_{ilj}) ~\cdot \nonumber\\
&&  \begin{bmatrix}
X_{ilj}^{\bs{\beta},\bs{\beta}^T}(\ol{\bs{\theta}},\ol{\bs{\beta}}) -
X_{ilj}^{\bs{\beta},\bs{\beta}^T}(\bs{\theta}^*,\bs{\beta}^*) &
(X_{ilj}^{\bs{\beta},\theta_i}(\ol{\bs{\theta}},\ol{\bs{\beta}})
- X_{ilj}^{\bs{\beta},\theta_i}(\bs{\theta}^*,\bs{\beta}^*))\mathbf{e}_{i-1}^T \\
\mathbf{e}_{i-1} (X_{ilj}^{\theta_i,\bs{\beta}^T}(\ol{\bs{\theta}},\ol{\bs{\beta}})
- X_{ilj}^{\theta_i,\bs{\beta}^T}(\bs{\theta}^*,\bs{\beta}^*)  )  &
(X_{ilj}^{\theta_i,\theta_i}
(\ol{\bs{\theta}},\ol{\bs{\beta}})- X_{ilj}^{\theta_i,\theta_i}
(\bs{\theta}^*,\bs{\beta}^*)) \mathbf{e}_{i-1}\mathbf{e}_{i-1}^T
\end{bmatrix}.
\end{eqnarray}
First break the last summation in the last term of
(\ref{eq:H_diff_expansion}) into two parts -- one corresponding to
$\Delta_{ilj}$'s and the other corresponding to
$\varepsilon_{ilj}$'s. Then, using (\ref{eq:X_path_bias}),
(\ref{eq:X_diff_theta_theta_bound}),
(\ref{eq:X_diff_theta_beta_bound}) and
(\ref{eq:X_diff_beta_beta_bound}), we conclude that the sum
involving $\Delta_{ilj}$ is $O(\alpha_N M^{5/2-p} \ol{N}\ol{m})$.
The summation involving $\varepsilon_{ilj}$'s can be expressed as a
linear function of $\bs\varepsilon$ with coefficients that are
functions of $\bs{a},\bs{T}$ and $\bs{\gamma}$, and depend smoothly
on $\bs{\gamma}$. From this, conditionally on $(\bs{a},\bs{T})$,
this term is coordinatewise normally distributed with standard
deviation $O(\sigma_\varepsilon \alpha_N M^{5/2}
(\ol{N}\ol{m})^{1/2})$ for each fixed $\bs{\gamma}$. We can conclude
from this by an entropy argument (similar to the one used in the proof
of Lemma A.1) that the supremum of this term over all $\bs{\gamma}
\in \mathbb{S}^{M+n-1}$ is $O(\sigma_\varepsilon
\alpha_N M^{3} (\ol{N}\ol{m})^{1/2}\sqrt{\log(\ol{N}\ol{m})})$
with probability tending to 1.

Next, using (\ref{eq:X_Hess_repr}) we express the first term of
(\ref{eq:H_diff_expansion}) as
\begin{eqnarray}\label{eq:X_Hess_repr_first}
&&  \sum_{i=1}^n\sum_{l=1}^{N_i}\sum_{j=1}^{m_i} (X_{ilj}(\bs{\theta}^*,\bs{\beta}^*)
- X_{ilj}(\ol{\bs{\theta}},\ol{\bs{\beta}})) \left[U_{il}(t_{ilj}) + U_{il}(t_{ilj})^T + V_{il}(t_{ilj})\right]
\nonumber\\
&& + \sum_{i=1}^n\sum_{l=1}^{N_i}\sum_{j=1}^{m_i} (X_{ilj}(\bs{\theta}^*,\bs{\beta}^*)
- X_{ilj}(\ol{\bs{\theta}},\ol{\bs{\beta}}))  \cdot \nonumber\\
&&  \begin{bmatrix}
X_{ilj}^{\bs{\beta},\bs{\beta}^T}(\ol{\bs{\theta}},\ol{\bs{\beta}}) -
X_{ilj}^{\bs{\beta},\bs{\beta}^T}(\bs{\theta}^*,\bs{\beta}^*) &
(X_{ilj}^{\bs{\beta},\theta_i}(\ol{\bs{\theta}},\ol{\bs{\beta}})
- X_{ilj}^{\bs{\beta},\theta_i}(\bs{\theta}^*,\bs{\beta}^*))\mathbf{e}_{i-1}^T \\
\mathbf{e}_{i-1} (X_{ilj}^{\theta_i,\bs{\beta}^T}(\ol{\bs{\theta}},\ol{\bs{\beta}})
- X_{ilj}^{\theta_i,\bs{\beta}^T}(\bs{\theta}^*,\bs{\beta}^*)  )  &
(X_{ilj}^{\theta_i,\theta_i}
(\ol{\bs{\theta}},\ol{\bs{\beta}})- X_{ilj}^{\theta_i,\theta_i}
(\bs{\theta}^*,\bs{\beta}^*)) \mathbf{e}_{i-1}\mathbf{e}_{i-1}^T
\end{bmatrix}.
\end{eqnarray}
The second sum is $O(\alpha_N^2 M^{3} \ol{N}\ol{m})$ by
(\ref{eq:X_diff_bound}), (\ref{eq:X_diff_theta_theta_bound}),
(\ref{eq:X_diff_theta_beta_bound}) and
(\ref{eq:X_diff_beta_beta_bound}). Again, by
(\ref{eq:X_diff_bound}), the contribution in the first sum for the
term involving $V_{il}(t_{ilj})$'s is $O(\alpha_N M^{1/2}
\ol{N}\ol{m})$. Finally, by Cauchy-Schwarz inequality, and
(\ref{eq:X_Hess_repr}), and the facts that $ g_{\bs\beta}$ is
bounded both above and below (for $x\geq x_0$),  we have for all $0
\leq t \leq 1$,
\begin{eqnarray*}
|\bs\gamma^T U_{il}(t) \bs\gamma|^2
&\leq& c_{g} \int_0^t
\left(\bs\gamma^T \begin{bmatrix} X_{il}^{\bs{\beta}}(s) \\ X_{il}^{\theta_i}(s) \mathbf{e}_{i-1} \\
\end{bmatrix}\right)^2  ds  \int_0^1
\left(\bs\gamma^T \begin{bmatrix} \bs{\phi}'(X_{il}(s))\\ \bs{0}_{n-1} \\ \end{bmatrix}
\right)^2 ds,
\end{eqnarray*}
for some constant $c_{g} > 0$ depending on $g_{\bs{\beta}^*}$,
$\bs{\theta}$ and $x_0$. Notice that $\int \bs{\phi}'(x) (\bs{\phi}'(x))^T dx \leq
c_7 M^2 I_M$ for some $c_7 > 0$. Furthermore,
$$
\sup_{\bs\gamma \in \mathbb{S}^{M+n-1}}
\sup_{t \in[0,1]} \int_0^t \left(\bs\gamma^T \begin{bmatrix} X_{il}^{\bs{\beta}}(s) \\ X_{il}^{\theta_i}(s) \mathbf{e}_{i-1} \\
\end{bmatrix}\right)^2  ds  = O(1).
$$
These, together with an application of Cauchy-Schwarz inequality,
and using manipulations as in the proof of Lemmas A.2 and A.3, shows
that the sum involving the terms $U_{il}(t_{ilj})$'s in the
expression (\ref{eq:X_Hess_repr_first}) is
$O(\alpha_N M^{3/2}(\ol{N}\ol{m})^{1/2}) \sqrt{\bs{\gamma}^T {\cal
G}_* \bs{\gamma}}$. Lemma A.4 now follows.

\vskip.1in\noindent{\bf Proof of Lemma A.5 :} Suppose that the result is false.
Then for any sequence of positive constant $\delta_{n}$ decreasing to zero,
we can find a polynomial $p_{n}$ with coefficients $\beta_{n,j}$, $j=0,...,d$,
such that $|p_{n}|_{\infty} \leq\delta_{n}\max_{0\leq j\leq d}|\beta_{n,j}|$.
Let $q_{n}$ be the polynomial whose coefficients $\{\gamma_{n,j}\}$ are obtained by dividing
$\beta_{j}$'s by $\max_{0\leq j\leq d}|\beta_{n,j}|$. Note that $\max
|\gamma_{n,j}|=1$ for any $n$. By the usual compactness argument we can
find a subsequence of $\{\gamma_{n,j}\}$, which we continue to denote by
$\{\gamma_{n,j}\}$, such that $\gamma_{n,j}\rightarrow\gamma_{j}$,
$j=0,...,d$, where $\max_{j}|\gamma_{j}|=1$. However the supremum norm of the
limiting polynomial $q$ of degree $d$ with coefficients $\gamma_{j}$ is zero
which implies that $\gamma_{j}=0$ for all $j$. This leads to a contradiction.

\vskip.1in\noindent{\bf Proof of Lemma A.6 :} Note that for any interval
$A=[a,b]$, $a<b$, we can write $\int_{A}p^{2}d\mu=\int_{0}^{1}q^{2}d\mu_{A}$,
where $q(z)=p(a+(b-a)z)$ and $d\mu_{A}(z)=d\mu(a+(b-a)z)$.
Since $|q|_{\infty}=\sup_{u\in A}|p(u)|$, we may
take $|q|_{\infty}=1$. Let $x^{*}$ be a point in $[0,1]$ at which
$q(x^{*})$ equals $\pm1$. Then for any $0\leq x\leq1$, $q^{2}(x)=1+(x-x^*)^{2}
q(x^{**})q^{\prime\prime}(x^{**})$ for some
$x^{**}$ in $[0,1]$. Using Lemma A.5, we see that there is a constant
$c_{14}$ such that $|p^{\prime\prime}|_{\infty}\leq c_{14}$ for all polynomials
with $|p|_{\infty}=1$. So $|q^{2}(x)-1|\leq c_{14}(x-x^{\ast})^{2}$. So we can
find an interval $I\subset [0,1]$ of length at least $L=(2c_{14})^{-1/2}$
containing $x^{*}$ such that $q^{2}(x)\geq 1/2$. Let $B=a+(b-a)I$. Then
$length(B)/length(A)=length(I)\geq L$. Consequently,
\begin{eqnarray*}
\int_{A}p^{2}d\mu &=& \int_{0}^{1}q^{2}d\mu_{A} ~\geq~ \frac{1}{2}\int_{I}d\mu_{A} \\
&=& \frac{1}{2}\mu(B) ~\geq~ \frac{1}{2} C(L)\mu(A).
\end{eqnarray*}
The result now follows.

\vskip.1in\noindent{\bf Proof of Lemma A.7 :} This result is clearly true for $d=0$.
We will prove it for the case $d \geq 1$. Let $\bs{\beta} \in \mathbb{R}^{k}$
and let $s(x)=\bs{\beta}^T \bs{\psi}(x)$. Since $s$ is a convex combination of
$\beta_{i-d},...,\beta_{i}$ on the interval $A_{i}$, we have
\begin{equation*}
\int s^{2}d\mu = \sum_{d+1\leq i\leq M}\int_{t_{i}}^{t_{i+1}} s^{2}d\mu\leq
\sum_{d+1\leq i\leq M}\sum_{i-d\leq j\leq i}\beta_{j}^{2}\mu(A_{i}).
\end{equation*}
This establishes the upper bound for the largest eigenvalue of $\int\bs{\psi}
\bs{\psi}^T$. We will now establish the result on the lower bound of the
smallest eigenvalue.

Using property (viii) in chapter XI in de Boor (1978), we know that
$\sup_{t_{i+1}\leq x\leq t_{i+d+1}}|s(x)|\geq c_{15}|\beta_{i}|$ for all $i$
for some constant $c_{15}>0$. Denote  $m_0 =\min_{d+1\leq i\leq M}\mu(A_{i})$.
Hence for any $d\leq i\leq M$, by Lemma A.6 we have
\begin{eqnarray*}
\int_{t_{i+1}}^{t_{i+d+1}}s^{2}d\mu &=& \sum_{i+1\leq j\leq i+d+1}\int_{t_{j}}^{t_{j+1}}
s^{2}d\mu\\
&\geq& c\sum_{i+1\leq j\leq i+d+1}\sup_{t_{j}\leq x\leq t_{j+1}}|s(x)|^{2}
\mu(A_{j})\\
&\geq& c\sup_{t_{i+1}\leq x\leq t_{i+d+1}}|s(x)|^{2}m_0 ~\geq~ c_{16}\beta_{i}^{2}m_0.
\end{eqnarray*}
Incidentally, the same type of inequality holds for any $i=1,..,d-1$.
Consequently we have$\,\int s^{2}d\mu\geq c_{17}\sum\beta_{i}^{2}m_0$ for
some constant $c_{17}>0.$ This completes the proof.

\vskip.1in\noindent{\bf Proof of Lemma A.8:}  This follows from Lemma A.7 once
we take $d\mu(x) = h(x) dx$.

\section*{Appendix F : Perturbation of Differential Equations}

For nonparametric estimation of the gradient function $g$, we need
to control the effect of lack of fit to $g$ (meaning that $g$ may
not be exactly represented in the given basis $\{\phi_k(\cdot)\}$)
on the sample paths $\{X_{il}(t): t \in [0,1]\}$. It is
convenient to do this study under a general setting of first order
differential equations where the state variable $x(\cdot)$ is
$d$-dimensional for $d\geq 1$. Our aim is to control the
perturbation of the sample paths and its derivatives
with respect to the parameters governing the differential
equation when the \textit{true} gradient
function $g$ is perturbed by an arbitrary function $\delta
g(\cdot)$.

We present two different results about the perturbation of the
solution paths of the initial value problem:
\begin{equation}\label{eq:ODE_general}
x' = f(t,x),~~~x(t_0) = x_0,
\end{equation}
where $x \in \mathbb{R}^d$, when the function $f$ is perturbed by a
smooth function.

\vskip.1in \noindent{\bf Theorem F.1 (Deuflhard and Bornemann, 2002,
p.80) :} \textit{On the augmented phase space $\Omega$ let the
mappings $f$ and $\delta f$ be continuous and continuously
differentiable with respect to the state variable. Assume that for
$(t_0,x_0) \in \Omega$, the initial value problem
(\ref{eq:ODE_general}), and the perturbed problem
\begin{equation}\label{eq:ODE_perturbed}
x' = f(t,x) + \delta f(t,x),~~~x(t_0) = x_0,
\end{equation}
have the solutions $x$ and $\ol x = x + \delta x$, respectively.
Then for $t_1$ sufficiently close to $t_0$, there exists a
continuous matrix-valued mapping $M : \Delta \to \mathbb{R}^{d\times
d}$ on $\Delta = \{(t,s) \in \mathbb{R}^2: t\in [t_0,t_1], s \in
[t_0,t]\}$ such that the perturbation $\delta x$ is represented by
\begin{equation}\label{eq:solution_perturbed}
\delta x(t) = \int_{t_0}^t M(t,s) \delta f(s,\ol{x}(s)) ds,
~~~\mbox{for all}~t \in [t_0,t_1].
\end{equation}
}

Note that, the point $t_1$ can be chosen so that, the one-parameter
family of initial value problems
\begin{equation}\label{eq:ODE_perturbed_lambda}
x' = f(t,x) + \lambda \cdot \delta f(t,x),~~~x(t_0) = x_0,
\end{equation}
has a corresponding solution $\phi(\cdot;\lambda) \in
C^1([t_0,t_1],\mathbb{R}^d)$ for each parameter value $\lambda \in
[0,1]$. In particular, $\phi(\cdot;0) = x(\cdot)$ and $\phi(\cdot;1)
= \ol x(\cdot)$.

\subsubsection*{Propagation matrix and its relationship to
perturbation}

Let $\Phi^{t,t_0}$ denote the map such that $x(t) = \Phi^{t,t_0}
x_0$ is the unique solution of the initial value problem
(\ref{eq:ODE_general}). The following result (Theorem 3.1 in
Deuflhard and Bornemann, 2002, p.77) describes the dependence of the
map on the gradient function $f$.

\vskip.1in \noindent{\bf Theorem F.2 :} \textit{On the extended state
space $\Omega$ let $f$ be continuous and $p$-times continuously
differentiable, $p \geq 1$, with respect to the state variable.
Moreover, suppose that for $(t_0,x_0) \in \Omega$ the unique
solution of the initial value problem (\ref{eq:ODE_general}) exists
up to some time $t > t_0$. Then there is a neighborhood of the the
state $x_0$ where for all $s \in [t_0,t_1]$, the evolution
\begin{equation*}
x \to \Phi^{s,t_0} x
\end{equation*}
is $p$-times continuously differentiable with respect to the state
variable. In other words, the evolution inherits from the right side
the smoothness properties with respect to the state variable.}

\vskip.1in

Then, the linearized perturbation of the state, due to a
perturbation $\delta x_s$ of the state at time $s$, namely,
\begin{equation*}
\delta x(t) \approx \Phi^{t,s} (x(s) + \delta x_s) - \Phi^{t,s}x(s)
\end{equation*}
is given by $\delta x(t) = W(t,s) \delta x_s$ where
\begin{equation}\label{eq:W_t_s}
W(t,s) = D_\xi \Phi^{t,s}\xi \left|_{\xi = \Phi^{s,t_0}x_0}\right.
\in \mathbb{R}^{d\times d}
\end{equation}
is the \textit{Jacobi matrix}. Note that, $W(t,s)$ satisfies the
differential equation:
\begin{equation}\label{eq:W_t_s_ODE}
\frac{d}{dt}W(t,s) = f_x(t,\Phi^{t,t_0}x_0)W(t,s),
\end{equation}
with initial condition $W(s,s) = I$. $W(t,s)$ is called the
\textit{propagation matrix} belonging to $x$.

In general, we can express the matrix $M(t,s)$ appearing in Theorem
F.1 as
\begin{equation}\label{eq:M_t_s}
M(t,s) = \int_0^1 W(t,s;\lambda) d\lambda,
\end{equation}
where $W(t,s;\lambda)$ is the propagation matrix belonging to
$\phi(\cdot;\lambda)$, and hence solves the \textit{homogeneous}
differential equation
\begin{equation*}
\frac{d}{dt} W(t,s;\lambda) = f_x(t,\phi(t;\lambda))
W(t,s;\lambda),~~~~W(s,s) = I.
\end{equation*}
From this, the following corollary follows easily.

\vskip.1in\noindent{\bf Corollary F.1 :} \textit{If the
limit $\delta f \to 0$ is uniform in a neighborhood of the graph of
the solution $x$, then the linearization
\begin{equation*}
\delta x(t) \approx \int_{t_0}^t W(t,s) \delta f(s,x(s))
ds,~~~\mbox{for all}~ t\in [t_0,t_1]
\end{equation*}
holds.}

\subsubsection*{Gronwall's Lemma and its implications}

{\bf Lemma F.1 (Gronwall's Lemma):} \textit{Let $\psi, \chi \in
C([t_0,t_1],\mathbb{R})$ be nonnegative functions and $\rho \geq 0$.
Then the integral inequality
\begin{equation*}
\psi(t) \leq \rho + \int_{t_0}^t \chi(s) \psi(s) ds, ~~~~\mbox{for
all}~t \in [t_0,t_1]
\end{equation*}
implies
\begin{equation*}
\psi(t) \leq \rho \exp\left(\int_{t_0}^t \chi(s)\psi(s)ds \right)
~~~~\mbox{for all}~t \in [t_0,t_1].
\end{equation*}
In particular, $\psi \equiv 0$ holds for $\rho =0$.}

\vskip.1in An immediate application of Lemma 1 is that, it gives a
bound for $\parallel W(t,s;\lambda)\parallel$. Indeed, if
\begin{equation}\label{eq:f_x_bound}
\parallel f_x(t;\phi(t;\lambda))\parallel \leq \chi(t), ~~\mbox{for
all}~~\lambda \in [0,1],
\end{equation}
then taking $\psi(t) = \parallel W(t,s;\lambda)\parallel$ (note that
$\psi(\cdot)$ depends on $s$) and $\rho =
\parallel W(s,s;\lambda) \parallel =
\parallel I \parallel = 1$, we obtain
\begin{equation}\label{eq:W_t_s_norm_bound}
\parallel W(t,s;\lambda) \parallel \leq \exp\left(\int_s^t \chi(u)
du\right), ~~~~\mbox{for all}~ t_0 \leq s < t \leq t_1, ~~\mbox{for
all}~~\lambda \in [0,1].
\end{equation}
Condition (\ref{eq:f_x_bound}) holds in particular if $\parallel
f_x(t,\cdot) \parallel_{\infty} \leq \chi(t)$, and then, from
Theorem F.1, we obtain the important result,

\vskip.1in\noindent{\bf Corollary F.2 :} \textit{If $f$ is such that
$\parallel f_x(t,\cdot) \parallel_{\infty} \leq \chi(t)$ for a
function $\chi(\cdot)$ bounded on $[t_0,t_1]$, and $\parallel \delta
f(t, \cdot) \parallel_\infty \leq \tau(t)$ for some nonnegative
function $\tau(\cdot)$ on $[t_0,t_1]$, then
\begin{equation}\label{eq:solution_perturb_norm_bound}
\parallel \delta x(t) \parallel \leq \int_{t_0}^t \exp\left(\int_s^t \chi(u)
du\right) \tau(s) ds, ~~~~\mbox{for all}~t \in [t_0,t_1].
\end{equation}
}

Note that, even though $M(t,s)$ in (\ref{eq:solution_perturbed}) in
general depends on $x_0$, the bound in
(\ref{eq:solution_perturb_norm_bound}) does not. This has the
implication that if one can prove the existence of solutions
$\{\phi(\cdot;\lambda):\lambda\in [0,1]\}$ on an interval
$[t_0,t_1]$ for an arbitrary collection of initial conditions $x_0$,
and the conditions of Corollary F.2 hold, then the same perturbation
bound (\ref{eq:solution_perturb_norm_bound}) applies uniformly to
each one of them.

\end{document}